%% file: vnpaper.tex
\documentclass[reprint,amsmath,amssymb,aps,prc,showpacs,nofootinbib,floatfix,numerical,twocolumn,preprintnumbers]{revtex4-1}
\usepackage{atlasphysics}
\usepackage{afterpage}
\usepackage{setspace}
\usepackage{rotating}
\usepackage[colorlinks,breaklinks,pdftitle={Measurement of the azimuthal anisotropy for charged particle production in sqrt(s_NN)=2.76 TeV lead-lead collisions with the ATLAS detector},pdfauthor={The ATLAS Collaboration}]{hyperref}
\hypersetup{linkcolor=blue,citecolor=blue,filecolor=black,urlcolor=blue}

\usepackage{preprintcover} 
\PreprintCoverPaperTitle{Measurement of the azimuthal anisotropy for charged particle production in $\sqrt{s_{\mathrm{NN}}}=2.76$ TeV lead-lead collisions with the ATLAS detector}  
\PreprintIdNumber{CERN-PH-EP-2012-035}  
\PreprintCoverAbstract{Differential measurements of charged particle azimuthal anisotropy are presented for lead-lead collisions at $\sqrt{s_{\mathrm{NN}}}=2.76$ TeV with the ATLAS detector at the LHC, based on an integrated luminosity of approximately 8~$\mu\mathrm{b}^{-1}$. This anisotropy is characterized via a Fourier expansion of the distribution of charged particles in azimuthal angle relative to the reaction plane, with the coefficients $v_n$ denoting the magnitude of the anisotropy. Significant $v_2$--$v_6$ values are obtained as a function of transverse momentum ($0.5<\pT<20$ GeV), pseudorapidity ($|\eta|<2.5$) and centrality using an event plane method. The $v_n$ values for $n\ge3$ are found to vary weakly with both $\eta$ and centrality, and their $\pT$ dependencies are found to follow an approximate scaling relation, $v_n^{1/n}(\pT)\propto v_2^{1/2}(\pT)$, except in the top 5\% most central collisions. A Fourier analysis of the charged particle pair distribution in relative azimuthal angle ($\Delta\phi=\phi_{\mathrm{a}}-\phi_{\mathrm{b}}$) is performed to extract the coefficients $v_{n,n}=\langle\cos n\Delta\phi \rangle$. For pairs of charged particles with a large pseudorapidity gap ($|\Delta\eta=\eta_{\mathrm{a}}-\eta_{\mathrm{b}}|>2$) and one particle with $\pT<3$ GeV, the $v_{2,2}$--$v_{6,6}$ values are found to factorize as $v_{n,n}(\pT^{\mathrm a},\pT^{\mathrm b}) \approx v_n(\pT^{\mathrm a})v_n(\pT^{\mathrm b})$ in central and mid-central events. Such factorization suggests that these values of $v_{2,2}$--$v_{6,6}$ are primarily due to the response of the created matter to the fluctuations in the geometry of the initial state. A detailed study shows that the $v_{1,1}(\pT^{\mathrm a},\pT^{\mathrm b})$ data are consistent with the combined contributions from a rapidity-even $v_1$ and global momentum conservation. A two-component fit is used to extract the $v_1$ contribution. The extracted $v_1$ is observed to cross zero at $\pT\approx1.0$ GeV, reaches a maximum at 4--5 GeV with a value comparable to that for $v_3$, and decreases at higher $\pT$.
}  
\PreprintJournalName{Physical Review C}  

\begin{document}
\title{Measurement of the azimuthal anisotropy for charged particle production in $\sqrt{s_{\mathrm{NN}}}=2.76$ TeV lead-lead collisions with the ATLAS detector}
\author{The ATLAS Collaboration}\altaffiliation{Full author list given at the end of the article.}
\begin{abstract}
Differential measurements of charged particle azimuthal anisotropy are presented for lead-lead collisions at $\sqrt{s_{\mathrm{NN}}}=2.76$ TeV with the ATLAS detector at the LHC, based on an integrated luminosity of approximately 8~$\mu\mathrm{b}^{-1}$. This anisotropy is characterized via a Fourier expansion of the distribution of charged particles in azimuthal angle relative to the reaction plane, with the coefficients $v_n$ denoting the magnitude of the anisotropy. Significant $v_2$--$v_6$ values are obtained as a function of transverse momentum ($0.5<\pT<20$ GeV), pseudorapidity ($|\eta|<2.5$) and centrality using an event plane method. The $v_n$ values for $n\ge3$ are found to vary weakly with both $\eta$ and centrality, and their $\pT$ dependencies are found to follow an approximate scaling relation, $v_n^{1/n}(\pT)\propto v_2^{1/2}(\pT)$, except in the top 5\% most central collisions. A Fourier analysis of the charged particle pair distribution in relative azimuthal angle ($\Delta\phi=\phi_{\mathrm{a}}-\phi_{\mathrm{b}}$) is performed to extract the coefficients $v_{n,n}=\langle\cos n\Delta\phi \rangle$. For pairs of charged particles with a large pseudorapidity gap ($|\Delta\eta=\eta_{\mathrm{a}}-\eta_{\mathrm{b}}|>2$) and one particle with $\pT<3$ GeV, the $v_{2,2}$--$v_{6,6}$ values are found to factorize as $v_{n,n}(\pT^{\mathrm a},\pT^{\mathrm b}) \approx v_n(\pT^{\mathrm a})v_n(\pT^{\mathrm b})$ in central and mid-central events. Such factorization suggests that these values of $v_{2,2}$--$v_{6,6}$ are primarily due to the response of the created matter to the fluctuations in the geometry of the initial state. A detailed study shows that the $v_{1,1}(\pT^{\mathrm a},\pT^{\mathrm b})$ data are consistent with the combined contributions from a rapidity-even $v_1$ and global momentum conservation. A two-component fit is used to extract the $v_1$ contribution. The extracted $v_1$ is observed to cross zero at $\pT\approx1.0$ GeV, reaches a maximum at 4--5 GeV with a value comparable to that for $v_3$, and decreases at higher $\pT$.
\end{abstract}
\maketitle
\section{Introduction}
\label{sec:intro}
The primary goal of high-energy heavy ion physics is to understand the properties of the hot and dense matter created in nuclear collisions at facilities such as the Relativistic Heavy Ion Collider (RHIC) and the Large Hadron Collider (LHC). An important observable towards this goal is the azimuthal anisotropy of particle emission. At low $\pT$ ($\lesssim3$--4 GeV), this anisotropy results from a pressure-driven anisotropic expansion of the created matter, with more particles emitted in the direction of the largest pressure gradients~\cite{Ollitrault:1992bk}. At higher $\pT$, this anisotropy is understood to result from the path-length dependent energy loss of jets as they traverse the matter, with more particles emitted in the direction of smallest path-length~\cite{Gyulassy:2000gk}. These directions of maximum emission are strongly correlated, and the observed azimuthal anisotropy is customarily expressed as a Fourier series in azimuthal angle $\phi$~\cite{Voloshin:1994mz,Poskanzer:1998yz}:
\small{
\begin{eqnarray}
\nonumber
E\frac{d^3N}{dp^3} =\frac{d^2N}{2\pi\pT d\pT d\eta}\left(1+2\sum_{n=1}^{\infty}v_n(\pT,\eta) \cos n\left(\phi-\Phi_n\right)\right),\\\label{eq:1}
\end{eqnarray}}\normalsize
where $\pT$ is the transverse momentum, $\eta$ is the pseudorapidity, and $v_n$ and $\Phi_n$ represent the magnitude and direction of the $n^{\mathrm{th}}$-order harmonic, respectively (see Section~\ref{sec:m1}). The $n^{\mathrm{th}}$-order harmonic has $n$-fold periodicity in azimuth, and the coefficients at low $\pT$ are often given descriptive names, such as ``directed flow'' ($v_1$), ``elliptic flow'' ($v_2$), or ``triangular flow'' ($v_3$).

In typical non-central heavy ion collisions where the nuclear overlap region has an ``elliptic'' shape (or quadrupole asymmetry) on average, the azimuthal anisotropy is expected to be dominated by the $v_2$ component~\cite{Voloshin:2008dg,Heinz:2009xj,Teaney:2009qa}. However, it was recently pointed out that the positions of the nucleons in the overlap region can fluctuate to create matter distributions with additional shape components, such as dipole ($n=1$) and sextupole ($n=3$) asymmetries~\cite{Alver:2010gr,Alver:2010dn,Staig:2010pn,Teaney:2010vd}. Due to strong final-state interactions, manifested as either pressure or jet energy loss, these spatial asymmetries can be converted into final-state momentum anisotropies, leading to non-zero first-order and higher-order harmonic coefficients~\cite{Teaney:2010vd,Luzum:2010fb}.  

The observation of large $v_2$ for $\pT\lesssim$ 3--4 GeV at the RHIC~\cite{Ackermann:2000tr,RHIC} and LHC~\cite{Aamodt:2010pa,Collaboration:2011yk} has led to the conclusion that the hot and dense medium behaves like a ``perfect fluid''~\cite{RHIC,Gyulassy:2004vg,shuryakview}. This is because the large $v_2$ values require hydrodynamic models~\cite{Teaney:2003kp,Romatschke:2007mq,Song:2011hk} with a shear viscosity to entropy density ratio that is close to the conjectured lower bound of $1/4\pi$~\cite{Policastro:2001yc,Kovtun:2004de}. Precise determination of this ratio using only $v_2$ data is limited by many model uncertainties~\cite{Song:2008hj}. Because the shear viscosity tends to dampen the harmonics, with more damping for larger $n$~\cite{Teaney:2010vd,Qin:2010pf,Schenke:2011bn}, measurements of harmonic coefficients beyond $v_2$ can provide stronger constraints for the shear viscosity of the medium. Extending these measurements to higher $\pT$ is also valuable for discriminating between jet-quenching models, as high-$\pT$ $v_n$ is sensitive to the path-length dependence of the jet energy loss~\cite{Bass:2008rv,Adare:2010sp,Betz:2011tu}. These coefficients can also help to distinguish between different models of the initial geometry~\cite{Adare:2011tg,Aamodt:2011vk,Qiu:2011hf,Jia:2010ee}, and provide insights into the granularity of the initial state fluctuations~\cite{Staig:2011wj,Schenke:2011bn,Qin:2011uw,Petersen:2012qc}. 

Another related observable for studying the properties of the medium is the correlation function between two particles in relative azimuthal angle $\Delta\phi=\phi_{\mathrm{a}}-\phi_{\mathrm{b}}$ and pseudorapidity $\Delta\eta=\eta_{\mathrm{a}}-\eta_{\mathrm{b}}$~\cite{Adare:2008cqb}. The distribution of pairs in $\Delta\phi$ can be expanded into a Fourier series:
\begin{eqnarray}
\label{eq:2a}
 \frac{dN_{\mathrm{pairs}}}{d\Delta\phi} \propto 1+2\sum_{n=1}^{\infty}v_{n,n}(\pT^{\mathrm{a}},\pT^{\mathrm b}) \cos n\Delta\phi\;,
\end{eqnarray} 
where the coefficients $v_{n,n}$ are symmetric functions with respect to $\pT^{\mathrm{a}}$ and $\pT^{\mathrm b}$. The harmonics defined in Eq.~\ref{eq:1} also contribute to this distribution:
\begin{eqnarray}
\label{eq:2}
\frac{dN_{\mathrm{pairs}}}{d\Delta\phi}\propto 1+2\sum_{n=1}^{\infty}v_n(\pT^{\mathrm{a}})v_n(\pT^{\mathrm b}) \cos n\Delta\phi\;,
\end{eqnarray}
where the global direction $\Phi_n$ drops out in the convolution, and $v_n$ is assumed to be independent of $\eta$ (which is approximately true within $|\eta|<2.5$ at the LHC, see Section~\ref{sec:re1}). Thus if the anisotropy is driven by collective expansion, $v_{n,n}$ should factorize into the product of two single-particle harmonic coefficients~\cite{Adare:2008cqb}:
\begin{eqnarray}
\label{eq:fac}
v_{n,n}(\pT^{\mathrm a},\pT^{\mathrm b}) = v_n(\pT^{\mathrm a})v_n(\pT^{\mathrm b})\;.
\end{eqnarray}
Such factorization may also be valid if the anisotropies of the two particles are independently driven by collective expansion and path-length dependent jet energy loss (both are associated with the same initial spatial asymmetries). This factorization relation has been used to calculate the single-particle $v_n$~\cite{Adcox:2002ms,Aamodt:2011by,CMS:2012wg}. On the other hand, autocorrelations induced by resonance decays or fragmentation of back-to-back jets, are expected to break the factorization. Therefore, Eq.~\ref{eq:fac} can be used to identify the regions of $\pT^{\mathrm {a}}$ and $\pT^{\mathrm b}$ where correlations are dominated by effects controlled by the initial spatial asymmetries.

The study of the structures of two-particle correlation in $\Delta\eta$ and $\Delta\phi$ has been the focus of major experimental and theoretical efforts in the last decade. In typical proton-proton collisions, where a medium is presumably not formed, the pair distributions are dominated by strong correlation peaks at $(\Delta\phi,\Delta\eta)\sim(0,0)$ and $\Delta\phi\sim\pi$. These peaks reflect mainly autocorrelations among particles from fragmentation of back-to-back jets. In heavy ion collisions, additional structures have been observed for $\pT<3$--4 GeV and large $\Delta\eta$ at $\Delta\phi\sim 0$ (known as the ``ridge'')~\cite{Abelev:2009qa,Alver:2009id} and $|\Delta\phi-\pi|\sim1.1$ (known as the ``double-hump'')~\cite{Adler:2005ee,Adare:2008cqb}. These unexpected structures have been interpreted as the response of the medium to the energy deposited by quenched jets~\cite{CasalderreySolana:2004qm,Armesto:2004pt}. However, similar structures can also be generated by the flow harmonics, as they all contribute constructively at $\Delta\phi\sim0$ but tend to cancel on the away-side according to Eq.~\ref{eq:2}~\cite{Alver:2010gr}. Therefore, a detailed comparison between the measured pair distribution (Eq.~\ref{eq:2a}) and that expected from anisotropic flow (Eq.~\ref{eq:2}) can determine whether the structures in two-particle correlations are a consequence of the so-called ``jet-induced medium response'', or whether they are a consequence of a sum of the flow harmonics.

The $v_2$ coefficient has been extensively studied at the RHIC~\cite{Ackermann:2000tr,Abelev:2008ed,Alver:2006wh,Adams:2003zg,Adare:2010ux,Adare:2010sp} and LHC~\cite{Aamodt:2010pa,Collaboration:2011yk}. Results for higher-order $v_n$ for $n\ge3$ also became available recently~\cite{Adare:2011tg,Aamodt:2011vk,CMS:2012wg}. In contrast, no experimental measurement of $v_1$ including systematic uncertainties exists at the LHC, although an estimate has recently been performed by a theoretical group~\cite{Retinskaya:2012ky} based on published ALICE data~\cite{Aamodt:2011by}. A primary complication for $v_1$ measurements is global momentum conservation, which induces a significant dipole component~\cite{Borghini:2000cm,Borghini:2002mv}. A ``sideward'' deflection of colliding ions can also lead to a small rapidity-odd (i.e. changes sign crossing $\eta=0$) dipole component~\cite{Back:2005pc,Abelev:2008jga}. Therefore, the extraction of $v_1$ values associated with the initial dipole asymmetry requires careful separation of these contributions, which generally break the factorization relation given by Eq.~\ref{eq:fac}.

This paper presents comprehensive results for $v_1$--$v_6$ over broad ranges of centrality, pseudorapidity and $\pT$ for charged particles in lead-lead (Pb-Pb) collisions at $\sqrt{s_{\mathrm{NN}}}=2.76$ TeV with the ATLAS detector at the LHC. The $v_n$ values are measured directly using an ``event plane'' (EP) method for $n=2$--6, and are also derived from the $v_{n,n}$ measured using a two-particle correlation (2PC) method for $n=1$--6. These detailed measurements provide new insights into the hydrodynamic picture at low $\pT$, the jet energy loss picture at high $\pT$, and the nature of the fluctuations in the initial geometry. They also allow a detailed study of the factorization relation (Eq.~\ref{eq:fac}) over broad ranges of centrality, $\Delta\eta$, $\pT^{\mathrm a}$ and $\pT^{\mathrm b}$. Together, these measurements should shed light on the physics underlying the structures observed in two-particle correlation functions.

The paper is organized as follows. Sections~\ref{sec:det} and \ref{sec:sel} give a brief overview of the ATLAS detector, trigger, and selection criteria for events and tracks. Section~\ref{sec:m} discusses the details of the EP method and the 2PC method used to measure the $v_n$. Section~\ref{sec:re1} presents results for $v_2$--$v_6$ from the EP method as a function of $\pT$, $\eta$ and centrality. Section~\ref{sec:re2} presents a detailed Fourier analysis of the two-particle correlation functions to measure $v_{n,n}$ as a function of $\pT^{\mathrm a}$, $\pT^{\mathrm b}$, $\Delta\eta$ and centrality, which are then used to calculate $v_2$--$v_6$ via the factorization relation (Eq.~\ref{eq:fac}). These $v_n$ values are compared with those obtained from the EP method in Section~\ref{sec:re3}, with a focus on understanding the structures of the 2PC in terms of single-particle $v_n$. Section~\ref{sec:re4} presents results for $v_1$ based on a two-component fit of the $v_{1,1}$ data with a modified functional form of Eq.~\ref{eq:fac} that includes the contribution of global momentum conservation. Section~\ref{sec:con} gives a summary of the results and main observations.

\section{ATLAS Detector and trigger}
\label{sec:det}

The ATLAS detector~\cite{Aad:2008zzm} provides nearly full solid angle coverage of the collision point with tracking detectors, calorimeters and muon chambers, well suited for measurements of azimuthal anisotropies over a large pseudorapidity range~\footnote{ATLAS uses a right-handed coordinate system with its origin at the nominal interaction point (IP) in the center of the detector and the $z$-axis along the beam pipe. The $x$-axis points from the IP to the center of the LHC ring, and the $y$-axis points upward. Cylindrical coordinates $(r,\phi)$ are used in the transverse plane, $\phi$ being the azimuthal angle around the beam pipe. The pseudorapidity is defined in terms of the polar angle $\theta$ as $\eta=-\ln\tan(\theta/2)$.}. This analysis primarily uses three subsystems for $v_n$ measurement: the inner detector (ID), the barrel and endcap electromagnetic calorimeters (ECal) and the forward calorimeter (FCal). The ID is contained within the 2~T field of a superconducting solenoid magnet, and measures the trajectories of charged particles in the pseudorapidity range $|\eta|<2.5$ and over the full azimuth. A charged particle passing through the ID typically traverses three modules of the silicon pixel detector (Pixel), four double-sided silicon strip modules of the semiconductor tracker (SCT) and, for $|\eta|<2$, a transition radiation tracker composed of straw tubes. The electromagnetic energy measurement of the ECal is based on a liquid-argon sampling technology. The ECal covers the pseudorapidity range $|\eta|<3.2$, and is used as a reference detector in the event plane measurements. The FCal consists of three longitudinal sampling layers and extends the calorimeter coverage to $|\eta|< 4.9$. It uses tungsten and copper absorbers with liquid argon as the active medium, and has a total thickness of about 10 interaction lengths. The centrality measurement uses towers in all three layers of the FCal, while the event plane measurements use towers in the first two layers of the FCal excluding those at the edge of the FCal $\eta$ acceptance. These selection criteria are found to minimize the effect of fluctuations in the reaction plane measurement.

The minimum-bias Level-1 trigger used for this analysis requires signals in two zero-degree calorimeters (ZDC), each positioned at 140~m from the collision point, detecting neutrons and photons with $|\eta|>8.3$, or either one of the two minimum-bias trigger scintillator (MBTS) counters, covering $2.1<|\eta|<3.9$ on each side of the nominal interaction point. The ZDC Level-1 trigger thresholds on each side are set below the peak corresponding to a single neutron, e.g. as produced from Coulomb dissociation of the lead ion~\cite{Chiu:2001ij}. A Level-2 timing requirement based on signals from each side of the MBTS is imposed to remove beam backgrounds.

\section{Event and track selections}
\label{sec:sel}
This paper is based on approximately 8~$\mu\mathrm{b}^{-1}$ of Pb-Pb data collected in 2010 at the LHC with a nucleon-nucleon center-of-mass energy $\sqrt{s_{\mathrm{NN}}}=2.76$~TeV. An offline event selection requires a reconstructed vertex and a time difference $|\Delta t| < 3$ ns between the MBTS trigger counters on either side of the interaction point to suppress non-collision backgrounds. A coincidence between the ZDCs at forward and backward pseudorapidity is required to reject a variety of background processes, while maintaining high efficiency for non-Coulomb processes.

Events satisfying these conditions are then required to have a reconstructed primary vertex within $|z_{\mathrm{vtx}}|<150$~mm of the nominal center of the ATLAS detector for the EP analysis. A more stringent vertex cut of 100~mm is required for the 2PC analysis, such that enough events can be found in the same $z_{\mathrm{vtx}}$ bin for the event mixing procedure (see discussion in Section~\ref{sec:m2}). About 48 million and 43 million events pass the requirements for the EP and 2PC analysis, respectively. Pile-up probability is estimated to be at the $10^{-4}$ level and is therefore negligible.

The Pb-Pb event centrality is characterized using the total transverse energy ($\sum \eT$) deposited in the FCal over the pseudorapidity range $3.2 < |\eta| < 4.9$ at the electromagnetic energy scale. An analysis of this distribution after all trigger and event selections gives an estimate of the fraction of the sampled non-Coulomb inelastic cross-section to be $98\pm2$\%~\cite{Collaboration:2011yr}. This estimate is obtained from a shape analysis of the measured FCal $\sum \eT$ distributions compared with a convolution of proton-proton data with a Monte Carlo Glauber calculation~\cite{Miller:2007ri}. The FCal $\sum \eT$ distribution is then divided into a set of 5\% or 10\% percentile bins, together with a bin defined for the 1\% most central events. The uncertainty associated with the centrality definition is evaluated by varying the effect of trigger and event selection inefficiencies as well as background rejection requirements in the most peripheral FCal $\sum \eT$ interval~\cite{Collaboration:2011yr}.

Tracks are reconstructed within the full acceptance of the ID, requiring $\pT>0.5$ GeV and $|\eta|<2.5$. To improve the reliability of track reconstruction in the high-multiplicity environment of heavy ion collisions, more stringent requirements on track quality, compared to those defined for proton-proton collisions~\cite{Aad:2010rd}, are used. At least nine hits in the silicon detectors (out of a typical value of 11) are required for each track, with no missing Pixel hits and not more than one missing SCT hit, in both cases where such hits are expected. In addition, the point of closest approach is required to be within 1~mm of the primary vertex in both the transverse and longitudinal directions~\cite{Collaboration:2011yk}. This selection is varied in the analysis to check the influence of both the acceptance and fake tracks. The tracking efficiency for charged particles is studied by comparing data to Monte Carlo calculations based on the HIJING event generator~\cite{Gyulassy:1994ew} and a full GEANT4~\cite{Agostinelli:2002hh} simulation of the detector. This efficiency is estimated to be about 72\% near mid-rapidity in central events. However, this analysis is found to be insensitive to variations in the tracking efficiency, as found previously~\cite{Collaboration:2011yk}. Fake tracks from random combinations of hits are generally negligible, reaching only 0.1\% for $|\eta|<1$ for the highest multiplicity events. This rate increases slightly at large $\eta$.
\section{Data analysis}
\label{sec:m}
Equation~\ref{eq:1} implies that each harmonic component of the final-state momentum distribution is represented by its magnitude $v_n$ and azimuthal direction $\Phi_{n}$. In general, any distribution can be expanded into a set of Fourier components. However, the distinguishing feature of correlation due to initial geometry, as opposed to other sources of correlations, is that it is a ``global'' correlation. That is, $\Phi_n$ specifies a common direction, independent of the particle species, $\pT$ and $\eta$, and it drops out in the two-particle correlations (Eq.~\ref{eq:2}). This feature is quite different from the correlations expected from jet fragmentation or resonance decays, which typically involve a subset of particles correlated over a finite range in $\Delta\eta$ with no preferred global direction. Thus, $v_n$ can be measured either by correlating tracks with the $\Phi_{n}$ estimated in the forward direction, or it can be measured from two-particle correlations with a large $\Delta\eta$ gap. In the following, the details of these two methods are discussed.
\subsection{Event plane method}
\label{sec:m1}
The azimuthal angle $\Phi_{n}$ and the beam direction define the $n^{\mathrm{th}}$-order reaction plane~\footnote {If the shape of Pb nuclei is approximated by the smooth Woods-Saxon function without fluctuations, $\Phi_{n}$ coincides with the azimuthal angle of the reaction plane defined by the beam axis and the impact parameter (the vector separating the barycenters of the two nuclei).}. However, due to incomplete detector acceptance and finite event multiplicity, the true reaction plane angle $\Phi_{n}$ can not be determined. Instead it is approximated by the event plane angle $\Psi_n$, which is defined as the direction of the ``flow vector'' $\overrightarrow{Q}_n$, calculated in this analysis from the $\eT$ deposited in the FCal towers in each event:
\begin{eqnarray}
\label{eq:mep0}
\nonumber
 \overrightarrow{Q}_n &=&(Q_{x,n},Q_{y,n})\\\nonumber
&=&\left(\vphantom{A^{A^A}_{A_A}}\textstyle\sum \eT\cos n\phi-\langle\sum \eT \cos n\phi\rangle,\right.\\\nonumber \textstyle\;\;\;\;\;\;\;\;\; &&\left.\vphantom{A^{A^A}_{A_A}}\textstyle\sum \eT \sin n\phi-\langle\sum \eT \sin n\phi\rangle\right)\;, \\ \tan n\Psi_n &=& \frac{Q_{y,n}}{Q_{x,n}}\;,
\end{eqnarray}
where the sum ranges over towers in the first two layers of the FCal (see Section~\ref{sec:det}). Subtraction of the event-averaged centroid removes biases due to detector effects~\cite{Afanasiev:2009wq}. A standard flattening technique is then used to remove the residual non-uniformities in the event plane angular distribution~\cite{Barrette:1997pt}. These calibration procedures are similar to those used by the RHIC experiments~\cite{Afanasiev:2009wq,Back:2004mh}.

The coefficient $v_n$ is measured by correlating tracks with $\Psi_n$ to obtain the raw values $v_n^{\mathrm{obs}}=\langle \cos n\left(\phi-\Psi_n\right)\rangle$. The value of $v_n^{\mathrm{obs}}$ is then corrected by a resolution factor that accounts for the dispersion of $\Psi_n$ about $\Phi_{n}$~\cite{Poskanzer:1998yz}:
\begin{eqnarray}
\label{eq:mep1}
v_n&=&\frac{v_n^{\mathrm{obs}}}{\mathrm{Res}\{n\Psi_{n}\}}=  \frac{\langle \cos n\left(\phi-\Psi_n\right)\rangle}{\langle \cos n\left(\Psi_n-\Phi_{n}\right)\rangle}\;,
\end{eqnarray}
where the average is performed over all events for the denominator and all tracks and all events for the numerator. The EP resolution of the FCal, ${\mathrm{Res}\{n\Psi_{n}\}}$, is~\cite{Poskanzer:1998yz}:
\begin{eqnarray}
\nonumber
{\mathrm{Res}\{n\Psi_{n}\}}&=&\left\langle {\cos n (\Psi_n-\Phi_{n}) } \right\rangle \\
&=&\frac{{\chi_n\sqrt \pi }}{2} e^ {- \frac{{\chi_n^2 }}{2}}\left[\vphantom{\frac{\chi_n^{A^{A}} }{A_A}} {I_{0} (\frac{{\chi_n^2 }}{2}) + I_{1} (\frac{{\chi_n^2 }}{2})} \right]\;,\label{eq:mep2}
\end{eqnarray}
{\parindent0ex where $I_{\alpha}$ are the modified Bessel functions of the first kind, and $\chi_n$ (known as the ``resolution parameter'') is the fundamental variable that quantifies the precision of a detector for determining the event plane. The value of $\chi_n$ is proportional to the $\eT$-weighted harmonic coefficient $v_n^{\mathrm{FCal}}$ and the square-root of the total multiplicity $M$ in the FCal acceptance~\cite{Poskanzer:1998yz}:}
\begin{eqnarray}
\label{eq:mep2b}
\chi_n\propto v_n^{\mathrm{FCal}}\sqrt{M}\;.
\end{eqnarray}

The values of $\chi_n$ and ${\mathrm{Res}\{n\Psi_{n}\}}$ are obtained from a two-subevents method (2SE) and a three-subevents method (3SE)~\cite{Poskanzer:1998yz}. In the 2SE method, the signal of a detector used to measure the event plane is divided into two ``subevents'' covering equal pseudorapidity ranges in opposite hemispheres, such that the two subevents nominally have the same resolution. The FCal detectors located at positive and negative $\eta$, FCal$_{\mathrm P}$ and FCal$_{\mathrm N}$, provide such a division. The resolution of each FCal subevent is calculated directly from the correlation between the two subevents:
\small{
\begin{eqnarray}
\label{eq:mep3}
{\mathrm{Res}\{n\Psi_{n}^{\mathrm{P(N)}}\}}=\langle {\cos n(\Psi_n^{\mathrm{P(N)}}-\Phi_{n})}\rangle = \sqrt{\langle {\cos n(\Psi_n^{\mathrm P}-\Psi_n^{\mathrm N})}\rangle}.
\end{eqnarray}}\normalsize
The resolution parameter of the FCal subevent $\chi_{n,\mathrm{sub}}$ is determined by inverting Eq.~\ref{eq:mep2}. The resolution parameter for the full FCal is $\chi_n=\sqrt{2}\chi_{n,\mathrm{sub}}$, with $\sqrt{2}$ accounting for a factor of two increase in the total multiplicity (Eq.~\ref{eq:mep2b}). Finally, $\chi_n$ is incorporated into Eq.~\ref{eq:mep2} to obtain the resolution for the full FCal.

In the 3SE method, the ${\mathrm{Res}\{n\Psi_{n}\}}$ value for a given subevent A is determined from its correlations with two subevents B and C covering different regions in $\eta$:
\begin{eqnarray}
\nonumber
{\mathrm{Res}}\{n\Psi^{\mathrm A}_{n}\}=\sqrt{\frac{\left\langle {\cos n \left(\Psi_n^{\mathrm A} - \Psi_n^{\mathrm B}\right)} \right\rangle\left\langle {\cos n \left(\Psi_n^{\mathrm A} - \Psi_n^{\mathrm C}\right)} \right\rangle}{\left\langle {\cos n \left(\Psi_n^{\mathrm B} - \Psi_n^{\mathrm C}\right)} \right\rangle}}.\\\label{eq:mep4}
\end{eqnarray}
The large $\eta$ coverage of the ID and ECal, with their fine segmentation, allows for many choices for subevents B and C. The ID and ECal are divided into a set of 22 reference subevents each covering 0.5 units in $\eta$. The subevents B and C are chosen to ensure a minimum separation in $\eta$ of 1 unit between all three subevents. This separation in $\eta$ is required to suppress short range correlations~\cite{Afanasiev:2009wq}. Various 3SE combinations are studied to check the sensitivity to the size of the chosen pseudorapidity gaps, as well as potential systematic effects due to the explicit use of the correlation between FCal$_{\mathrm P}$ and FCal$_{\mathrm N}$ in the 2SE method.

\begin{figure}[t]
\begin{center}
\includegraphics[width=0.85\columnwidth]{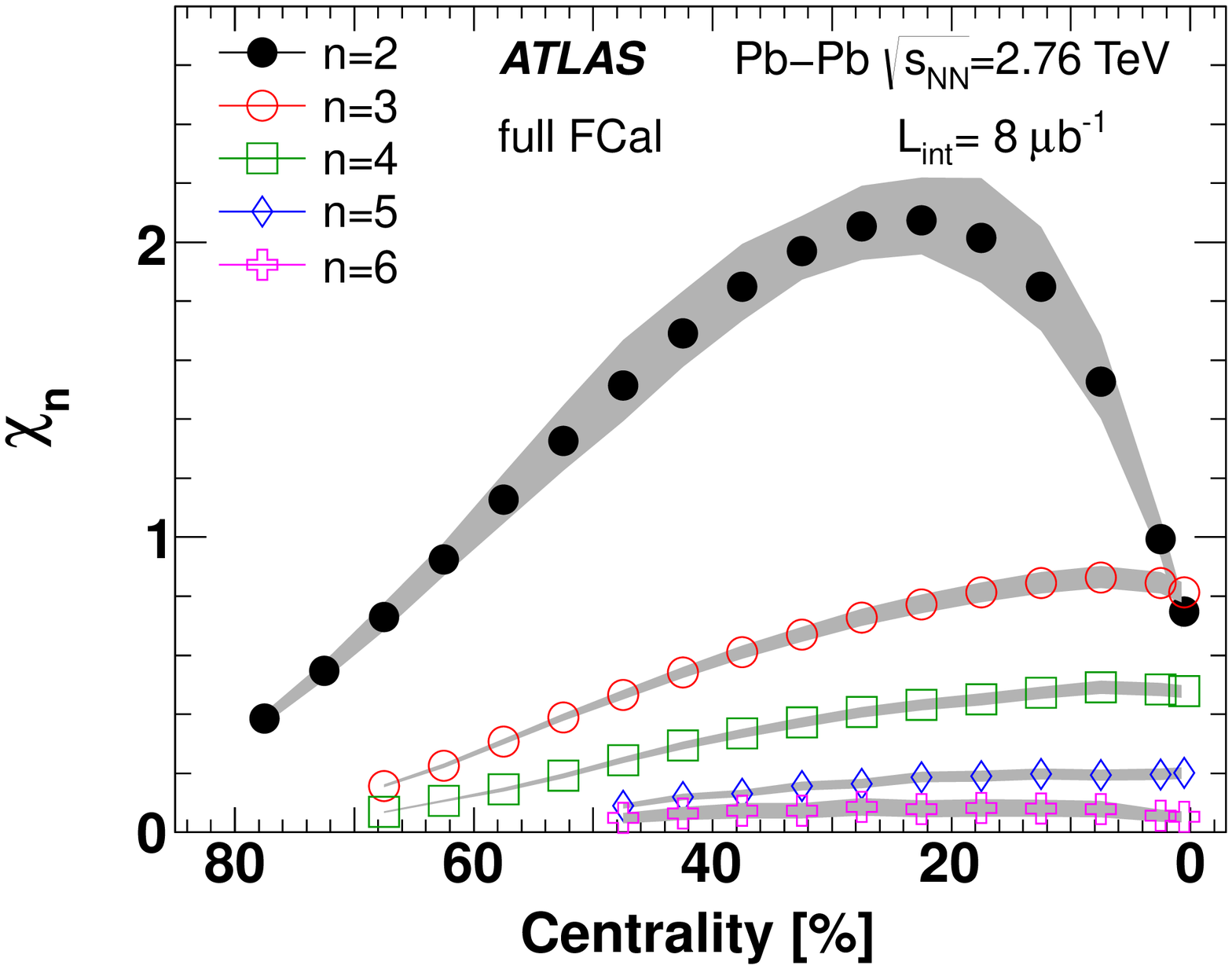}
\includegraphics[width=0.85\columnwidth]{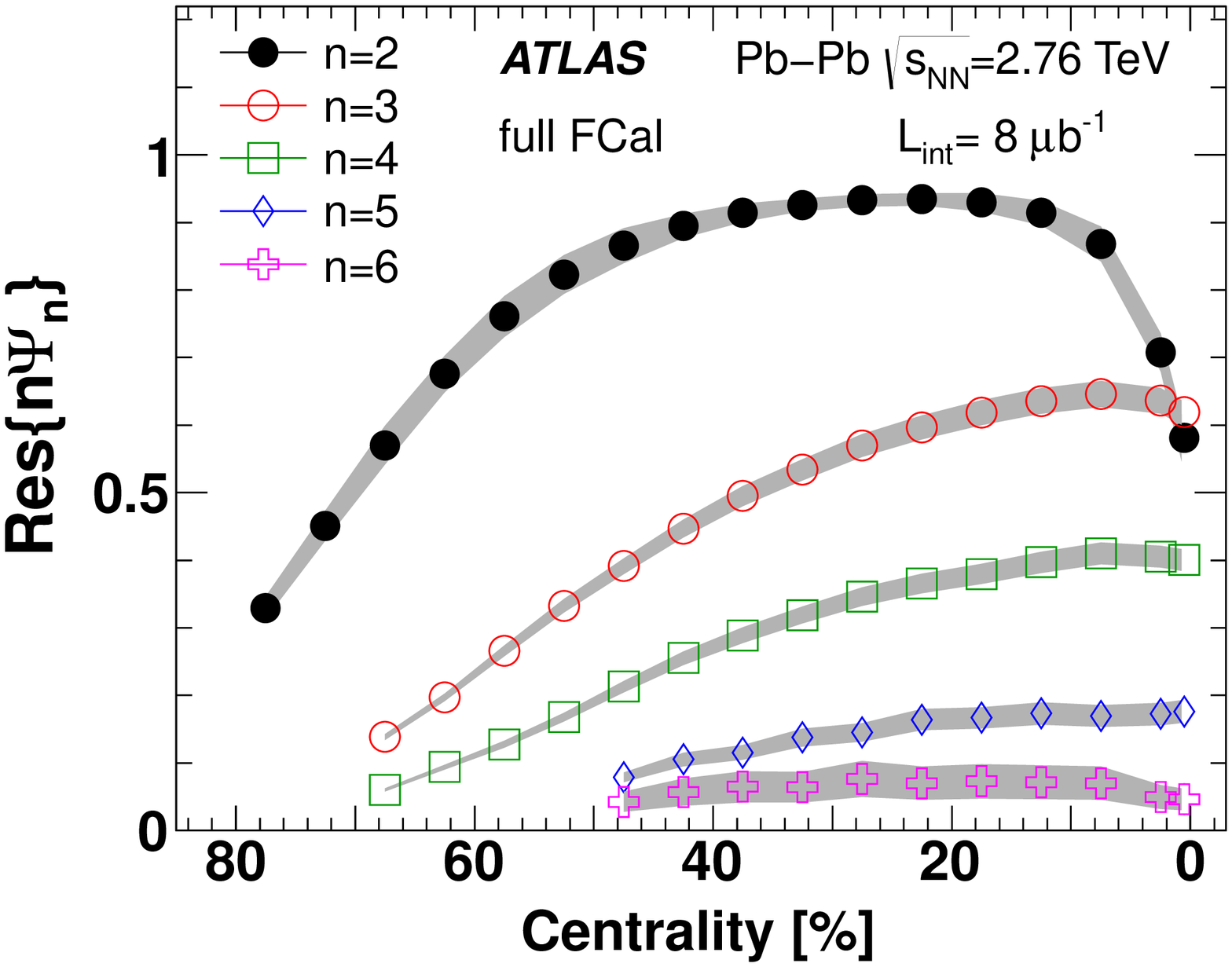}
\end{center}
\caption{\label{fig:reso1} (Color online) The $\chi_n$ (top) and EP resolution factor (bottom) vs. centrality (smaller value refers to more central events) for $n=2$--6, together with the systematic uncertainty as shaded bands. The EP is measured by both sides of the FCal detector (denote by ``full FCal''). Note that the Res$\{n\Psi_n\}$ value can not be greater than one (see Eq.~\ref{eq:mep2}), thus its systematic uncertainty shrinks as it approaches one.}
\end{figure}

Figure~\ref{fig:reso1} shows the values of $\chi_n$ and ${\mathrm{Res}}\{n\Psi_{n}\}$ measured as a function of centrality for $n=2$--6 using the full FCal. The data points and associated statistical uncertainties are calculated using the 2SE method. However, 5\% upward and 15\% downward centrality-independent corrections are applied to $n=5$ and $n=6$ respectively, to adjust to the average of the 2SE and the 3SE estimates for the ${\mathrm{Res}}\{n\Psi_{n}\}$. The differences between the two estimates are quoted as systematic uncertainties for ${\mathrm{Res}}\{n\Psi_{n}\}$, and they are propagated via Eq.~\ref{eq:mep2} to obtain a systematic uncertainty for $\chi_n$. In this analysis, the centrality range for each harmonic $n$ is chosen such that the relative statistical uncertainty for ${\mathrm{Res}}\{n\Psi_{n}\}$ is less than 30\% of its mean value, and the 2SE and 3SE estimations show good agreement. They are 0--80\% for $v_2$, 0--70\% for $v_3$ and $v_4$, 0--50\% for $v_5$ and $v_6$ as indicated in Fig.~\ref{fig:reso1}.

In the event plane analysis, two complementary methods are employed to measure $v_n$. The first (``full FCal'') method calculates $v_n^{\mathrm{obs}}$ by correlating tracks in the ID with the EP from the full FCal detector; the resolution correction for the full FCal is then applied to obtain the final $v_n$ (Eq.~\ref{eq:mep1}). In the second (``FCal subevent'' or simply FCal$_{\mathrm{P(N)}}$) method~\cite{Collaboration:2011yk}, tracks with $\eta\ge0$ ($\eta<0$) are correlated with the EP in the opposite hemisphere given by the FCal$_{\mathrm N}$ (FCal$_{\mathrm P}$). The resolution correction for the FCal$_{\mathrm{P(N)}}$ is then applied to obtain the final $v_n$. Note that the relative statistical and systematic uncertainties for ${\mathrm{Res}\{n\Psi_{n}\}}$ are almost identical for the two methods since they both rely on similar subevent correlations (Eqs.~\ref{eq:mep3} and \ref{eq:mep4}). However, the values of $v_n^{\mathrm {obs}}$ from the FCal subevent method are smaller than those from the full FCal method due to its poorer EP resolution. Therefore, these $v_n^{\mathrm {obs}}$ values have a larger fractional statistical uncertainty. The primary advantage of the FCal subevent method is that it increases the minimum (maximum) pseudorapidity separation between the track and the EP from about 0.8 (4.8) units for the full FCal method to about 3.3 (7.3) units. Thus the subevent approach is less affected by short range autocorrelations, stemming primarily from jet fragmentation and resonance decays. In this analysis, the FCal$_{\mathrm{P(N)}}$ method is used for the $\eta$ dependence of $v_n$ to minimize short range correlations, while the full FCal method is used for the $\pT$ and centrality dependence of $v_n$ to optimize the EP resolution (see Section~\ref{sec:re1}). However, the potential influence of short range correlations on the full FCal method is cross-checked with the FCal$_{\mathrm{P(N)}}$ method. Good agreements are always observed for $\eta$-integrated $v_n$, within the systematic uncertainties for the two methods.

The systematic uncertainty in $v_n^{\mathrm{obs}}$ is determined by varying the track quality cuts, comparing data for different running periods, varying the full centrality range by $\pm$ 2\% according to the uncertainty in the trigger and event selections, as well as by determining the value of $\langle\sin n(\phi-\Psi_n)\rangle$. The study of track quality cuts accounts for influences of background contaminations and tracking inefficiency. Finite sine terms can arise from detector biases that introduce a correlation between the ID and the FCal. Their magnitudes relative to the cosine terms are included in the uncertainty for $v_n^{\mathrm{obs}}$. All these uncertainties are generally small, except for $n=6$. They are also quite similar for the full FCal and FCal subevent methods, so the larger of the two is quoted as the main uncertainty. As a cross-check, $v_2$--$v_6$ values are also extracted using the EP measured either for the three layers of FCal individually or for two $\eta$ regions of FCal ($3.3<|\eta|<4.0$ and $4.0<|\eta|<4.8$). Although these five FCal subevents have up to a factor of four difference in their resolution corrections, the measured $v_2$--$v_6$ all agree to within 2\%--10\%.

Tables~\ref{tab:v2}-\ref{tab:v6} summarize the systematic uncertainties for $v_2$--$v_6$ in various centrality intervals. The total uncertainties are calculated as the quadrature sum of all sources in these tables. In most cases, they are specified for multiple 5\% wide centrality intervals. For example, 0--20\% in Tables~\ref{tab:v2}-\ref{tab:v6} refers to four bins: 0--5\%, 5--10\%, 10--15\%, and 15--20\%; a ``5.0--2.0'' notation indicates the values of relative systematic uncertainty in percentage at the beginning and at the end of the 0--20\% centrality interval. Tables~\ref{tab:v5} and \ref{tab:v6} also quote the uncertainty for the 0--1\% centrality interval, which generally has the same systematic uncertainty as that for the 0--5\% centrality interval, but with a larger statistical uncertainty. The systematic uncertainties only include those associated with the measurements themselves, and no attempt is made to disentangle potential contributions from various sources of autocorrelations, as their exact origin and quantitative effects on $v_n$ are not fully understood~\cite{Voloshin:2008dg}. Nevertheless, these autocorrelations should be largely suppressed by the large average $\eta$ gap between the ID and the detector used for determining the EP. 

\begin{table}[!h]
\begin{center}
\small{
\begin{ruledtabular}\begin{tabular}{l | c c c c}
Centrality        &0--20\%  & 20--50\% & 50--70\% & 70--80\% \tabularnewline\hline
Resolution[\%]        &5.0--2.0   & 1.0--2.0& 3.0--4.0& 4.0--6.0 \tabularnewline\hline
Track selection[\%]   & 2.0    & 0.5   & 0.5   & 1.0  \tabularnewline\hline 
Residual sine term[\%]&0.8   & 0.6    & 0.5    &0.2\tabularnewline\hline 
Running periods[\%] &0.2     &0.2       &0.5     &1.0\tabularnewline\hline 
Trigger \& event sel.[\%] &1.0      &1.0--0.5       &1.0     &1.5\tabularnewline\hline
Total[\%]      &5.6--3.2  &1.4--2.3 & 3.4--4.2 &4.6--6.4\tabularnewline
\end{tabular}\end{ruledtabular}}\normalsize
\end{center}
\caption{\label{tab:v2} Summary of relative systematic uncertainties in percentage for $v_2$ for both full FCal and FCal$_{\mathrm{P(N)}}$. See text for explanation of the arrangement of the uncertainties.}
\end{table}

\begin{table}[!h]
\begin{center}
\small{\begin{ruledtabular}\begin{tabular}{l | c c c}
Centrality[\%]         &0--20\%  & 20--50\% & 50--70\% \tabularnewline\hline
Resolution[\%]         &3.0     & 3.0     &  3.0--5.6 \tabularnewline\hline
Track selection[\%] &2.0    & 0.5   & 0.5--2.0   \tabularnewline\hline
Residual sine term[\%]   &1.0     & 1.0  & 1.5    \tabularnewline\hline 
Running periods[\%]  &0.5    &0.5--1.5   &2.0     \tabularnewline\hline 
Trigger \& event sel.[\%]   &0.4  &0.5--1.0   &1.5--3.5\tabularnewline\hline
Total[\%]  &3.8   & 3.5--3.9   &4.6--7.4  \tabularnewline
\end{tabular}\end{ruledtabular}}\normalsize
\end{center}
\caption{\label{tab:v3} Summary of relative systematic uncertainties in percentage for $v_3$ for both full FCal and FCal$_{\mathrm{P(N)}}$.}
\end{table}
\begin{table}[!h]
\begin{center}
\small{\begin{ruledtabular}\begin{tabular}{l | c c c } 
Centrality               &0--20\%  & 20--50\% & 50--70\% \tabularnewline\hline
Resolution[\%]              &4.0&4.0   &  4.4--16.0 \tabularnewline\hline
Track selection[\%]   &1.0     & 1.0--2.0    & 4.0  \tabularnewline\hline
Residual sine term[\%]       &2.0     & 2.0      & 3.0--5.0    \tabularnewline\hline 
Running periods[\%]     &1.0     &1.5--2.0      & 4.0   \tabularnewline\hline 
Trigger \& event sel.[\%] &0.6   &0.7    &1.0--2.0\tabularnewline \hline
Total[\%]                    &4.9   &4.9--5.4    &7.9--17.5\tabularnewline
\end{tabular}\end{ruledtabular}}\normalsize
\caption{\label{tab:v4} Summary of relative systematic uncertainties in percentage for $v_4$ for both full FCal and FCal$_{\mathrm{P(N)}}$.}
\end{center}
\end{table}
\begin{table}[!h]
\begin{center}
\small{\begin{ruledtabular}\begin{tabular}{l | c c c c }
Centrality            &0--1\% &0--20\%  & 20--40\% & 40--50\%  \tabularnewline\hline
Resolution[\%]            &10.8 & 10.2    & 10.2--10.4  &11.2--22.4 \tabularnewline\hline
Track selection[\%]    &1.0 & 1.0    & 1.0   &  2.0   \tabularnewline\hline
Residual sine term[\%] & \multicolumn{4}{c}{5.0}\tabularnewline\hline
Running periods[\%]  &2.0&2.0     &2.0   &4.0  \tabularnewline\hline 
Trigger \& event sel.[\%]   &\multicolumn{4}{c}{1.0}\tabularnewline \hline
Total[\%]               & 12.1 & 11.6   & 11.6--12.1    & 13.0--23.0 \tabularnewline
\end{tabular}\end{ruledtabular}}\normalsize
\end{center}
\caption{\label{tab:v5} Summary of relative systematic uncertainties in percentage for $v_5$ for both full FCal and FCal$_{\mathrm{P(N)}}$.}
\end{table}
\begin{table}[!h]
\begin{center}
\small{\begin{ruledtabular}\begin{tabular}{l | c c c c } 
Centrality         &0--1\% &0--20\%   & 20--40\% & 40--50\%  \tabularnewline\hline
Resolution[\%]    & 58&34--31   & 31  &32--38 \tabularnewline\hline
Track selection[\%] &\multicolumn{4}{c}{10}  \tabularnewline\hline
Residual sine term[\%]&\multicolumn{4}{c}{10}   \tabularnewline\hline 
Running periods[\%]&\multicolumn{4}{c}{10}  \tabularnewline\hline 
Trigger \& event sel.[\%]  &\multicolumn{4}{c}{1 }  \tabularnewline\hline
Total[\%]             &61  & 38--35& 36 & 37--42   \tabularnewline
\end{tabular}\end{ruledtabular}}\normalsize
\end{center}
\caption{\label{tab:v6} Summary of relative systematic uncertainties in percentage for $v_6$ for both full FCal and FCal$_{\mathrm{P(N)}}$.}
\end{table}

\subsection{Two-particle correlation method}
\label{sec:m2}

The two-particle correlation function is generally defined as the ratio of the same-event pair (foreground) distribution to the combinatorial pair (background) distribution in two-particle phase space $(\phi_{\mathrm a},\phi_{\mathrm b},\eta_{\mathrm a},\eta_{\mathrm b})$:
\begin{eqnarray} 
C(\phi_{\mathrm a},\phi_{\mathrm b},\eta_{\mathrm a},\eta_{\mathrm b}) = \frac{\frac{d^4N}{d\phi_{\mathrm a}d\eta_{\mathrm a}d\phi_{\mathrm b}d\eta_{\mathrm b}}}{\frac{d^2N}{d\phi_{\mathrm a}d\eta_{\mathrm a}}\times\frac{d^2N}{d\phi_{\mathrm b}d\eta_{\mathrm b}}}\;.
\end{eqnarray}
In practice, the correlation function is usually studied as a function of relative azimuthal angle ($\Delta\phi$) and relative pseudorapidity ($\Delta\eta$), by averaging pair distributions over the detector acceptance:
\begin{eqnarray} 
C(\Delta\phi,\Delta\eta) = \frac{S(\Delta\phi,\Delta\eta)}{B(\Delta\phi,\Delta\eta)}\;,
\end{eqnarray}
where
\begin{eqnarray} 
S(\Delta\phi,\Delta\eta) =\int d\phi_{\mathrm a}d\eta_{\mathrm a}d\phi_{\mathrm b}d\eta_{\mathrm b}\delta_{\mathrm {ab}}\frac{d^4N}{d\phi_{\mathrm a}d\eta_{\mathrm a}d\phi_{\mathrm b}d\eta_{\mathrm b}}\;,\\
B(\Delta\phi,\Delta\eta) =\int d\phi_{\mathrm a}d\eta_{\mathrm a}d\phi_{\mathrm b}d\eta_{\mathrm b}\delta_{\mathrm {ab}}\frac{d^2N}{d\phi_{\mathrm a}d\eta_{\mathrm a}}\frac{d^2N}{d\phi_{\mathrm b}d\eta_{\mathrm b}}\;.
\end{eqnarray}
The $\delta_{\mathrm {ab}}$ is a shorthand notation for $\delta(\phi_{\mathrm a}-\phi_{\mathrm b}-\Delta\phi)\delta(\eta_{\mathrm a}-\eta_{\mathrm b}-\Delta\eta)$. 

For an ideal detector, the combinatorial pair distribution is uniform in $\Delta\phi$, and has a nearly triangular shape in $\Delta\eta$ due to the weak dependence of the single-particle distribution on $\eta$~\cite{Collaboration:2011yr}. In reality, both same-event and combinatorial pair distributions are modulated by detector inefficiencies and non-uniformity. These detector effects influence the two distributions in the same way so they cancel in the ratio. Therefore, $B(\Delta\phi,\Delta\eta)$ is often referred to as the pair acceptance function~\cite{Adare:2008cqb}. In this analysis, $B(\Delta\phi,\Delta\eta)$ is estimated from track pairs from two events with similar centrality (matched within 5\%) and $z_{\mathrm{vtx}}$ (matched within 1 mm). The two particles in the pair are typically selected with different conditions, such as different $\pT$ ranges, pseudorapidities and charge signs. In this analysis, charged particles measured by the ID with a pair acceptance extending up to $|\Delta\eta|=5$ are used.

\begin{figure*}[!t]
\begin{tabular}{lr}
\begin{minipage}{0.58\linewidth}
\begin{flushleft}
\includegraphics[width=0.98\linewidth]{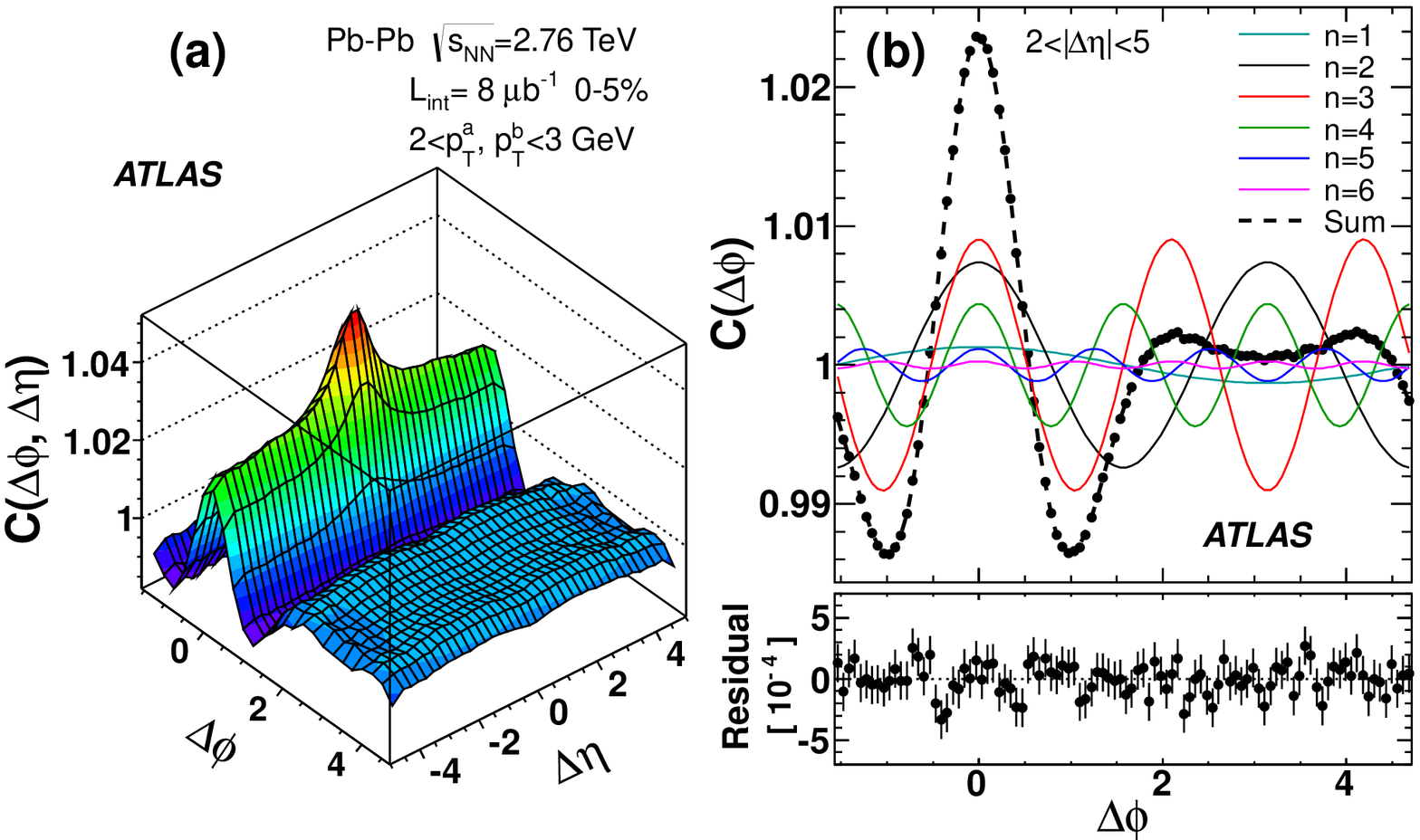}
\includegraphics[width=1\linewidth]{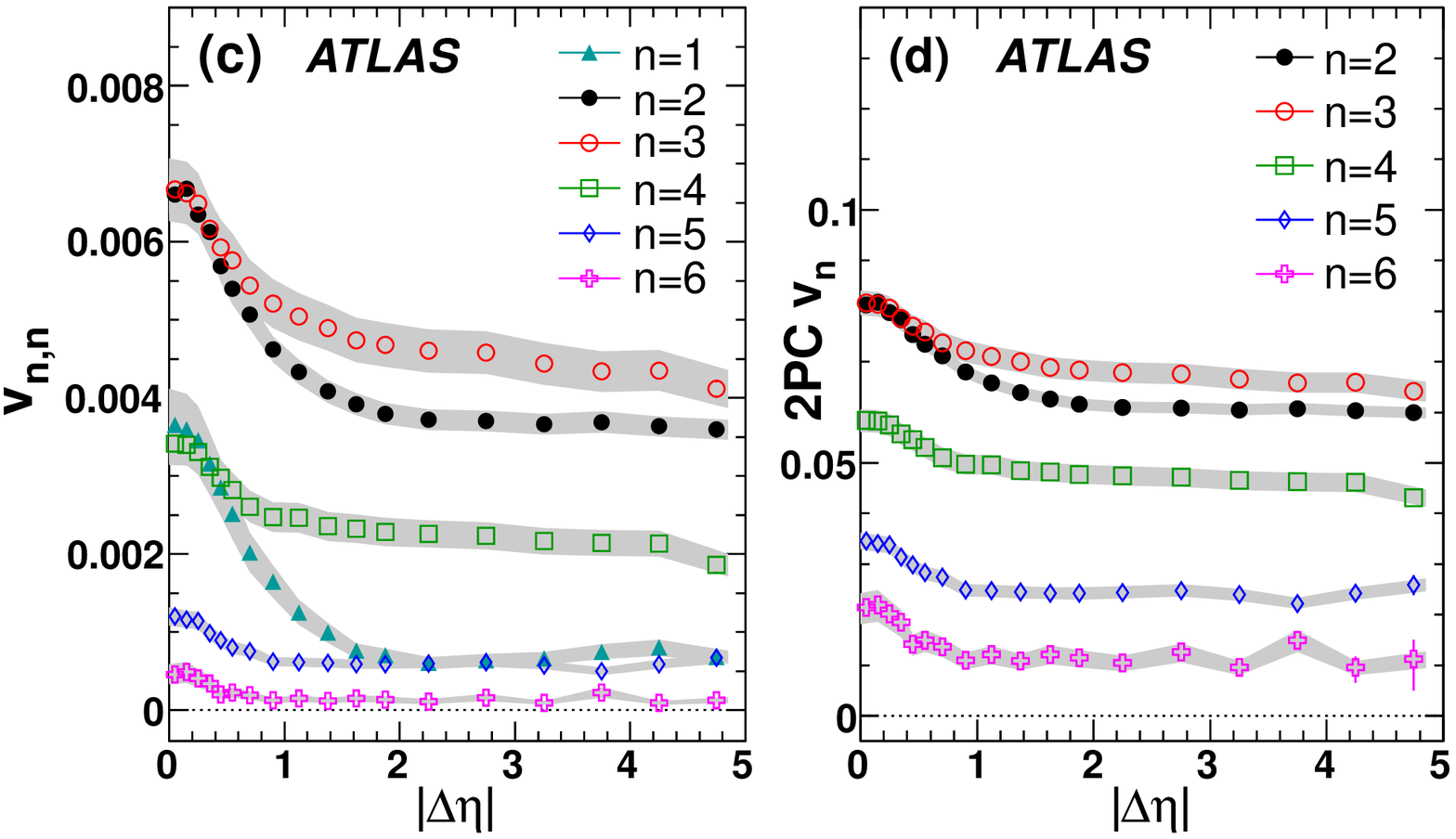}
\end{flushleft}
\end{minipage}
\begin{minipage}{0.35\linewidth}
\begin{flushright}
\caption{\label{fig:method1} (Color online) The steps involved in the extraction of $v_n$ values for 2--3 GeV fixed-$\pT$ correlations in the 0--5\% centrality interval: (a) two-dimensional correlation function, (b) the one-dimensional $\Delta\phi$ correlation function for $2<|\Delta\eta|<5$ (re-binned into 100 bins), overlaid with contributions from the individual $v_{n,n}$ components and their sum, as well as the residual difference between the data and the sum, (c) Fourier coefficient $v_{n,n}$ vs. $|\Delta\eta|$ for $n=1$--6, and (d) $v_n$ vs. $|\Delta\eta|$ for $n=2$--6. The shaded bands in (c) and (d) indicate the systematic uncertainties as described in the text.}
\end{flushright}
\end{minipage}
\end{tabular}
\end{figure*}

Figure~\ref{fig:method1}(a) shows the two-dimensional (2-D) correlation function for pairs from a representative $\pT$ range of 2--3 GeV and 0--5\% centrality interval. It reveals the characteristic long-range near-side ``ridge'' and away-side ``double-hump'' structures that were observed previously in heavy ion collisions at the RHIC for a similar $\pT$ range~\cite{Abelev:2009qa,Alver:2009id,Adler:2005ee}. A narrow short range correlation peak is also seen at $(\Delta\phi,\Delta\eta)\sim(0,0)$, presumably due to autocorrelations from jet fragmentation and resonance decays. From the 2-D correlation function, a one-dimensional (1-D) $\Delta\phi$ correlation function can be constructed for a given $\Delta\eta$ interval:
\begin{eqnarray}
\label{eq:cf}
C(\Delta\phi) = A\times \frac{\int S(\Delta\phi,\Delta\eta) d\Delta\eta }{\int B(\Delta\phi,\Delta\eta)d\Delta\eta }\;.
\end{eqnarray}
The normalization constant $A$ is determined by scaling the number of pairs in $2<|\Delta\eta|<5$ to be the same between the foreground (S) and background (B). This normalization is then applied to other $\Delta\eta$ intervals. Each 1-D correlation function is expanded into a Fourier series according to Eq.~\ref{eq:2a}, with coefficients $v_{n,n}$ calculated directly via a discrete Fourier transformation (DFT):
\begin{eqnarray} 
v_{n,n}= \langle\cos n\Delta\phi \rangle = \frac{\sum_{m=1}^{N} \cos (n\Delta\phi_m) C(\Delta\phi_m)}{\sum_{m=1}^N C(\Delta\phi_m)}\;,
\end{eqnarray}
where $n=1$--15, and $N$ = 200 is the number of $\Delta\phi$ bins. A small upward relative correction is applied ($\sim0.15\%$ for $n=6$ and increasing to $1\%$ for $n=15$) to account for the finite $\Delta\phi$ bin width. Figure~\ref{fig:method1}(b) shows one such 1-D correlation function for $2<|\Delta\eta|<5$, overlaid with the corresponding contributions from individual $v_{n,n}$ components. The shape of the correlation function is well described by the sum of the first six $v_{n,n}$ components.

According to Eq.~\ref{eq:fac}, if the correlations are dominated by those arising from asymmetry of the initial geometry such as flow, $v_{n,n}$ should factorize into the product of two single-particle harmonic coefficients. This is found to be the case for $n\geq2$ at low $\pT$ for pairs with a large $\Delta\eta$ gap, but is not true for $n=1$ (see Sections~\ref{sec:re2} and \ref{sec:re3}), similar to what was also found in other measurements~\cite{Aamodt:2011by,CMS:2012wg}. Thus if the two particles are selected from the same $\pT$ interval (``fixed-$\pT$'' correlations) as in Fig.~\ref{fig:method1}, the single-particle $v_n$ for $n\geq2$ can be calculated as $v_n=\sqrt{v_{n,n}}$. When $v_{n,n}<0$, $v_n$ is defined as $v_n = -\sqrt{|v_{n,n}|}$ (or $v_n = v_{n,n}/\sqrt{|v_{n,n}|}$ in general). This calculation is repeated for all 1-D correlation functions in each $|\Delta\eta|$ slice. The resulting full $|\Delta\eta|$ dependence of $v_{n,n}$ and $v_{n}$ are shown in Figs.~\ref{fig:method1}(c) and \ref{fig:method1}(d). 

\begin{figure}[!b]
\includegraphics[width=0.95\linewidth]{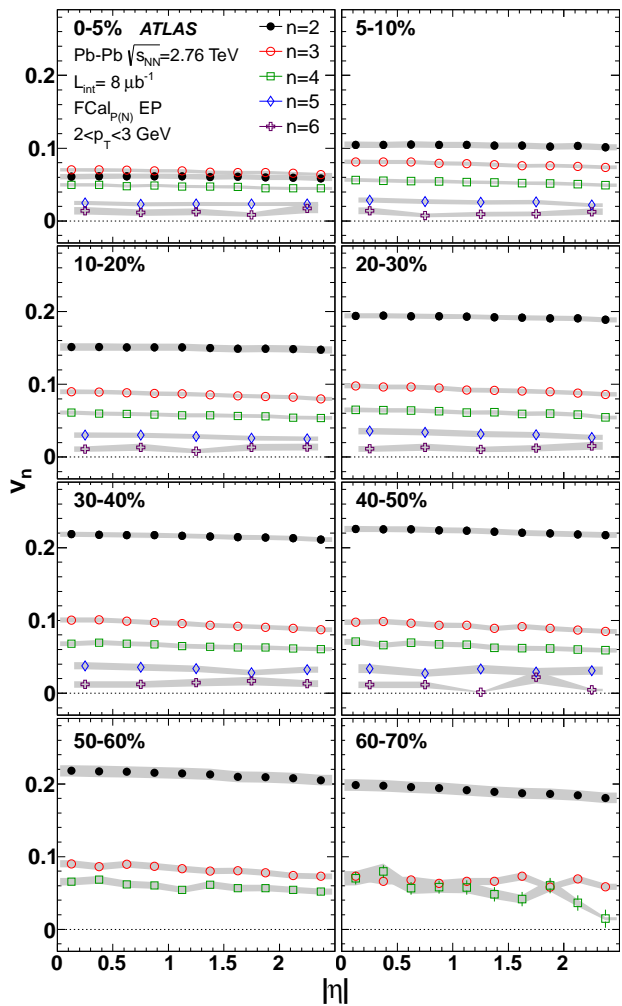}
\caption{\label{fig:reseta2b} (Color online) $v_n$ vs. $\eta$ for $2<\pT<3$ GeV from the FCal$_{\mathrm{P(N)}}$ method (i.e the EP is measured by either FCal$_{\mathrm N}$ or FCal$_{\mathrm P}$) with each panel representing one centrality interval. The shaded bands indicate systematic uncertainties from Tables~\ref{tab:v2}--\ref{tab:v6}.}
\end{figure}

The $v_{n,n}$ and $v_n$ values are found to vary rapidly for $|\Delta\eta|\lesssim1$, presumably reflecting the influence of the short range correlation at $(\Delta\phi,\Delta\eta)\sim(0,0)$ (Fig.~\ref{fig:method1} (a)), but they decrease much more slowly for larger $|\Delta\eta|$. This slow decrease is expected since the single-particle $v_n$ also decreases very slowly with $\eta$ (see Fig.~\ref{fig:reseta2b}), and the factorization relation Eq.~\ref{eq:fac} is valid for the present $\pT$ range (see Section~\ref{sec:re2}). These behaviors suggest that the autocorrelations from near-side jet fragmentation and resonance decays can be largely eliminated by requiring a large $\Delta\eta$ gap (e.g. $|\Delta\eta|>2$).

Each ``fixed-$\pT$'' correlation function provides a reference $v_n$ for a chosen $\pT$ range (denoted by superscript ``a''). Tracks from this $\pT$ range are then correlated with those from a target $\pT$ range (denoted by superscript ``b''), and this ``mixed-$\pT$'' correlation is used to calculate $v_{n,n}$ and to obtain the $v_n$ in the target $\pT$ via Eq.~\ref{eq:fac}. Since factorization is expected to be valid for the anisotropies driven by the initial geometry, but is broken by the presence of autocorrelations among the jet fragmentation products, the level of consistency between $v_n$ obtained from different reference $\pT$ ranges reveals whether the 2PC is dominated by anisotropies driven by the initial geometry. A detailed study of the factorization properties of $v_1$--$v_6$ is presented in Section~\ref{sec:re2}.

The correlation function relies on the pair acceptance function to reproduce and cancel the detector acceptance effects in the foreground distribution. Mathematically, the pair acceptance function in $\Delta\phi$ is simply a convolution of two single-particle azimuthal distributions, and should be uniform in $\Delta\phi$ without detector imperfections. A natural way of quantifying the influence of detector effects on $v_{n,n}$ and $v_n$ is to transform the single-particle and pair acceptance functions into the Fourier space. The resulting coefficients for pair acceptance $v_{n,n}^{\mathrm{det}}$ are the product of those for the two single-particle acceptances $v_{n}^{\mathrm{det,a}}$ and $v_{n}^{\mathrm{det,b}}$. In general, the pair acceptance function is quite flat: the maximum variation from its average is observed to be less than 0.001 for pairs integrated over $2<|\Delta\eta|<5$, and the corresponding $|v_{n,n}^{\mathrm{det}}|$ values are found to be less than $1.5\times 10^{-4}$. These $v_{n,n}^{\mathrm{det}}$ values are expected to mostly cancel in the correlation function, and only a small fraction contributes to the uncertainties of the pair acceptance function. Three possible residual effects for $v_{n,n}^{\mathrm{det}}$ are studied: 1) the time dependence of the pair acceptance, 2) the effect of imperfect centrality matching, and 3) the effect of imperfect $z_{\mathrm{vtx}}$ matching. In each case, the residual $v_{n,n}^{\mathrm{det}}$ values are evaluated by a Fourier expansion of the ratio of the pair acceptances before and after the variation. Overall, significant deviations are observed only for the effect of imperfect $z_{\mathrm{vtx}}$ matching, and they are generally larger for narrower $|\Delta\eta|$ ranges and higher $\pT$.

The systematic uncertainty of the pair acceptance is the quadrature sum of these three estimates, which is $\delta v_{n,n}=($2.5--8$)\times 10^{-6}$ depending on $n$, $\pT$, and the width of $|\Delta\eta|$ interval. This absolute uncertainty is propagated to the uncertainty in $v_n$, and it is the dominant uncertainty when $v_n$ is small, e.g. for $v_6$. Moreover, results for inclusive charged particles are compared to those obtained independently using same-charge and opposite-charge pairs. These two types of correlations have somewhat different pair acceptances due to different relative bending directions between the two tracks. They are found to give consistent results for $n\le6$, where the $v_{n,n}$ values are dominated by physics effects. However, small systematic deviations are observed for $n\ge8$, where the $v_{n,n}$ values are expected to be dominated by acceptance effects. Therefore, the systematic uncertainty also includes the RMS difference of the $v_{n,n}$ values averaged for $8\le n\le 15$ between the two types of correlations. This uncertainty is usually much smaller than those associated with $v_{n,n}^{\mathrm{det}}$, except for large $\pT$.

The second type of systematic uncertainty includes the sensitivity of the analysis to track quality cuts, variation between different running periods, trigger and event selection, as well as the ability to reproduce the input $v_n$ in fully simulated, digitized and reconstructed HIJING events with azimuthal anisotropy imposed on the generated particles. Most systematic uncertainties cancel for the correlation function when dividing the foreground and background distributions. The estimated residual effects are summarized in Table~\ref{tab:corr2}. 
\begin{table}[h]
\centering
\small{
\begin{ruledtabular}\begin{tabular}{l|c c c c c c}
&$v_1$ & $v_2$ & $v_3$ & $v_4$ & $v_5$ &$v_{n |n\ge 6}$ \\\hline
Track selection[\%]& 3.0   & 0.3 &0.3 &1.0 &2.0 &4 \\\hline
Running periods[\%]& 5.0       &0.3-1.0 &0.7-2.1 &1.2-3.1 &2.3     &7-11\\\hline
Trigger \& event sel.[\%] & 1.0    &0.5-1.0  &0.5-1.5&0.5-1  &1.0    &5 \\\hline
MC consistency[\%] &  2.0   & 1.0   & 1.5 & 2.0 & 3.5 & 5 \\\hline
Sum of above[\%] &6.3&1.2-1.8&1.8-3.0&2.6-3.9&4.8&11-14\\
\end{tabular}\end{ruledtabular}
}\normalsize
\caption{\label{tab:corr2} Relative systematic uncertainties for $v_n$ in percentage from tracking cuts, variation between different running periods, centrality variation, consistency between truth and reconstructed $v_n$ in HIJING simulation, and the quadrature sum of individual terms.}
\end{table}

The total systematic uncertainties are the quadrature sum of the uncertainties calculated from pair acceptance, the RMS difference of the $v_{n,n}$ averaged for $8\le n\le 15$ between the same-charge and opposite-charge correlations, and those listed in Table~\ref{tab:corr2}. They are then propagated to uncertainties for $v_n$. These uncertainties are plotted as shaded bands around the data points in Figs.~\ref{fig:method1}(c) and \ref{fig:method1}(d). Most of these uncertainties are correlated between different $\pT$ ranges. However, a fraction of them are found to be uncorrelated with $\pT$, coming mainly from the track selection, running period variation and MC comparison in Table~\ref{tab:corr2} and the pair acceptance. This fraction (point to point in $\pT$) is estimated to be about 30\% of the final systematic uncertainty, and the remaining uncertainty is treated as a $\pT$-correlated systematic uncertainty. They are used in the discussion of the $v_{1,1}$ results in Section~\ref{sec:re4}.

\section{Results}
\subsection{$v_2$--$v_6$ from the event plane method}
\label{sec:re1}
\begin{figure}[b]
\begin{center}
\includegraphics[width=0.95\linewidth]{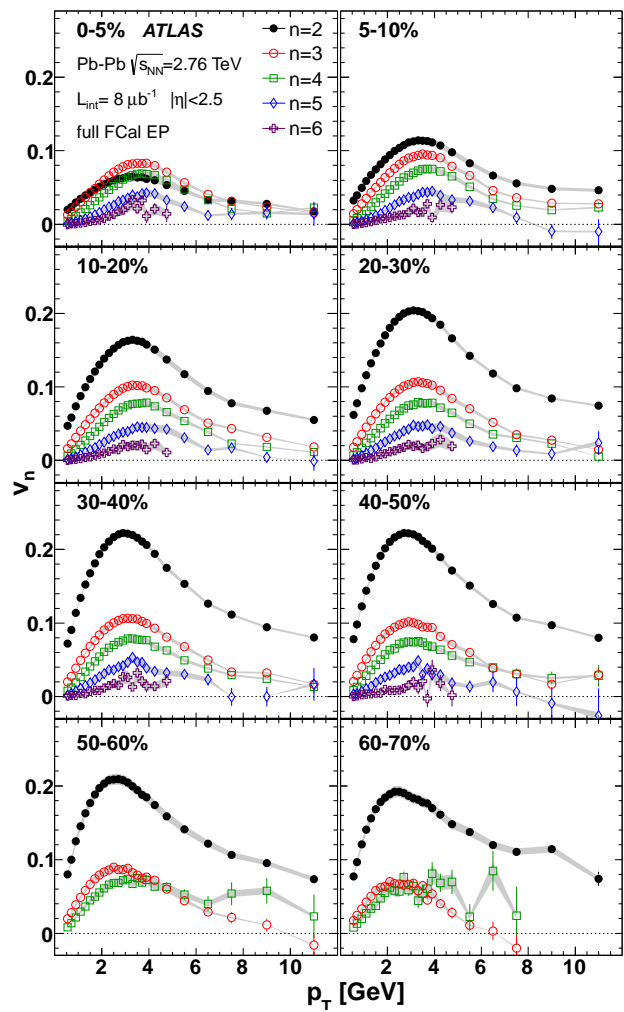}
\end{center}
\caption{\label{fig:respt1} (Color online) $v_n$ vs. $\pT$ for several centrality intervals. The shaded bands indicate the systematic uncertainties from Tables~\ref{tab:v2}--\ref{tab:v6}.}
\end{figure}
Figure~\ref{fig:reseta2b} shows the $\eta$ dependence of $v_n$ for several centrality intervals in the 2--3 GeV $\pT$ range from the FCal$_{\mathrm{P(N)}}$ EP method. Similar behaviors are observed in other $\pT$ ranges (see also~\cite{Collaboration:2011yk} for $v_2$). The $v_2$ values decrease by less than 5\% towards large $|\eta|$ for central and mid-central events, and the decrease is more pronounced both for $n\ge3$ and for peripheral events.

Figure~\ref{fig:respt1} shows the $\pT$ dependence of $v_2$--$v_6$ for several centrality intervals. All $v_n$ increase with $\pT$ in the range up to 3--4 GeV and then decrease. However, they remain positive even at the highest measured $\pT$, where occasional fluctuations to negative values do not exceed the statistical precision. This turn-over behavior in $\pT$ was also observed at RHIC for $v_2$~\cite{Adams:2004wz,Adare:2010sp}, and it is associated with the transition from anisotropy driven by the collective expansion to anisotropy driven by a path-length dependent jet energy loss~\cite{Gyulassy:2000gk,Betz:2011tu}. The overall magnitude of $v_n$ also decreases with increasing $n$, except in the most central events where $v_3$ is the largest. 

\begin{figure}[!t]
\begin{center}
\includegraphics[width=1\linewidth]{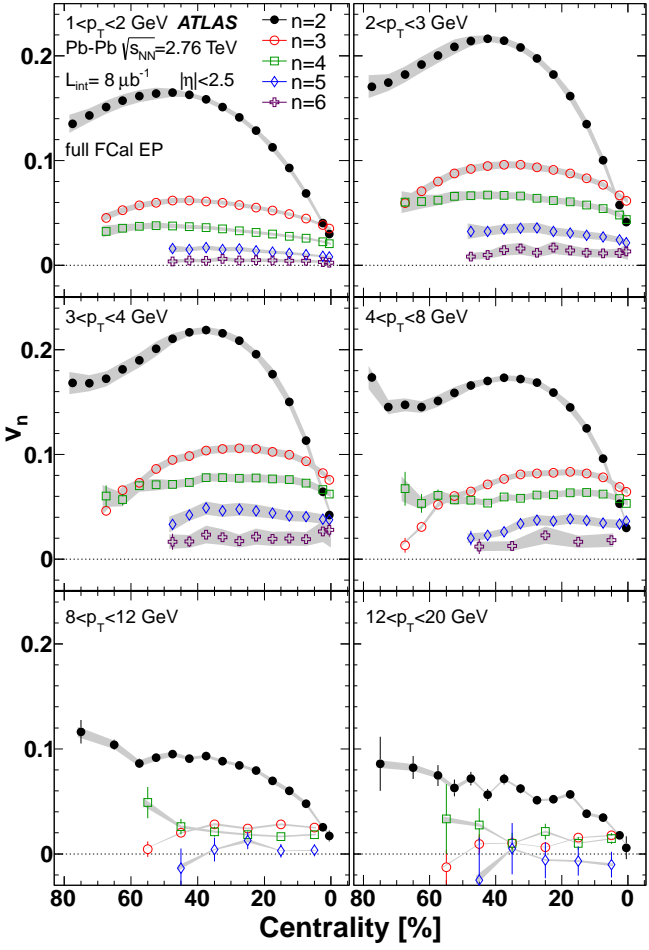}
\end{center}
\caption{\label{fig:rescent1} (Color online) $v_n$ vs. centrality for six $\pT$ ranges from the full FCal event plane method. The shaded bands indicate systematic uncertainties from Tables~\ref{tab:v2}--\ref{tab:v6}.}
\end{figure}

Figure~\ref{fig:rescent1} shows the centrality dependence of $v_n$ for several $\pT$ ranges.
 The centrality intervals are presented in $5\%$ or $10\%$ increments, with an additional interval for the 1\% most central events. Going from central to peripheral events (from right to left along the $x$-axis), $v_2$ first increases, reaching a maximum in the 30--50\% centrality range, and then decreases. The higher-order coefficients $v_3$--$v_6$ show a similar, but much weaker, centrality dependence, and this behavior is consistent with an anisotropy related to the fluctuations in the initial geometry~\cite{Qiu:2011hf}. For most of the measured centrality range, $v_2$ is much larger than the other harmonic coefficients. In central events, however, $v_3$ and/or $v_4$ becomes larger than $v_2$ for some $\pT$ ranges. At high $\pT$ ($>4$ GeV), $v_2$ increases towards more peripheral events, presumably reflecting the dominance of autocorrelations from di-jets.

\begin{figure}[!t]
\begin{center}
\includegraphics[width=1\linewidth]{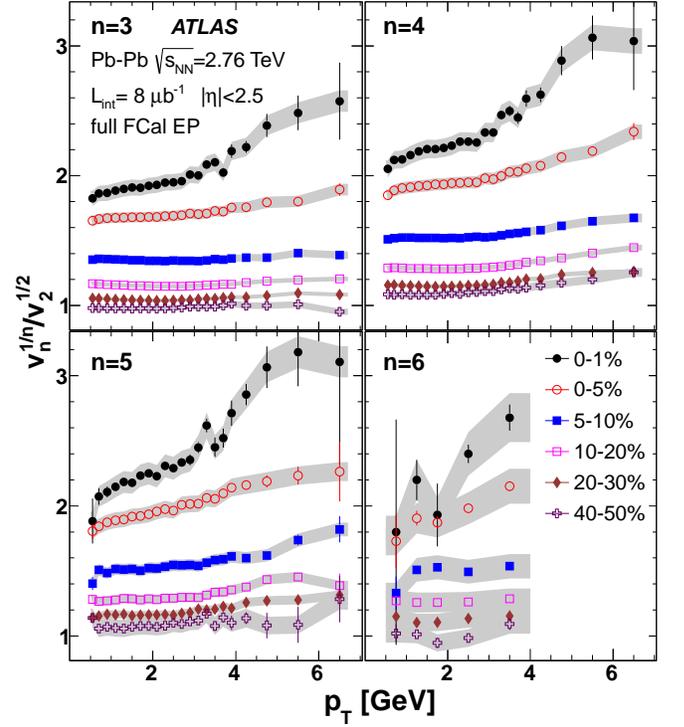}
\end{center}
\caption{\label{fig:respt2} (Color online) $v_n^{1/n}/v_2^{1/2}$ vs. $\pT$ for several centrality intervals. The shaded bands indicate the total systematic uncertainties.}
\end{figure}

In an ideal hydrodynamics scenario, $v_n$ at low $\pT$ is a power-law function of the radial expansion velocity of the fluid, leading to the qualitative expectation that $v_n(\pT)$ is a power-law function of $\pT$~\cite{Borghini:2005kd,Alver:2010dn}. Previous RHIC results have shown that $v_4/v_2^2$ (or equivalently $v_4^{1/4}/v_{2}^{1/2}$) is almost independent of $\pT$~\cite{Adams:2003zg,Adare:2010ux}~\footnote{This $v_4$ was measured relative to the $\Phi_2$ instead of the $\Phi_4$ reaction plane, and is known as mixed harmonics~\cite{Poskanzer:1998yz}. It can be regarded as a projection of $v_4$ measured in the $\Phi_4$, onto the $\Phi_2$.}. Figure~\ref{fig:respt2} shows $v_n^{1/n}/v_{2}^{1/2}$ vs. $\pT$ for various centrality intervals. These ratios vary weakly with $\pT$ except in the 5\% most central events, suggesting that such a scaling relation largely accounts for the $\pT$ dependence. However, the overall magnitudes of the ratios seem to vary with centrality and also vary slightly with $n$. 
\begin{figure}[!h]
\begin{center}
\includegraphics[width=1\linewidth]{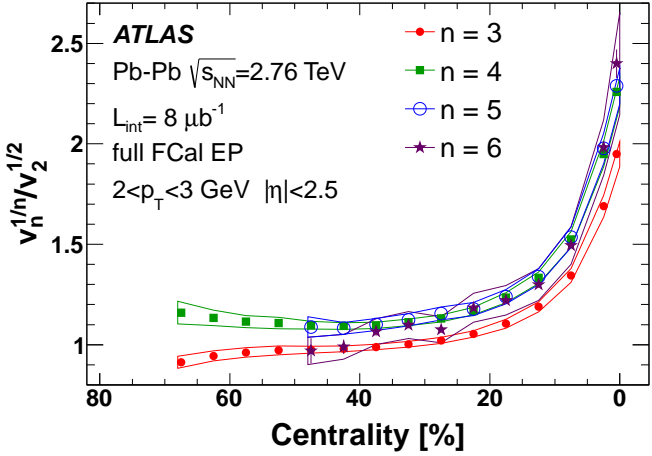}
\end{center}
\caption{\label{fig:rescent2} (Color online) $v_n^{1/n}/v_2^{1/2}$ vs. centrality for $2<\pT<3$ GeV. Lines indicate systematic uncertainty bands, calculated by assuming that the uncertainties for different $v_n$ are independent.}
\end{figure}

\afterpage{
\begin{figure*}[p]
\centering 
\includegraphics[width=0.95\linewidth]{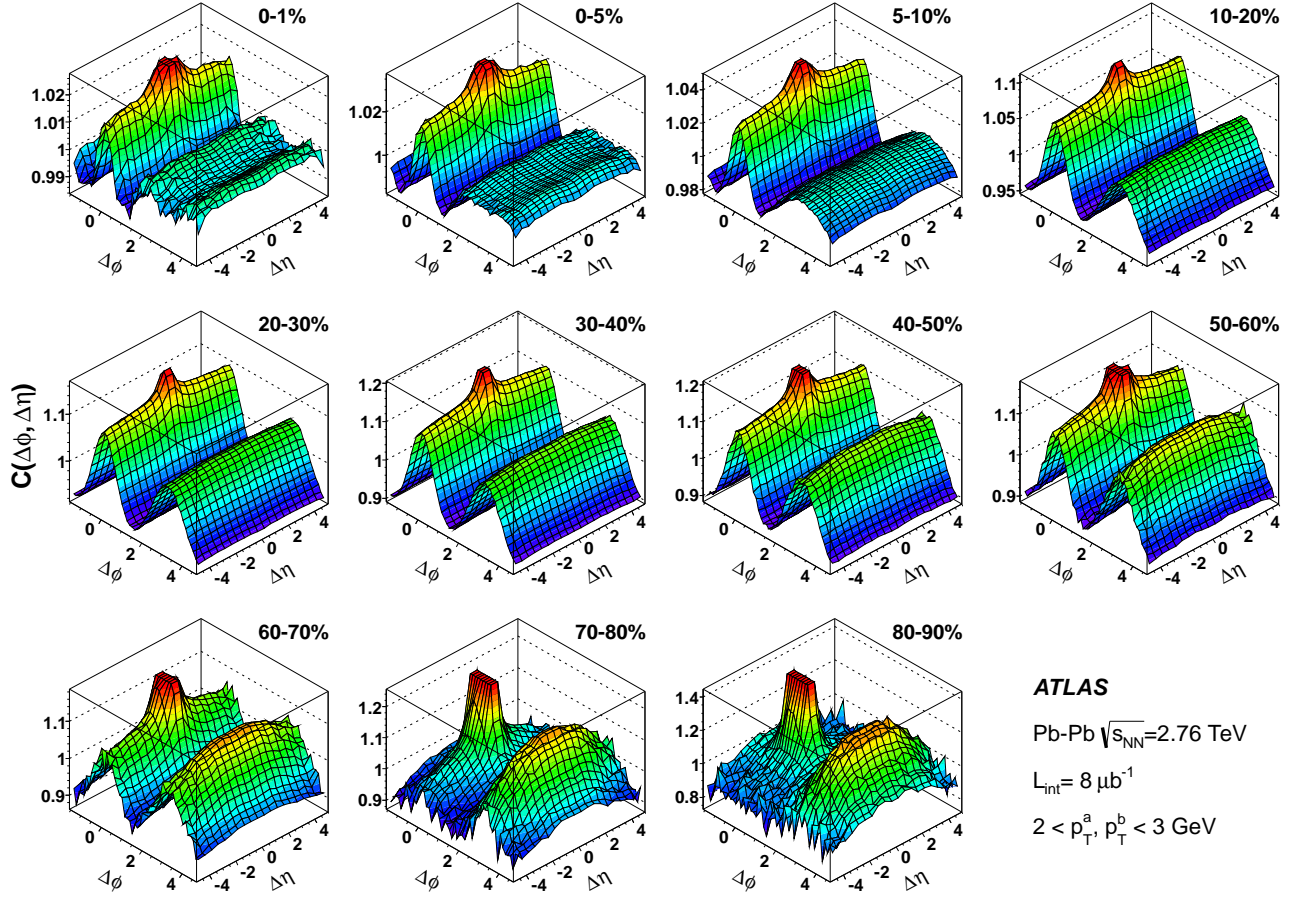}
\caption{\label{fig:reland1} (Color online) Two-dimensional correlations for $2<\pT^{\mathrm a},\, \pT^{\mathrm b}<3$ GeV in several centrality intervals. The near-side jet peak is truncated from above to better reveal the long-range structures in $\Delta\eta$.}
\end{figure*}
\begin{figure*}[p]
\begin{center}
\includegraphics[width=0.95\linewidth]{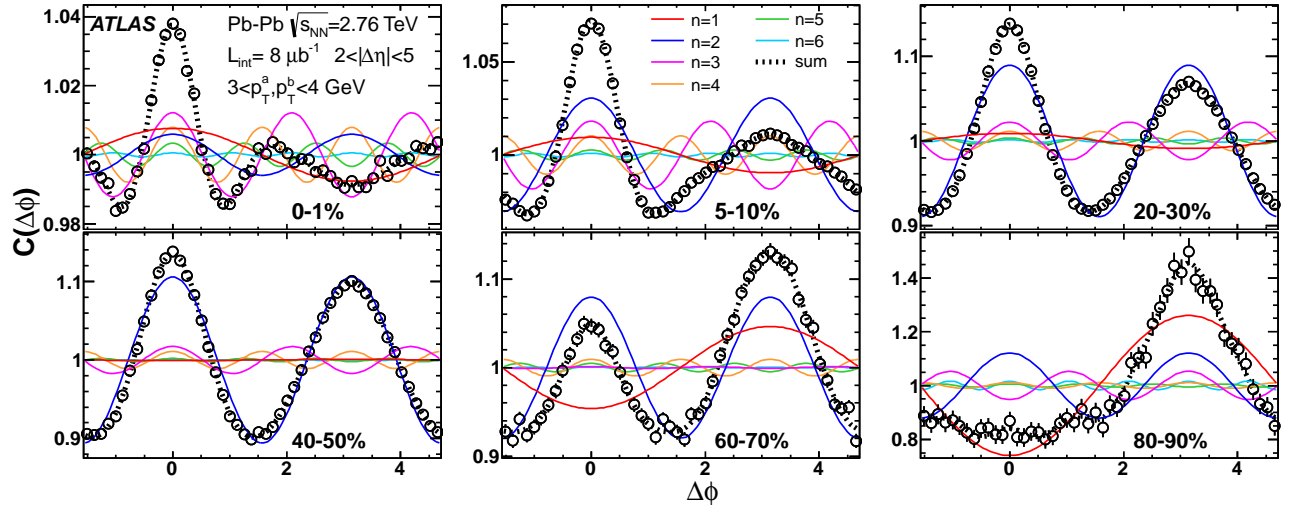}
\end{center}
\caption{\label{fig:reland1c} (Color online) Centrality dependence of $\Delta\phi$ correlations for $3< \pT^{\mathrm a},\pT^{\mathrm b}< 4$ GeV. A rapidity gap of $2<|\Delta\eta|<5$ is required to isolate the long-range structures of the correlation functions, i.e. the near-side peaks reflect the ``ridge'' instead of the autocorrelations from jet fragments. The error bars on the data points indicate the statistical uncertainty. The superimposed solid lines (thick-dashed lines) indicate contributions from individual $v_{n,n}$ components (sum of the first six components).}
\end{figure*}
}

Figure~\ref{fig:rescent2} shows the centrality dependence of $v_n^{1/n}/v_{2}^{1/2}$ for $2<\pT<3$ GeV. Given that the ratios vary weakly with $\pT$, the results for other $\pT$ ranges are similar. The ratios are almost independent of centrality in mid-central and peripheral events, but then increase sharply toward more central events, with a total change of almost a factor of two over the 0--20\% centrality range.  In addition, the ratios for $n=4$--6 are similar to each other, while they are systematically higher than those for $n=3$. A similar centrality dependence was observed for the $v_4/v_2^2$ ratio at the RHIC, and was argued to reflect the centrality dependence of fluctuations in the initial geometry~\cite{Adare:2010ux}.

\subsection{ $v_2$--$v_6$ from the two-particle correlation method}
\label{sec:re2}
Figure~\ref{fig:reland1} shows the evolution of the 2-D correlation function with centrality for particles with $2<\pT<3$ GeV. While central events show structures that are long range in $\Delta\eta$ (the ``ridge'' and ``double-hump''), the more peripheral events show a systematic disappearance of these long-range structures and the emergence of clear jet-related peaks on the away-side. The magnitude of the long-range structures, measured as deviation from unity, exhibits a characteristic centrality dependence. Figure~\ref{fig:reland1} shows that the near-side ``ridge'' (relative to unity) starts at about 0.015 in the 1\% most central events, increases to 0.12 in 30--50\% mid-central events, and then decreases and disappears in the most peripheral (80--90\%) events. Since the harmonics for different $n$ all contribute positively to the correlation function at $\Delta\phi=0$, this non-monotonic centrality dependence simply reflects the fact that $v_n$ for $n\geq2$ all reach their maxima in the 30--50\% centrality range for this $\pT$ selection, as shown in Fig.~\ref{fig:rescent1}. The away-side long-range structure exhibits a similar centrality dependence, but is complicated by the contribution from the recoil jet, which starts to dominate the away-side shape in the 60--90\% centrality range.
\begin{figure}[t!]
\begin{center}
\includegraphics[width=1\linewidth]{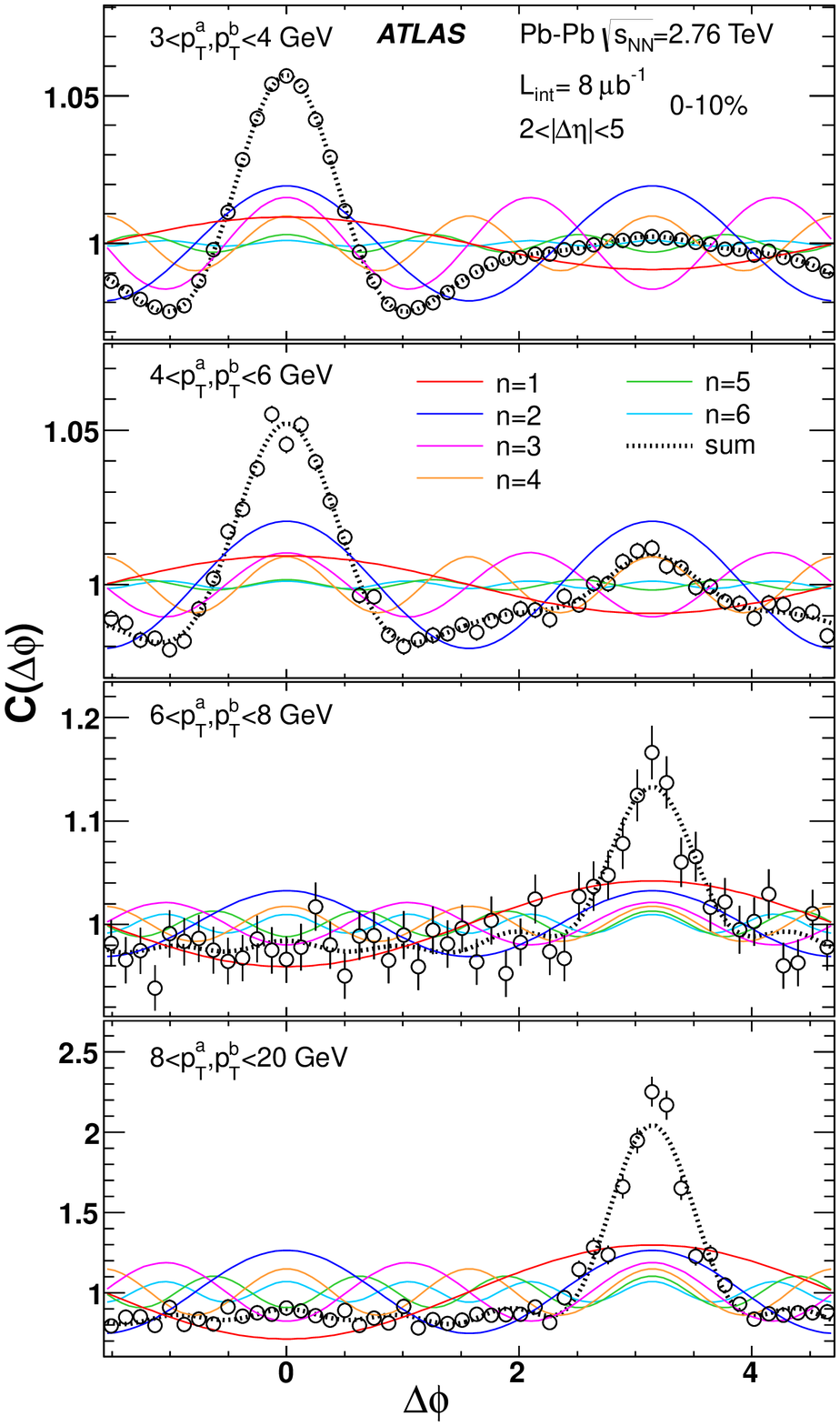}
\end{center}
\caption{\label{fig:jetb1} (Color online) Fixed-$\pT$ correlation function in the 0--10\% centrality interval for several $\pT$ ranges. A rapidity gap of $2<|\Delta\eta|<5$ is required to isolate the long-range structures of the correlation functions. The error bars on the data points indicate the statistical uncertainty. The superimposed solid lines (thick-dashed lines) indicate contributions from individual $v_{n,n}$ components (sum of the first six components). }
\end{figure} 
\begin{figure}[t!]
\begin{center}
\includegraphics[width=1\linewidth]{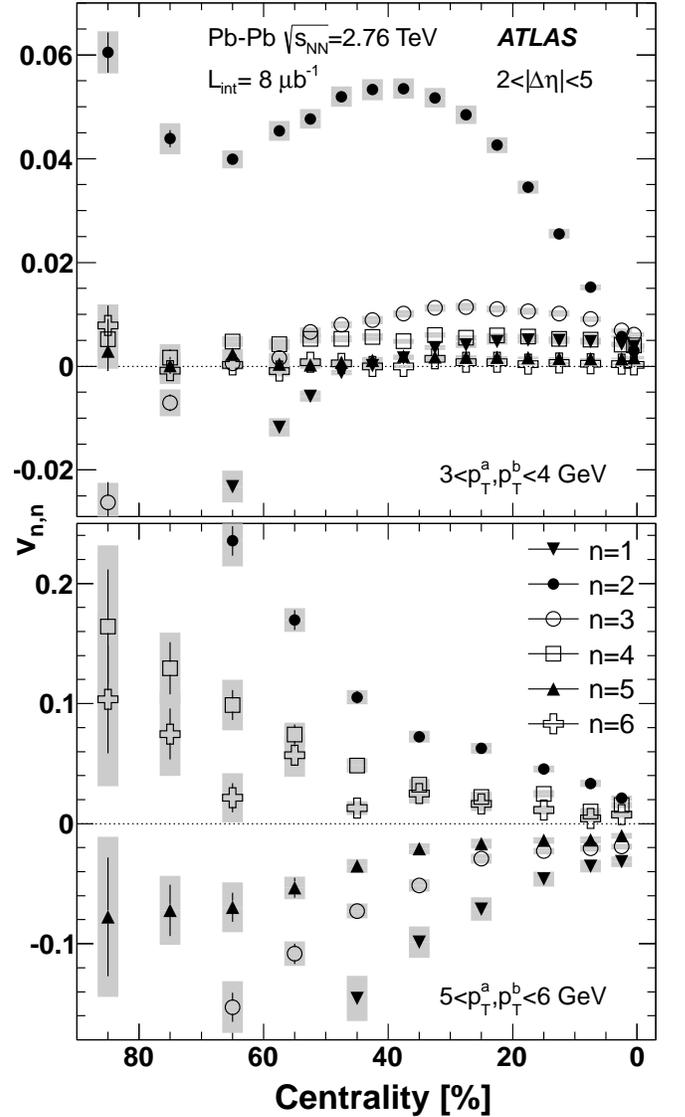}
\end{center}
\caption{\label{fig:jetb2} $v_{n,n}$ vs. centrality obtained from fixed-$\pT$ correlations with $2<|\Delta\eta|<5$ for a low $\pT$ interval (top panel) and a high $\pT$ interval (bottom panel). The error bars (shaded bands) indicate the statistical (systematic) uncertainties.}
\end{figure}
\begin{figure*}[h]
\includegraphics[width=1\linewidth]{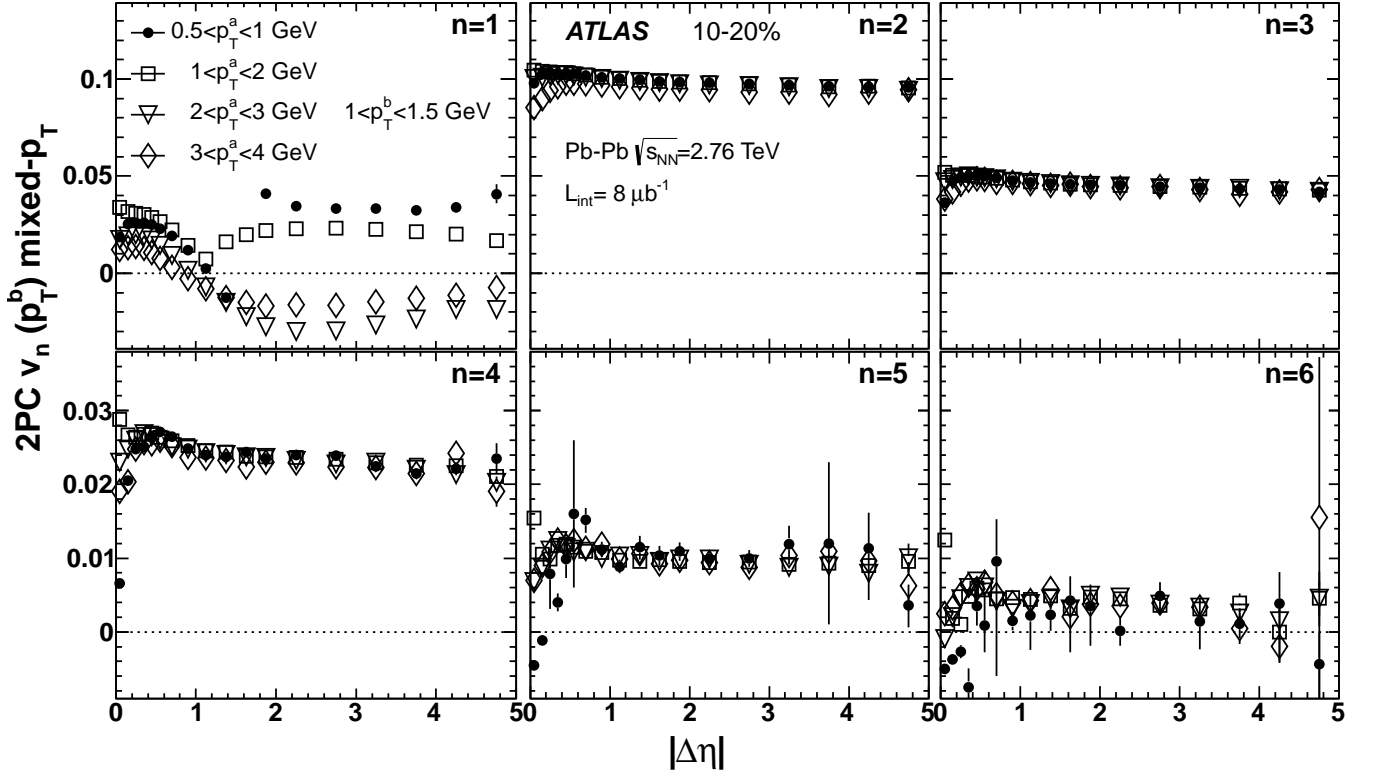}
\caption{\label{fig:refact1} $v_{n}(\pT^{\mathrm b})=\frac{v_{n,n}(\pT^{\mathrm a},\pT^{\mathrm b})}{v_n(\pT^{\mathrm a})}$ vs. $|\Delta\eta|$ for $1<\pT^{\mathrm b}<1.5$ GeV, calculated from a reference $v_n$ in four $\pT^{\mathrm a}$ ranges (0.5--1, 1--2, 2--3, and 3--4 GeV). The error bars indicate the statistical uncertainties. As discussed in the text, the $v_1$ defined this way is strongly biased by global momentum conservation, and should be distinguished from the collective $v_1$ discussed in Section~\ref{sec:re4}.}
\end{figure*}
\begin{figure*}[h]
\centering
\includegraphics[width=1\linewidth]{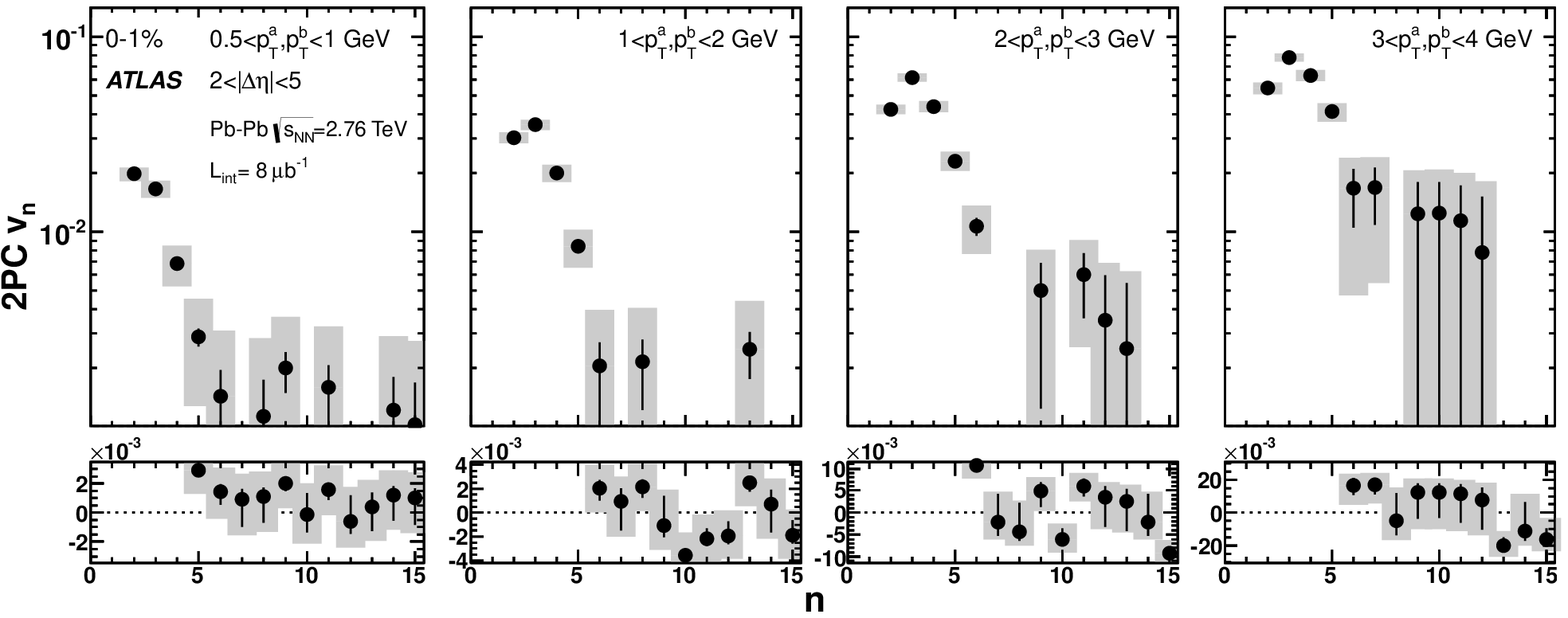}
\caption{\label{fig:ren0} $v_n$ vs. $n$ for $n\geq2$ in a 0--1\% centrality interval from four fixed-$\pT$ correlations (0.5--1, 1--2, 2--3 and 3--4 GeV from left to right). The error bars and shaded bands indicate the statistical and total systematic uncertainties, respectively. In the upper panels non-zero values of $v_n$ are presented using a logarithmic scale, while $v_n$ for large $n$ are shown more precisely using a linear scale in the lower panels.}
\end{figure*}

As discussed in Section~\ref{sec:m2} and shown in Fig.~\ref{fig:method1}, each 2-D correlation function is projected onto a set of 1-D $\Delta\phi$ correlation functions in slices of $|\Delta\eta|$, and the Fourier coefficients $v_{n,n}$ and $v_n$ are calculated from these distributions. Examples of such 1-D correlation functions are shown in Fig.~\ref{fig:reland1c} for pairs with $2<|\Delta\eta|<5$ and $3<\pT<4$ GeV, together with individual contributions from the first six $v_{n,n}$ components. In a scenario where the Fourier coefficients are dominated by anisotropic flow, the value of the correlation function at $\Delta\phi\sim0$ should be larger than its value at $\Delta\phi\sim\pi$ (see Eq.~\ref{eq:2}). This indeed is the case up to the 40--50\% centrality interval, but for centralities greater than 50\% the trend reverses. This reversing of the asymmetry between the near- and away-side amplitudes correlates with a continuous decrease of $v_{1,1}$, which eventually becomes negative at around the 40--50\% centrality interval (also see top panel of Fig.~\ref{fig:jetb2}). The correlation function in the 80--90\% centrality interval shows that a broad peak from the away-side jet predominantly generate a negative $v_{1,1}$ and a positive $v_{2,2}$. Therefore, they tend to cancel each other at the near-side but add up at the away-side. This behavior suggests that in peripheral collisions and at low $\pT$, the appearance of a large negative $v_{1,1}$ is a good indicator for a significant contribution of autocorrelations from jets to $v_{2,2}$.

Figure~\ref{fig:jetb1} shows Fourier decomposition of the correlation functions in the 0--10\% centrality interval for several $\pT$ ranges. Again, a large pseudorapidity gap of $|\Delta\eta|>2$ is required to suppress the near-side jet peak and to expose the long-range structures. At low $\pT$, the $v_{n,n}$ components are mainly driven by these long-range structures. However, for $\pT^{\mathrm a},\pT^{\mathrm b}>6$ GeV, they are dominated by the pronounced away-side jet correlation centered at $\Delta\phi=\pi$, leading to $v_{n,n}$ values with alternating sign: $(-1)^n$. Figure~\ref{fig:jetb2} contrasts the centrality dependence of the $v_{n,n}$ for a low-$\pT$ range (top panel) and a high-$\pT$ range (bottom panel). It shows that at low $\pT$, the sign-flipping of $v_{n,n}$ between even and odd $n$ happens only for peripheral events; at high $\pT$ the sign-flipping happens over the entire centrality range. More details on the variation of the correlation functions and $v_{n,n}$ values as a function of $\pT^{\mathrm{a}}$ and $\pT^{\mathrm{b}}$ are shown in Figs.~\ref{fig:land1}--\ref{fig:v6fac} in the Appendix.

As discussed in Section~\ref{sec:intro}, a necessary condition for the $v_{n,n}$ to reflect anisotropy associated with the initial spatial asymmetries is that it should factorize into the product of two single-particle $v_n$. A direct way to verify the factorization is to check whether the target $v_{n}(\pT^{\mathrm b})$ calculated as:
\small{
\begin{eqnarray}
\label{eq:ref}
 v_{n}(\pT^{\mathrm b})=\frac{v_{n,n}(\pT^{\mathrm a},\pT^{\mathrm b})}{v_{n}(\pT^{\mathrm a})}=\frac{\sqrt{|v_{n,n}(\pT^{\mathrm a},\pT^{\mathrm a})|}v_{n,n}(\pT^{\mathrm a},\pT^{\mathrm b})}{v_{n,n}(\pT^{\mathrm a},\pT^{\mathrm a})},
\end{eqnarray}}\normalsize
is independent of the reference $\pT^{\mathrm a}$. Figure~\ref{fig:refact1} shows one such study of $n=1$--6 for $1<\pT^{\mathrm b}<1.5$ GeV using four different reference $\pT^{\mathrm a}$ ranges. The $v_1(\pT^{\mathrm b})$ values calculated this way clearly change with the choice of $\pT^{\mathrm a}$, indicating a breakdown of the factorization over the measured $\pT$ range. This breakdown is mainly due to contributions from global momentum conservation for a system with finite multiplicity. More detailed discussions of the physics behind this breakdown and ways to separate different contributions and extract the $v_1$ associated with dipole asymmetry from $v_{1,1}$ are presented in Section~\ref{sec:re4}.

\begin{figure*}[!t] 
\begin{tabular}{lr}
\begin{minipage}{0.7\linewidth}
\begin{flushleft}
\includegraphics[width=1\linewidth]{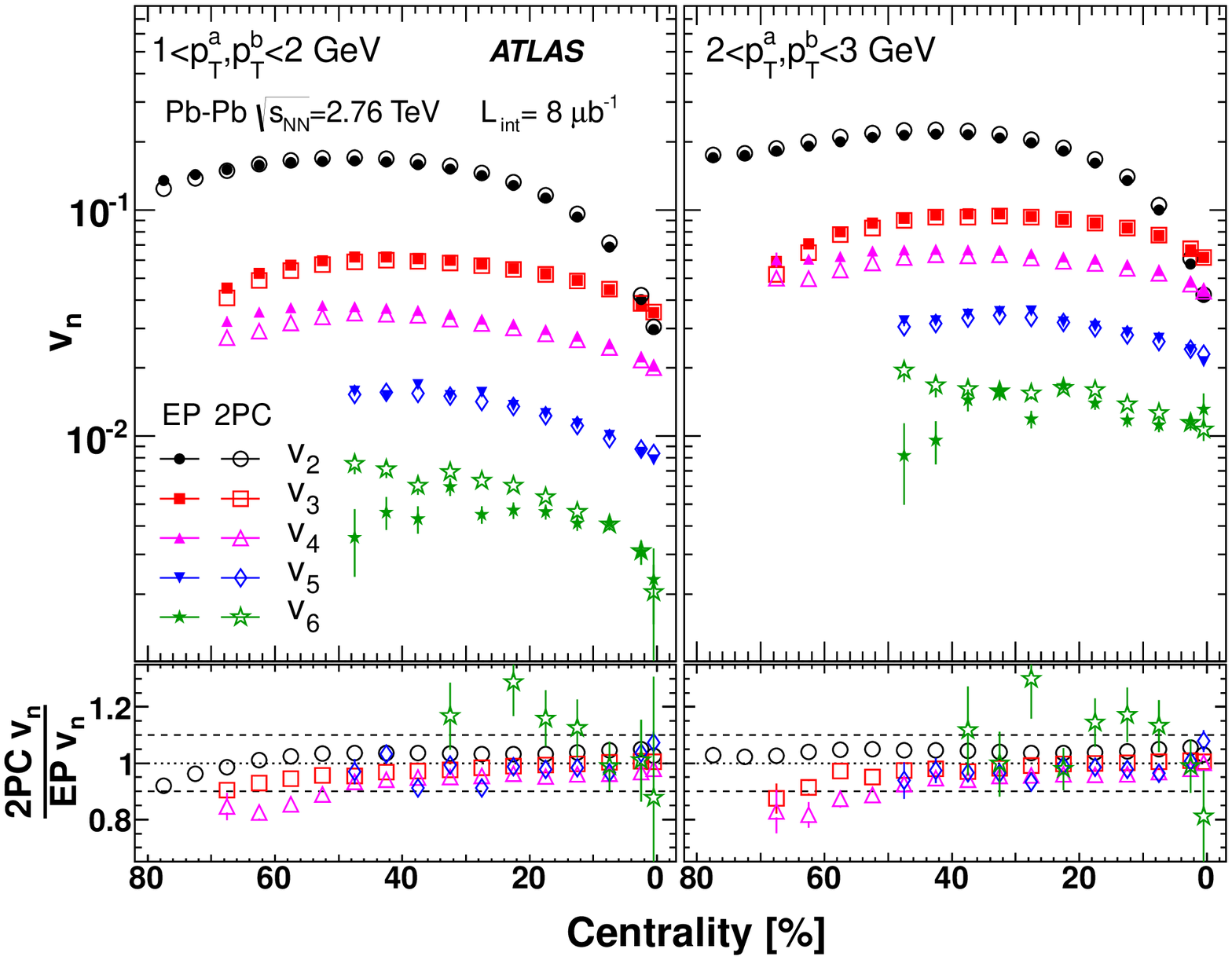}
\end{flushleft}
\end{minipage}
\begin{minipage}{0.28\linewidth}
\begin{flushright}
\caption{\label{fig:recmp1a} (Color online) Comparison of $v_n$ as a function of centrality between the fixed-$\pT$ 2PC method with $2<|\Delta\eta|<5$ (open symbols) and the full FCal EP method (solid symbols). Results are shown for $1<\pT<2$ GeV (left panels) and $2<\pT<3$ GeV (right panels). The error bars represent statistical uncertainties only. The ratios of the 2PC to EP methods are shown in the bottom panels, where the dashed lines indicate a $\pm10\%$ range to guide the eye.}
\end{flushright}
\end{minipage}
\end{tabular}
\end{figure*}
Figure~\ref{fig:refact1} shows that the factorization holds for $n=2$--6 for $|\Delta\eta|>1$. Despite a factor of three variation in the reference $v_n$ value for the $\pT$ ranges from 0.5--1 GeV to 3--4 GeV (see Fig.~\ref{fig:respt1}), the extracted $v_n(\pT^{\mathrm b})$ for $n=2$--6 is constant within statistical uncertainties for $|\Delta\eta|>1$. This factorization check is repeated for $n=2$--6 for various ranges of $\pT^{\mathrm a}$, $\pT^{\mathrm b}$ and centrality. For correlations with the default choice of $|\Delta\eta|>2$ used in this paper, factorization is found to hold at the 5\%--10\% level for $\pT<3$--4 GeV in the $70\%$ most central events. Further studies of this topic are presented in Section~\ref{sec:re3}.

Figure~\ref{fig:ren0} shows the extracted Fourier coefficients, $v_{n}$ vs. $n$, for several fixed-$\pT$ correlations with $|\Delta\eta|>2$ and for the 1\% most central events. The $\pT^{\mathrm {a,b}}$ ranges are restricted to below 4 GeV, where the factorization for $v_2$--$v_6$ works reasonably well. Significant signals are observed for $n\leq6$, while the signals for $n>6$ are consistent with zero within statistical and systematic uncertainties. Note that in cases where the uncertainty of $v_{n,n}$ is comparable to its own value, the $v_n$ uncertainty becomes asymmetric and highly non-Gaussian. A complete compilation of the Fourier coefficients for other centrality intervals is included in Figs~\ref{fig:ren1}--\ref{fig:ren7} in the Appendix. Recent model comparisons indicate that these results can provide important constraints on the nature of initial geometry fluctuations and the shear viscosity of the created matter~\cite{Staig:2011wj}.

\begin{figure}[p]
\centering
\includegraphics[width=0.95\linewidth]{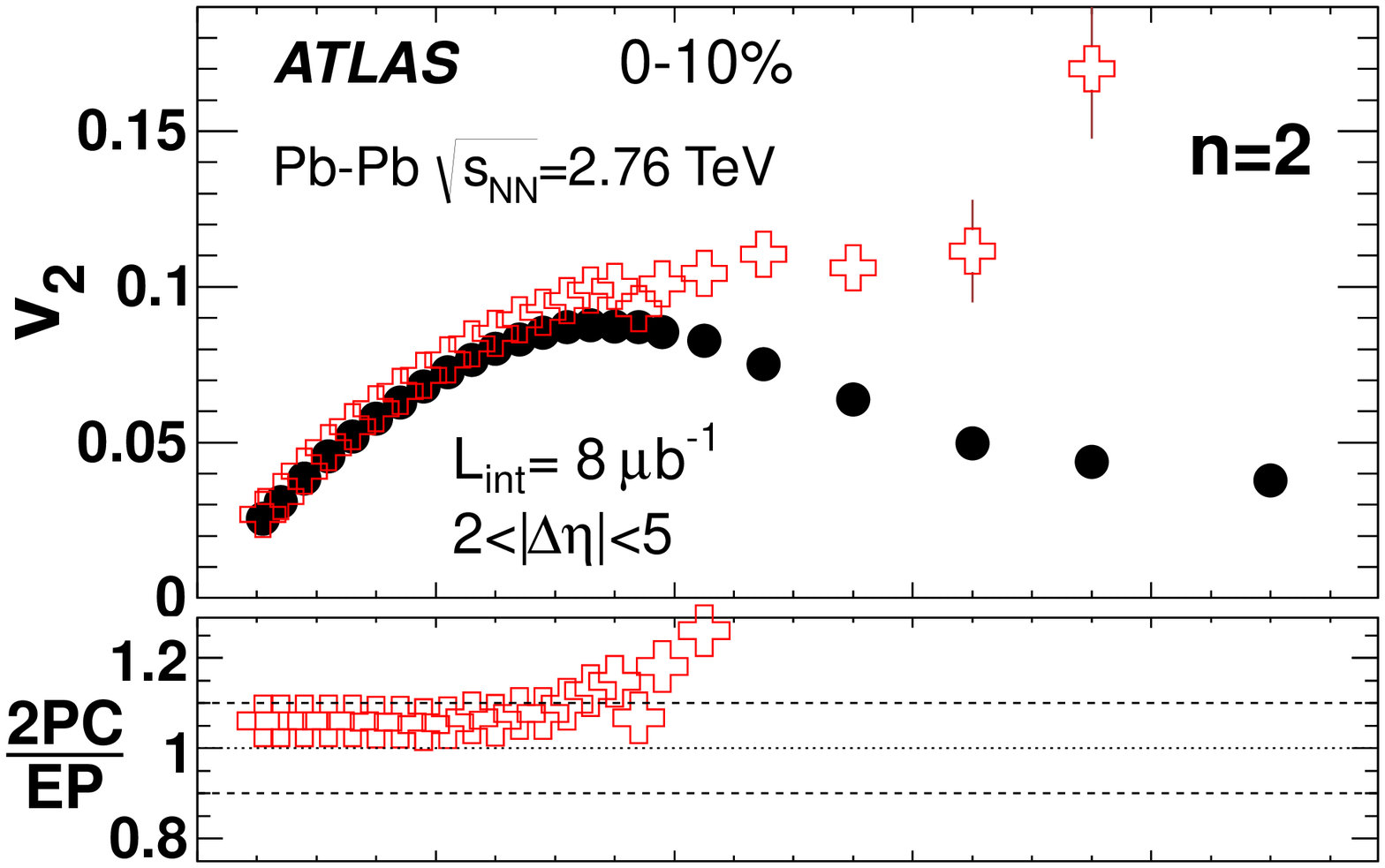}\vspace*{-0.83cm}
\includegraphics[width=0.95\linewidth]{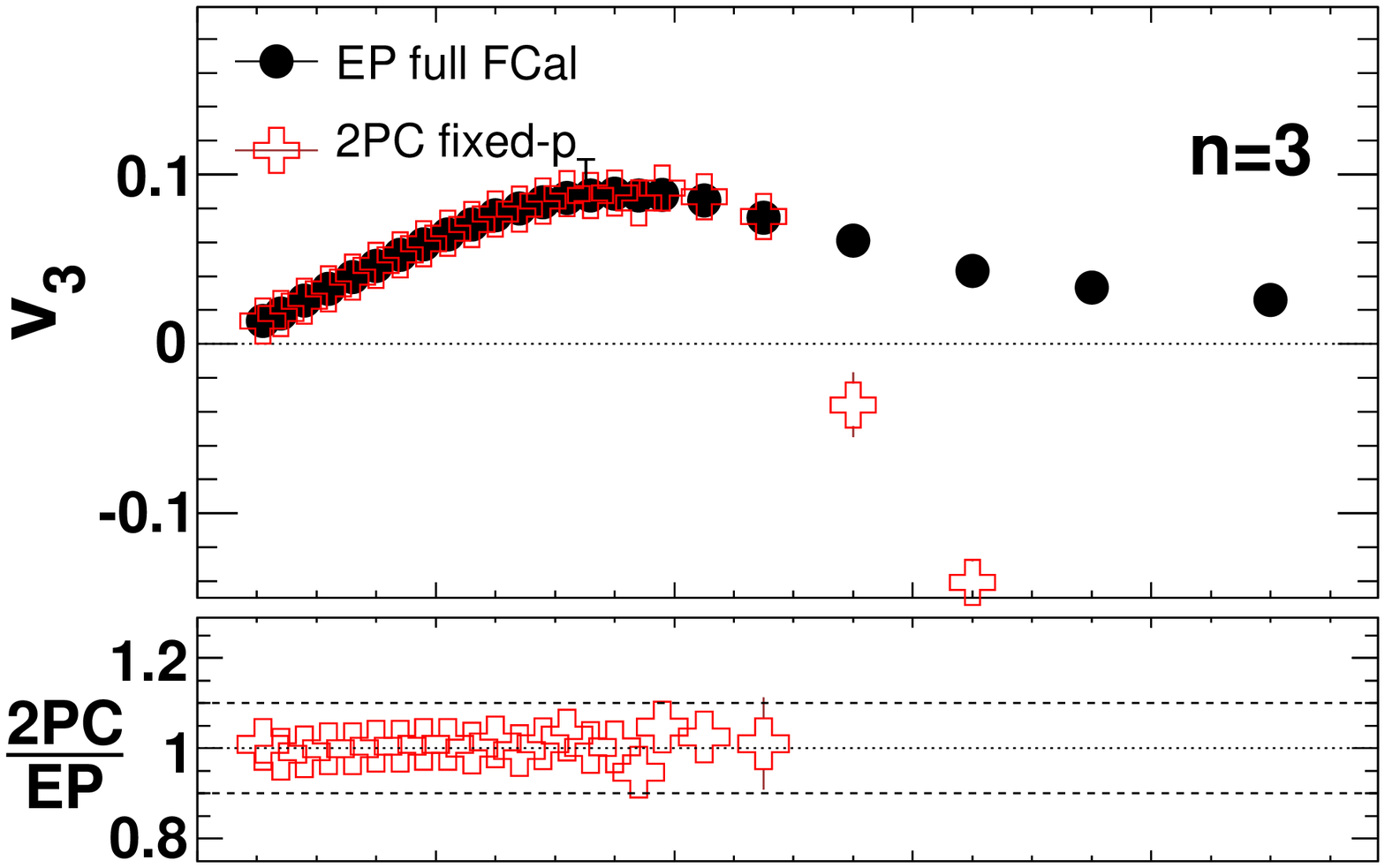}\vspace*{-0.83cm}
\includegraphics[width=0.95\linewidth]{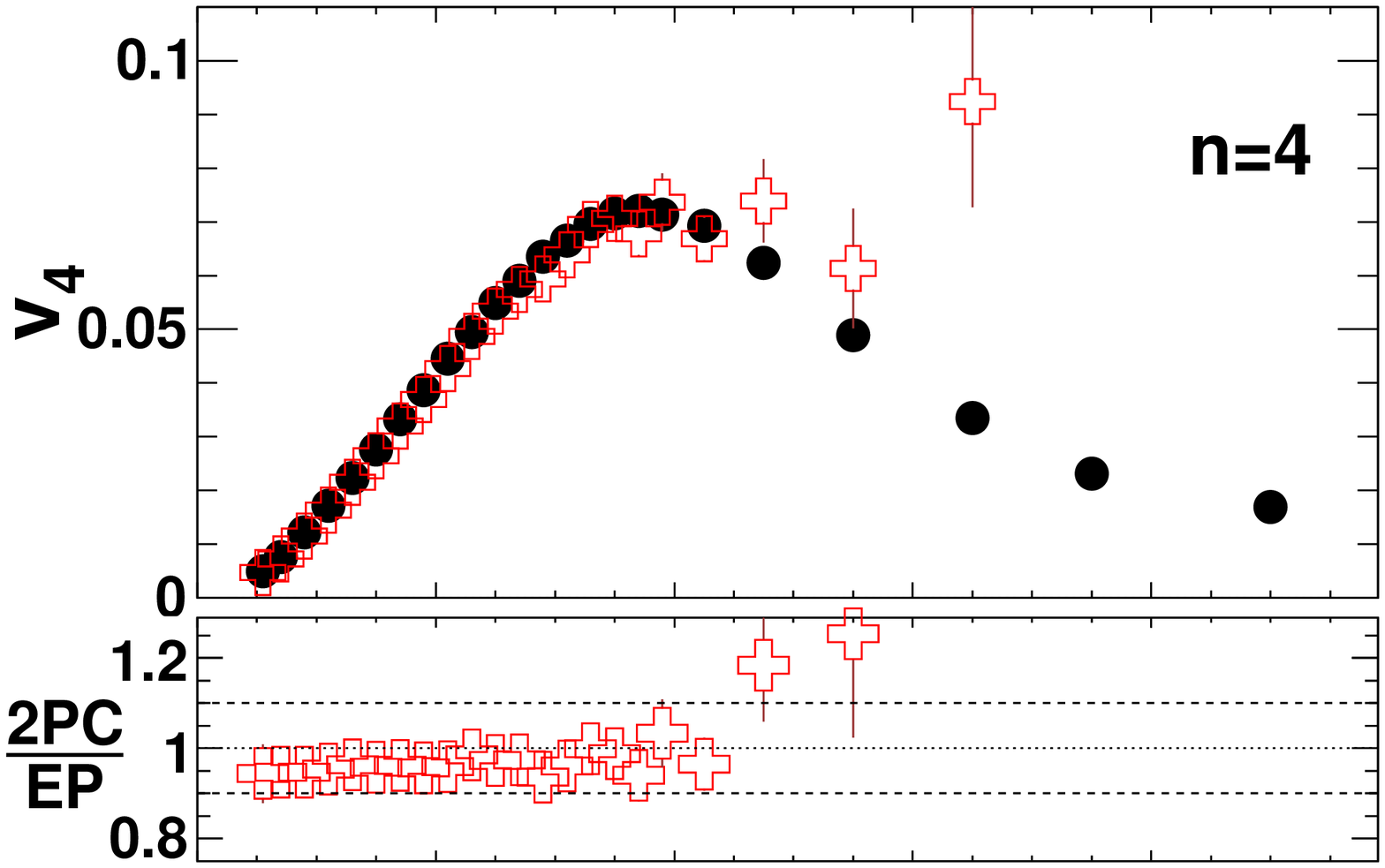}\vspace*{-0.83cm}
\includegraphics[width=0.95\linewidth]{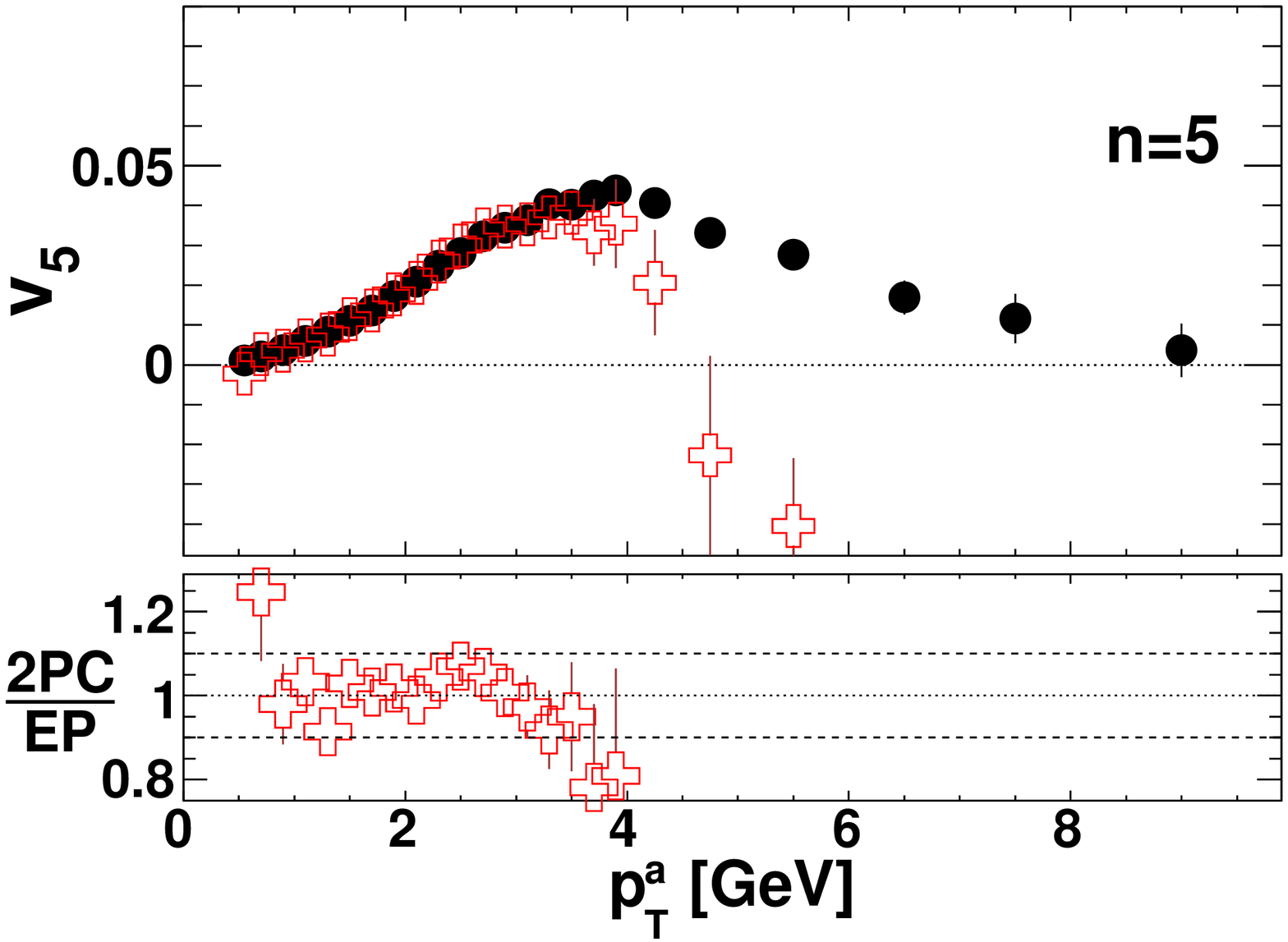}
\caption{\label{fig:disc32} (Color online)  Comparision of $v_n(\pT)$ and the ratios between the fixed-$\pT$ 2PC method and the EP method ($v_2$--$v_5$ from top row to bottom row) for the 0--10\% centrality interval. The error bars indicate the statistical uncertainties only. The dashed lines in the ratio plots indicate a $\pm10$\% band to guide eye.} 
\end{figure}
\begin{figure}[p]
\centering
\includegraphics[width=0.95\linewidth]{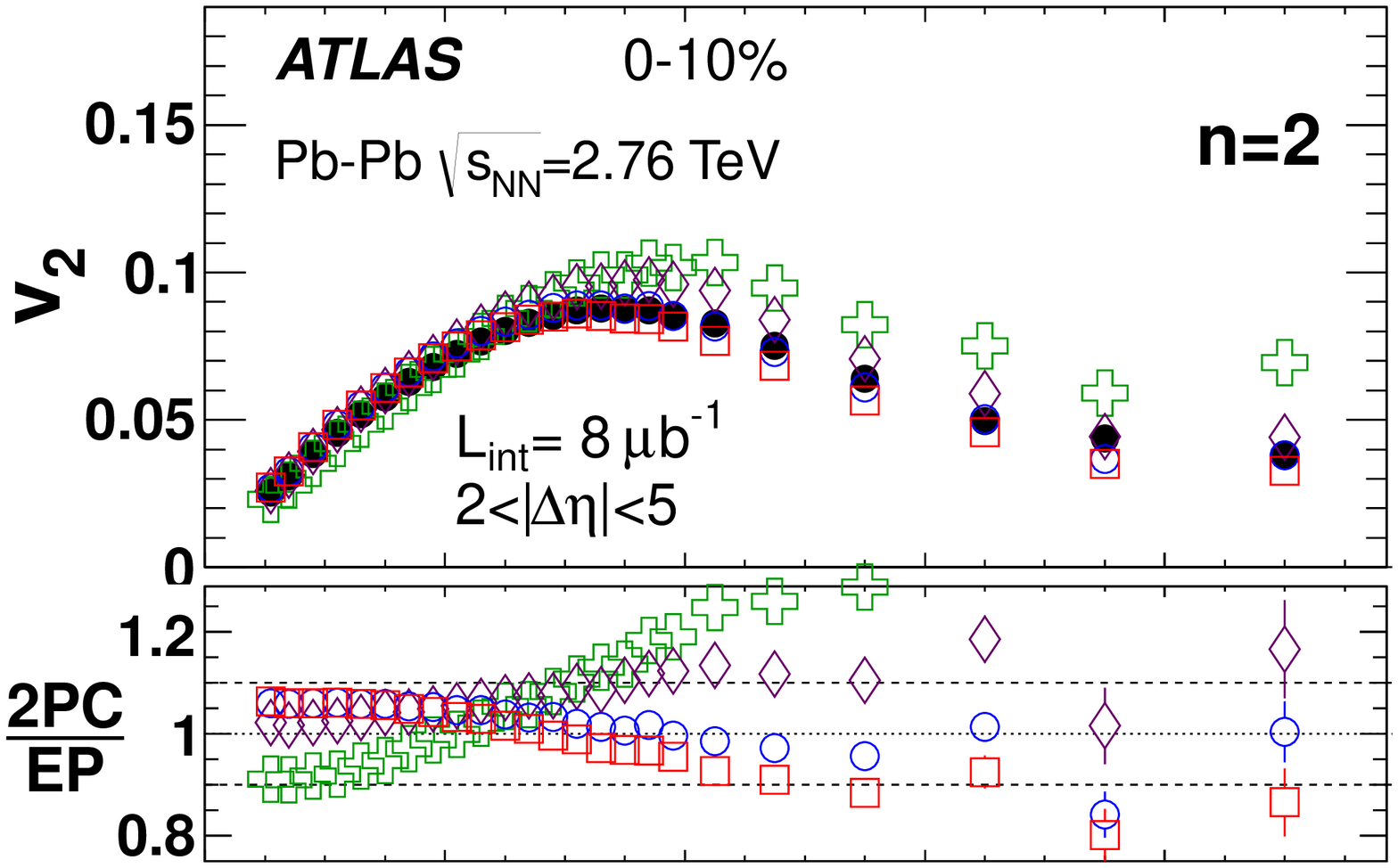}\vspace*{-0.83cm}
\includegraphics[width=0.95\linewidth]{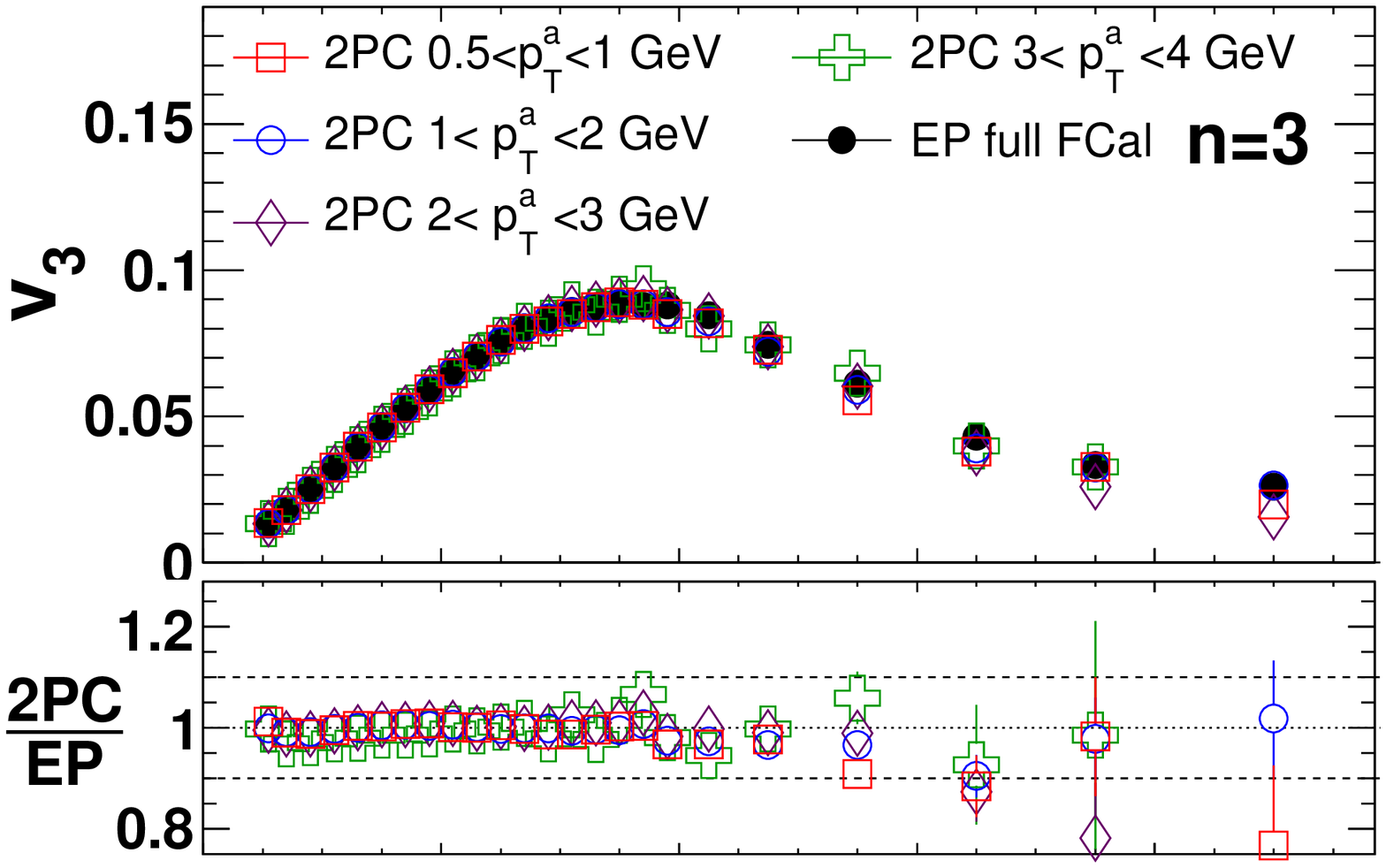}\vspace*{-0.83cm}
\includegraphics[width=0.95\linewidth]{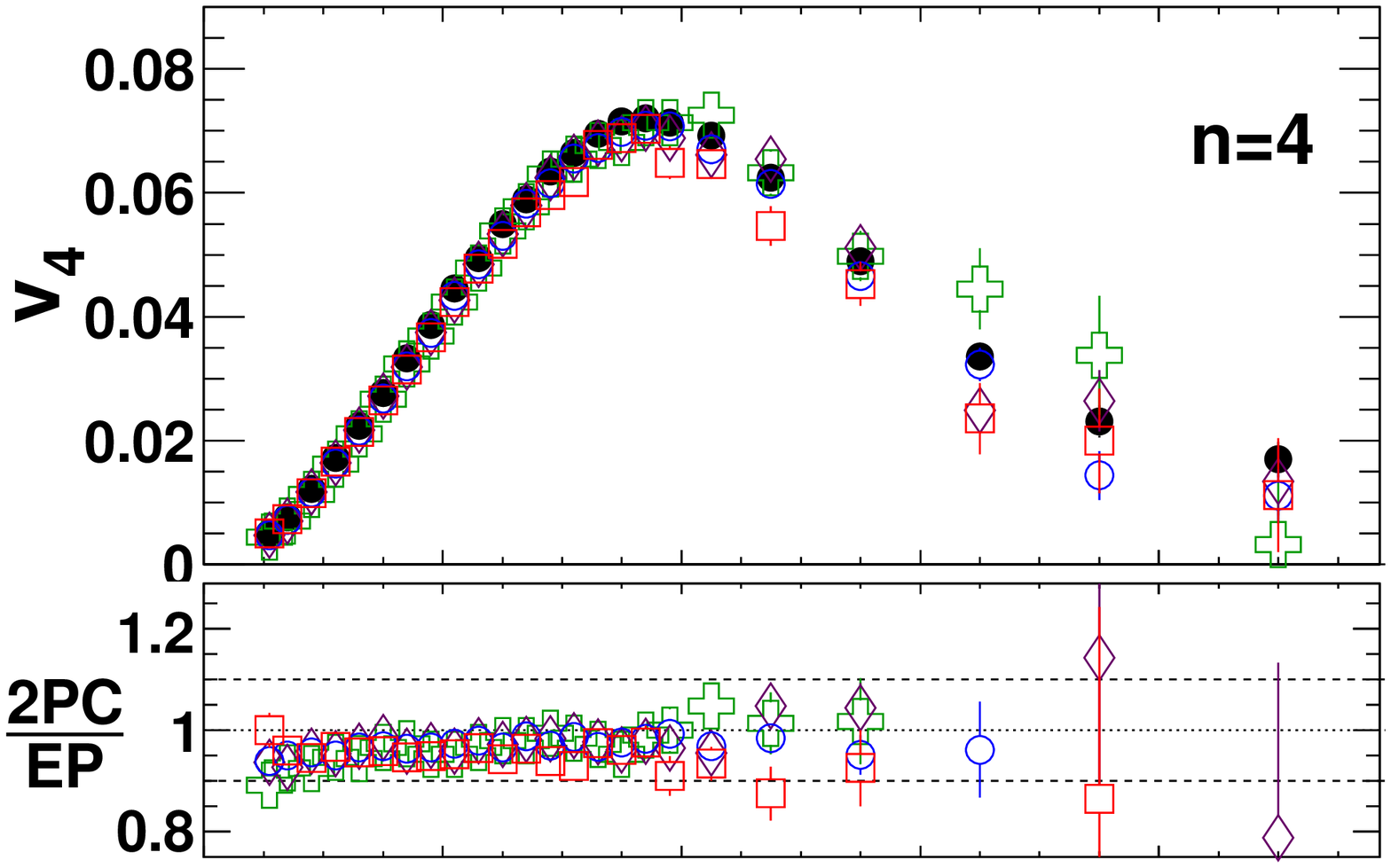}\vspace*{-0.83cm}
\includegraphics[width=0.95\linewidth]{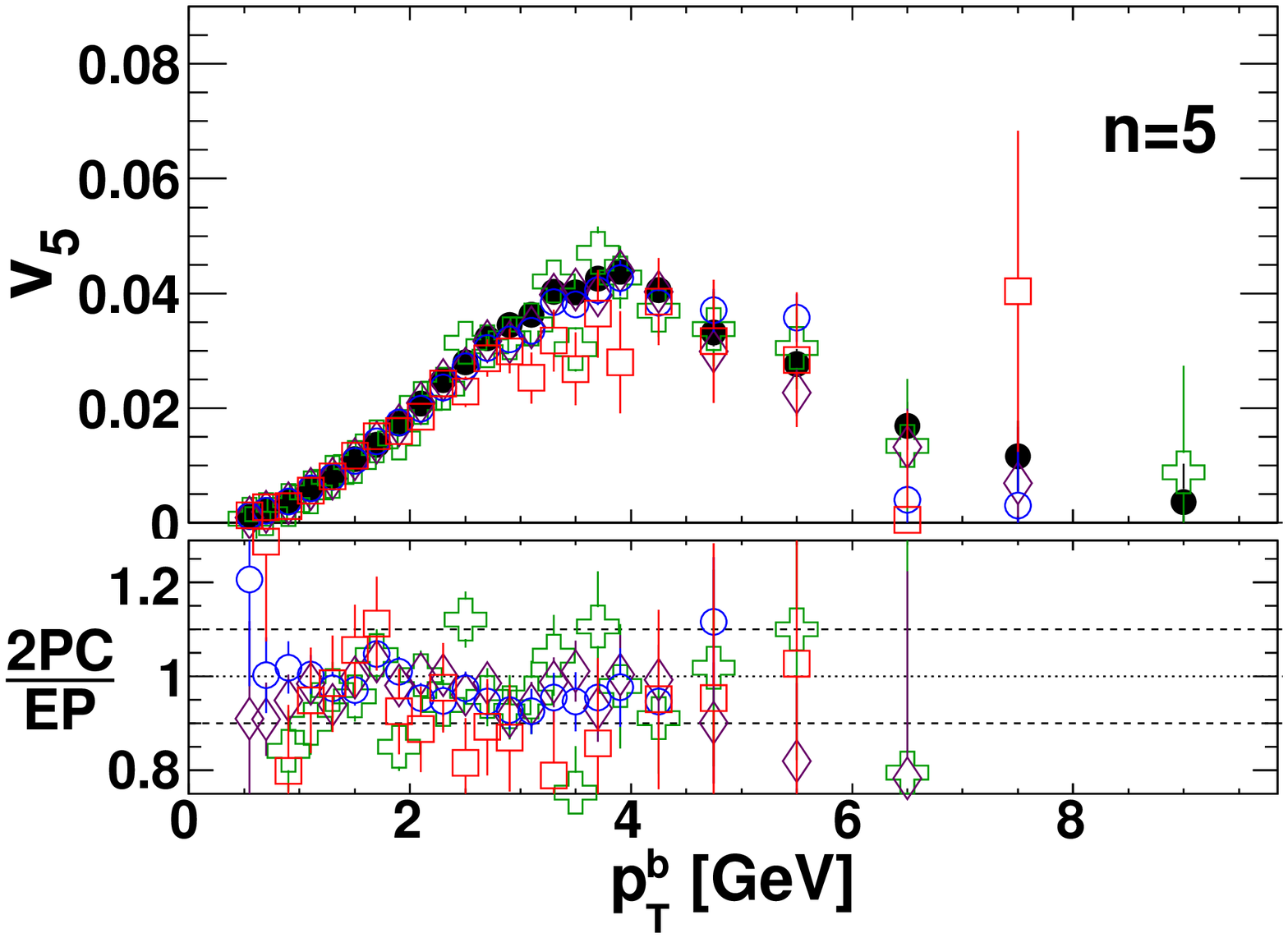}
\caption{\label{fig:disc33} (Color online) Comparison of $v_n(\pT)$ and ratios between the mixed-$\pT$ 2PC method obtained for four reference $\pT^{\mathrm{a}}$ ranges (0.5--1, 1--2, 2--3, 3--4 GeV) and the EP method ($v_2$--$v_5$ from top row to bottom row) for the 0--10\% centrality interval. The error bars indicate the statistical uncertainties only. The dashed lines in the ratio plots indicate a $\pm10$\% band to guide eye.}
\end{figure}

\subsection{Comparison of $v_2$--$v_6$ from the two methods}
\label{sec:re3}
\begin{figure*}[!t]
\begin{tabular}{lr}
\begin{minipage}{0.62\linewidth}
\begin{flushleft}
\includegraphics[width=1\linewidth]{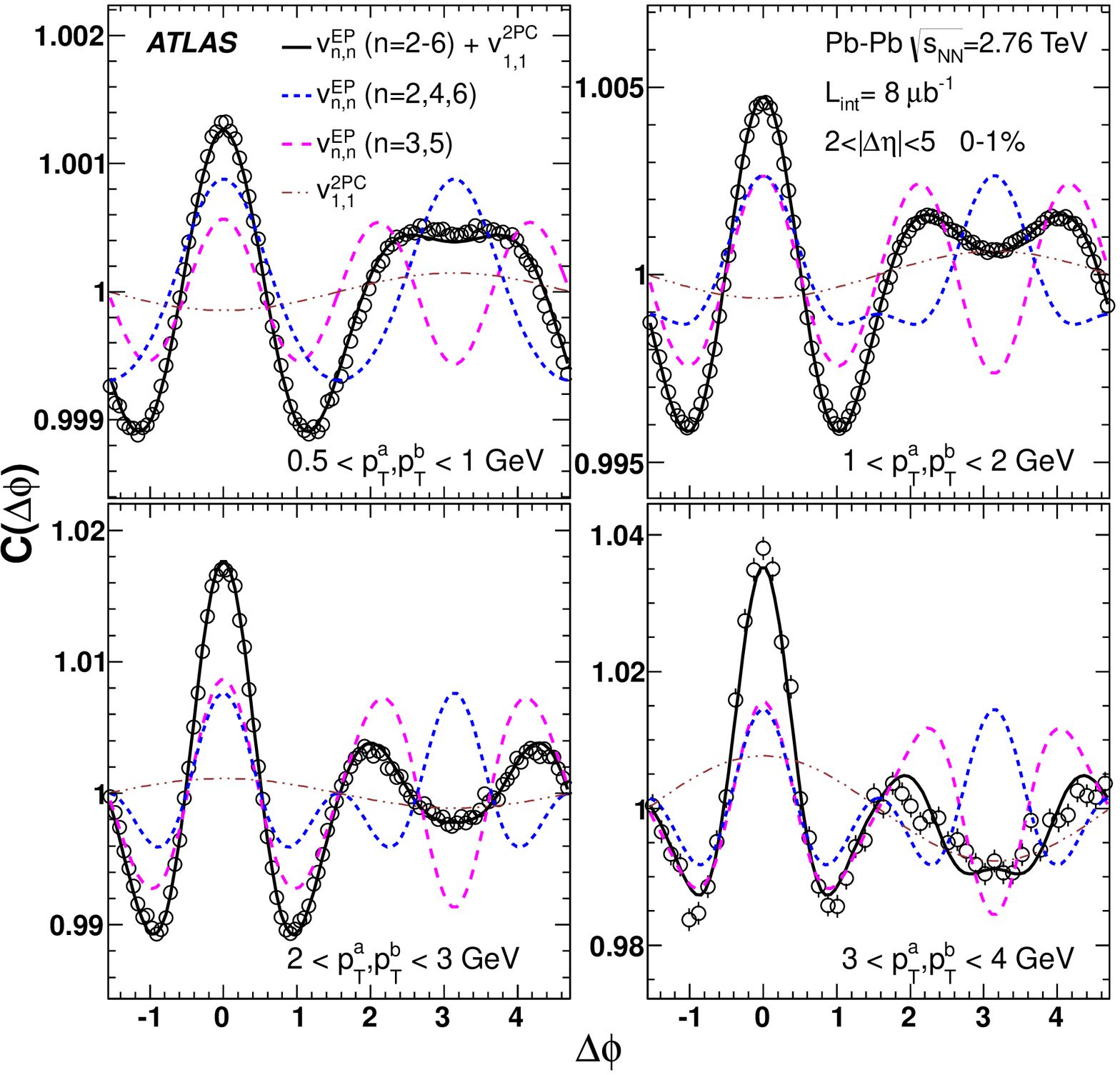}
\end{flushleft}
\end{minipage}
\begin{minipage}{0.35\linewidth}
\begin{flushright}
\caption{\label{fig:rec0b} (Color online) Measured correlation functions compared with those reconstructed from $v_2$--$v_6$ measured from the FCal$_{\mathrm{P(N)}}$ method and $v_{1,1}$ from the 2PC method in 0--1\% centrality interval for four $\pT$ ranges. The contributions from $n=1$, $n=3,5$ and $n=2,4,6$ are shown separately. The error bars indicate the statistical uncertainties.}
\end{flushright}
\end{minipage}
\end{tabular}
\end{figure*}

Figure~\ref{fig:recmp1a} compares the centrality dependence of $v_n$ obtained from the 2PC method and the EP method at low $\pT$. They agree within 5\% for $v_2$--$v_4$, 10\% for $v_5$ and 15\% for $v_6$ over a broad centrality range, well within the quoted systematic uncertainties for the two methods. 

As the contribution of the away-side jet increases with $\pT$, the $v_n$ values from the 2PC method are expected to deviate from the EP results. Figure~\ref{fig:disc32} compares $v_2$--$v_5$ from the fixed-$\pT$ 2PC method with those from the EP method in the 0--10\% centrality interval. The two methods agree within 5\%--15\% for $v_2$ and 5\%--10\% for $v_3$--$v_5$ for $\pT<4$ GeV. Significant deviations are observed for $\pT>4$--5 GeV, presumably due to contributions from the away-side jet (see Fig.~\ref{fig:jetb1}). 

Figure~\ref{fig:disc33} compares the $v_2$--$v_5$ obtained from the mixed-$\pT$ 2PC method with those from the EP method in the 0--10\% centrality interval. The layout of this figure is similar to Fig.~\ref{fig:disc32}, except that the $v_n(\pT^{\mathrm b})$ are calculated using Eq.~\ref{eq:ref} for four reference $\pT^{\mathrm a}$ ranges. Since only one particle is chosen from high $\pT$, the influence of autocorrelations from jets is expected to be less than when both particles are chosen from high $\pT$. Indeed, when $\pT^{\mathrm a}$ is below 3 GeV, the deviations from the EP method are less than 5\%--10\% out to much higher $\pT^{\mathrm b}$, when statistical precision allows. This behavior suggests that as long as the reference $\pT^{\mathrm a}$ is low, the factorization relation is valid and hence the $v_n$ can be measured to high $\pT$ via the 2PC method. It is also noticed that in central events, the agreement for $v_2$ is generally worse than that for the higher-order $v_n$: the differences for $v_2$ between the two methods reach $>20$\% for $\pT^{\mathrm b}>4$ GeV and $3<\pT^{\mathrm a}<4$ GeV, while the differences for the higher-order $v_n$ are small and show no visible dependence on $\pT^{\mathrm b}$ for $\pT^{\mathrm a}<4$ GeV. Comparisons for several other centrality intervals are included in Fig.~\ref{fig:disc33b} in the Appendix. Good agreement is always observed for $v_2$--$v_5$ in non-central events.

The agreement between the 2PC and EP methods implies that the structures of the two-particle correlation function at low $\pT$ and large $|\Delta\eta|$ mainly reflect collective flow. This is verified explicitly by reconstructing the correlation function as:
\begin{eqnarray}
\label{eq:reco}
\nonumber
C(\Delta\phi) &=& b^{\mathrm{2PC}}(1+2v_{1,1}^{\mathrm{2PC}}\cos \Delta\phi\\
&+&2\sum_{n=2}^{6}v_{n}^{\mathrm{EP,a}}v_{n}^{\mathrm{EP,b}}\cos n\Delta\phi)\;,
\end{eqnarray}
where $b^{\mathrm {2PC}}$ and $v_{1,1}^{\mathrm {2PC}}$ are the average of the correlation function and first harmonic coefficient from the 2PC analysis, and the remaining coefficients are calculated from the $v_n$ measured by the FCal$_{\mathrm{P(N)}}$ EP method. Figure~\ref{fig:rec0b} compares the measured correlation functions with those reconstructed using Eq.~\ref{eq:reco}, and they are found to agree well. In general, the correlation functions are reproduced within 5\% of their maximum variations for $\pT<3$ GeV, with slightly larger residual differences for $3<\pT<4$ GeV. The $v_{1,1}$ term, not measured by the EP method, also plays a significant role, but it is not large enough to generate the near- and away-side structures in the correlation function. 

Figure~\ref{fig:rec0b} also shows the separate contributions from odd harmonics $v_3, v_5$ and even harmonics $v_2, v_4, v_6$ measured by the FCal$_{\mathrm{P(N)}}$ EP method. The motivation for doing this is that at $\Delta\phi\sim\pi$, the even harmonics all give positive contributions while the odd harmonics all give negative contributions. Thus, the relative magnitudes of these two contributions control the away-side shape. Figure~\ref{fig:rec0b} shows that the ``double-hump'' structure can be explained by the interplay between the odd and even harmonics.
\subsection{$v_{1,1}$ from the two-particle correlation method and implications for the $v_1$ associated with the dipole asymmetry}
\label{sec:re4}

The previous results show that the factorization properties of $v_{1,1}$ are quite different from those of the higher-order coefficients. The main reason for this is that $v_{1,1}$ is strongly influenced by global momentum conservation, while all higher-order coefficients conserve momentum due to their multi-fold symmetries. One example of momentum conservation effects comes from di-jets in peripheral events, which tend to give a negative $v_{1,1}$ at large $|\Delta\eta|$ (see the $\cos\Delta\phi$ component in the bottom-right panel of Fig.~\ref{fig:reland1c}). The influence of global momentum conservation on the two-particle correlation has been studied extensively~\cite{Borghini:2000cm,Borghini:2002mv,Borghini:2006yk,Luzum:2010fb,Gardim:2011qn}, and was shown to explicitly break the factorization relation Eq.~\ref{eq:fac}~\cite{Borghini:2000cm,Borghini:2002mv}:
\begin{eqnarray}
\nonumber
v_{1,1}(\pT^{\mathrm a},\pT^{\mathrm b},\eta_{\mathrm a},\eta_{\mathrm b}) \approx v_1(\pT^{\mathrm a},\eta_{\mathrm a})v_1(\pT^{\mathrm b},\eta_{\mathrm b})-\frac{\pT^{\mathrm a}\pT^{\mathrm b}}{M\langle \pT^2\rangle}\;,\\\label{eq:v0}
\end{eqnarray}
where $M$ and $\langle \pT^2\rangle$ are the multiplicity and average squared transverse momentum for the whole event, respectively. The negative correction term is a leading-order approximation for momentum conservation. This approximation is expected to be valid when the correction term is much smaller than one. The global momentum conservation effect is important in peripheral events and for high $\pT$ particles, but is diluted in central events due to the large multiplicity. 

\begin{figure}[!t]
\centering
\includegraphics[width=0.9\linewidth]{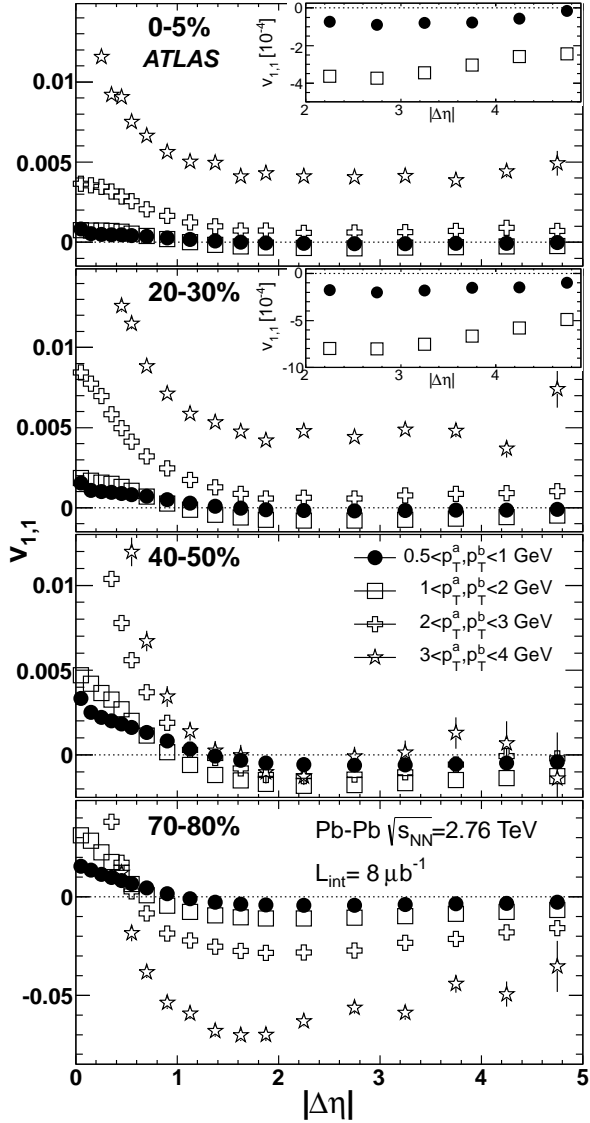}
\caption{\label{fig:v1fac1} $v_{1,1}$ vs. $|\Delta\eta|$ for four $\pT$ ranges in several centrality intervals. The error bars represent the statistical uncertainties. The insert panels show a zoomed-in view of the region $|\Delta\eta|>2$.}
\end{figure}
\begin{figure*}[!t]
\includegraphics[width=0.92\linewidth]{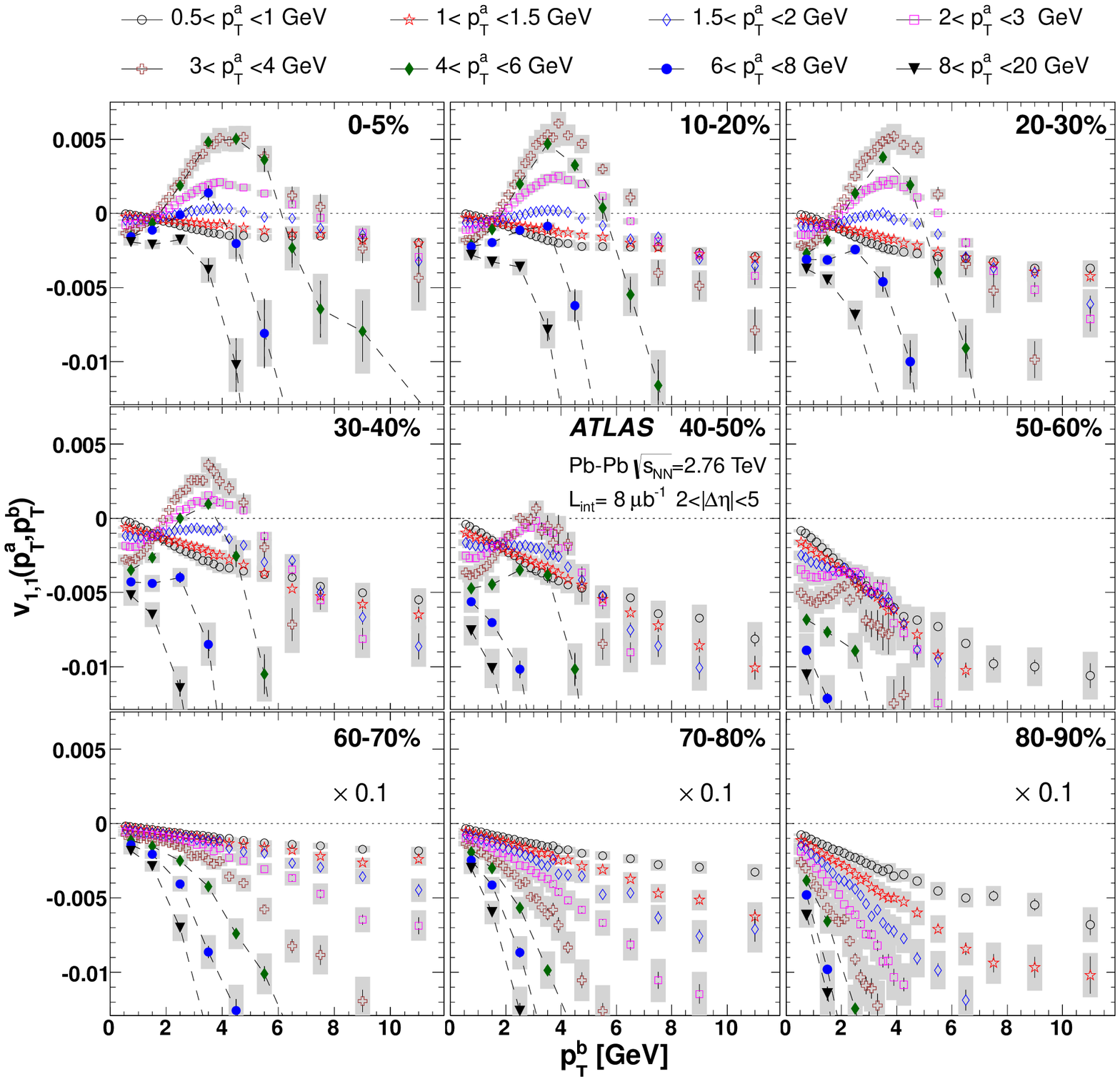}
\caption{\label{fig:v1fac3} (Color online) $v_{1,1}(\pT^{\mathrm a},\pT^{\mathrm b})$ for $2<|\Delta\eta|<5$ vs. $\pT^{\mathrm b}$ for different $\pT^{\mathrm a}$ ranges. Each panel presents results in one centrality interval. The error bars and shaded bands represent statistical and systematic uncertainties, respectively. The data points for the three highest $\pT^{\mathrm a}$ intervals have coarser binning in $\pT^{\mathrm b}$, hence are connected by dashed lines to guide the eye. The data in the bottom three panels are scaled down by a factor of ten to fit within the same vertical range.}
\end{figure*}

\begin{figure*}[!t]
\includegraphics[width=1\linewidth]{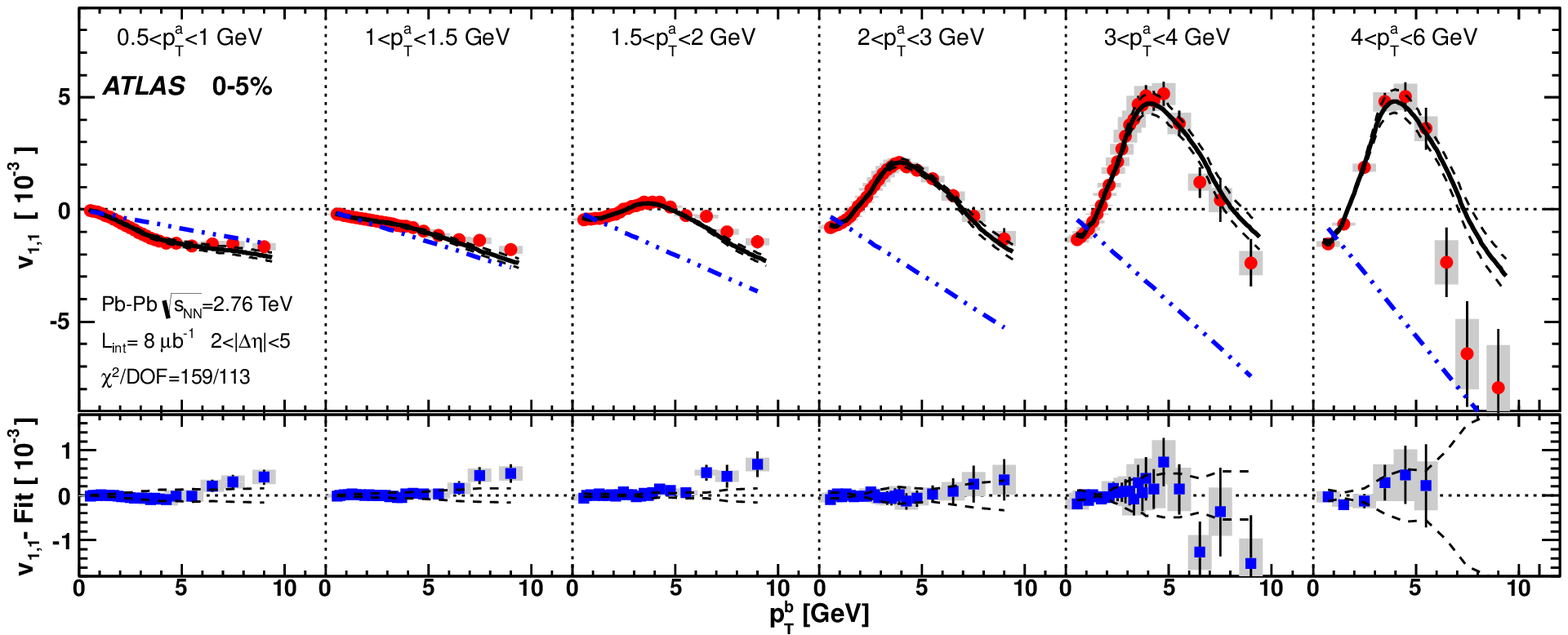}\\
\includegraphics[width=1\linewidth]{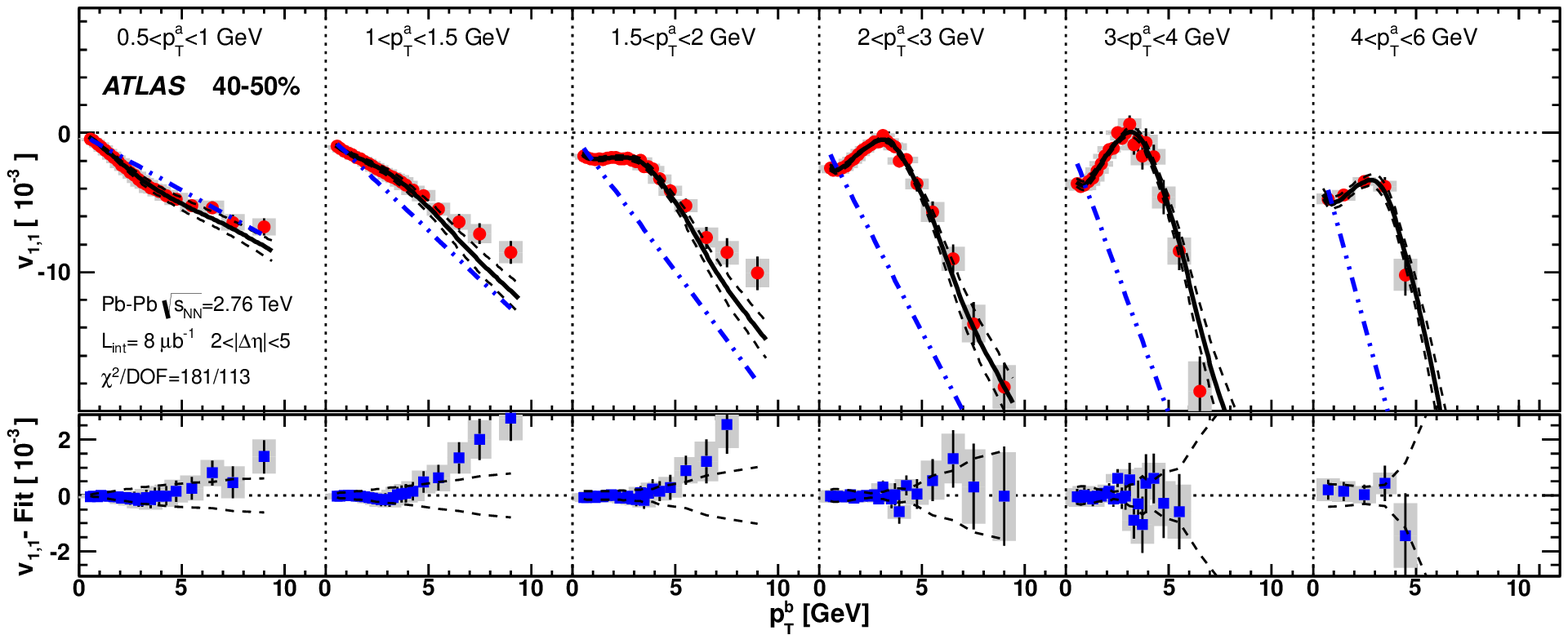}
\caption{\label{fig:v1fit1} (Color online) Global fit (via Eq.~\ref{eq:v2}) to the $v_{1,1}$ data for the 0--5\% centrality interval (top panel) and the (40--50\%) centrality interval (bottom panel). The fit is performed simultaneously over all $v_{1,1}$ data points in a given centrality interval, which are organized as a function of $\pT^{\mathrm b}$ for various $\pT^{\mathrm a}$ ranges (indicated at the top of each panel), with shaded bars indicating the correlated systematic uncertainties. The fit function and the associated systematic uncertainties are indicated by the thick-solid lines and surrounding dashed lines, respectively. The dot-dashed lines intercepting each dataset (with negative slope) indicate estimated contributions from the momentum conservation component. The lower part of each panel shows the difference between the data and fit (solid points), as well as the systematic uncertainties of the fit (dashed lines).}
\end{figure*}
\begin{figure*}[!t]
\includegraphics[width=1\linewidth]{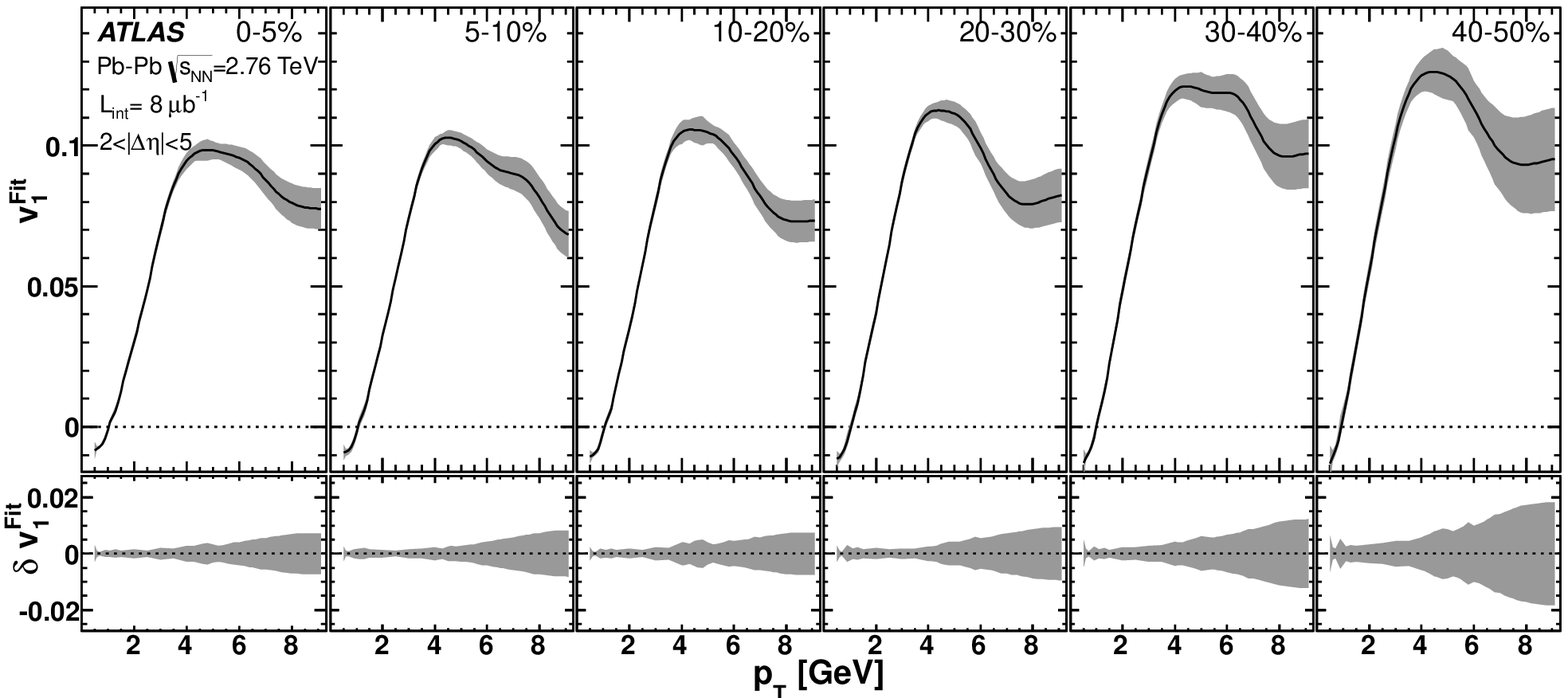}
\caption{\label{fig:v1fit2} $v_1^{\mathrm {Fit}}$ vs. $\pT$ for various centrality intervals. The shaded bands indicate the total uncertainty. The uncertainty bands are reproduced on their own at the bottom of the figure for clarity.}
\end{figure*}
The first term of Eq.~\ref{eq:v0} represents a contribution from $v_1$, whose dependence on pseudorapidity can be generally separated into a rapidity-odd component and a rapidity-even component. The rapidity-odd $v_1$ is thought to arise from the ``sideward'' deflection of the colliding ions~\cite{Voloshin:2008dg}, and changes sign going from negative $\eta$ to positive $\eta$~\cite{Back:2005pc,Abelev:2008jga}. Since pairs with large $|\Delta\eta|$ separation typically select two particles from opposite hemispheres, the rapidity-odd $v_1$ contribution to $v_{1,1}$ is negative at large $|\Delta\eta|$. However, the rapidity-odd $v_1$ is found to be less than 0.005 for $|\eta|<2$ at the LHC~\cite{Selyuzhenkov:2011zj}. The corresponding contribution to $v_{1,1}$ ($<2.5\times10^{-5}$) is negligible compared to the typical $v_{1,1}$ values seen in the data (see below). The rapidity-even $v_1$ signal is believed to arise from the dipole asymmetry of the nuclear overlap due to fluctuations in the initial geometry~\cite{Teaney:2010vd,Gardim:2011qn}. This spatial asymmetry results in a dipole anisotropy of the pressure gradient, which drives the $v_1$. If the rapidity-even $v_1$ depends weakly on $\eta$~\cite{Gardim:2011qn}, similar to the higher-order $v_n$, its contribution to $v_{1,1}$ should be positive and vary weakly with $\Delta\eta$. In this case, Eq.~\ref{eq:v0} can be simplified to:
\begin{eqnarray}
\label{eq:v1}
v_{1,1}(\pT^{\mathrm a},\pT^{\mathrm b}) \approx v_1(\pT^{\mathrm a})v_1(\pT^{\mathrm b})-\frac{\pT^{\mathrm a}\pT^{\mathrm b}}{M\langle \pT^2\rangle}\;.
\end{eqnarray}
A previous analysis of published RHIC data, supports the existence of a significant rapidity-even $v_1$ at lower beam energy~\cite{Luzum:2010fb}. Here the properties of the $v_{1,1}$ data are examined in detail, with the goal of understanding the relative contributions of $v_1$ and global momentum conservation.

Figure~\ref{fig:v1fac1} shows the $|\Delta\eta|$ dependence of the $v_{1,1}$ extracted from two-particle fixed-$\pT$ correlations for several centrality intervals (see discussions around Fig.~\ref{fig:method1}). The $v_{1,1}$ values for $|\Delta\eta|<0.5$ are always positive, presumably due to contributions from the near-side jet or other short range correlations. The $v_{1,1}$ values at large $|\Delta\eta|$, however, have a more complex dependence on centrality and $\pT$. In peripheral events, the $v_{1,1}$ values are always negative at $|\Delta\eta|>1$, and their absolute values increase up to $|\Delta\eta|\sim1.6$ and then decrease for larger $|\Delta\eta|$. This trend is consistent with the gradual falloff of the away-side jet contribution at large $|\Delta\eta|$ seen in 2-D correlation function, e.g. the last panel of Fig.~\ref{fig:reland1}. In central and mid-central events, the $v_{1,1}$ values at $|\Delta\eta|>1$ are negative for $\pT^{\mathrm a},\pT^{\mathrm b}<2$ GeV, but they become positive and relatively independent of $|\Delta\eta|$ at higher $\pT$, consistent with the contribution from a rapidity-even $v_{1}$ that varies weakly with $\eta$.

Given that the $|\Delta\eta|$ dependence of Fig.~\ref{fig:v1fac1} is generally weak at large $|\Delta\eta|$, $v_{1,1}$ is integrated over $2<|\Delta\eta|<5$ to obtain one value for each $\pT^{\mathrm a}$ and $\pT^{\mathrm b}$ combination. The $v_{1,1}(\pT^{\mathrm a},\pT^{\mathrm b})$ functions are then shown in Fig.~\ref{fig:v1fac3} for various centrality intervals. In peripheral events, the $v_{1,1}$ values are always negative and their absolute values increase nearly linearly with $\pT^{\mathrm a}$ and $\pT^{\mathrm b}$, presumably dominated by autocorrelations from the away-side jet. In more central events, the absolute value of this negative $v_{1,1}$ component is smaller, reflecting a dilution of the momentum conservation term by the large event multiplicity. Furthermore, Fig.~\ref{fig:v1fac3} clearly suggests that a positive $v_{1,1}$ component sets in for $2\lesssim\pT\lesssim6$ GeV. Its magnitude increases for more central events and eventually drives $v_{1,1}$ into the positive region. The $v_{1,1}$ values reach a maximum at around 4 GeV, similar to those for the higher-order harmonic coefficients (see Fig.~\ref{fig:respt1}).

To validate the two-component model of Eq.~\ref{eq:v1} as a reasonable interpretation of the $v_{1,1}$ data and to extract the $v_1$, a least-squares fit of the $v_{1,1}$ data is performed for each centrality interval with the following form:
\begin{eqnarray}
\nonumber\chi^2 = \sum_{\mathrm {a,b}}\frac{\left(v_{1,1}(\pT^{\mathrm a},\pT^{\mathrm b}) - [v_1^{\mathrm{Fit}}(\pT^{\mathrm a})v_1^{\mathrm{Fit}}(\pT^{\mathrm b})-c\pT^{\mathrm a}\pT^{\mathrm b}]\right)^2}{\left(\sigma_{\mathrm {a,b}}^{\mathrm{stat}}\right)^2+\left(\sigma_{\mathrm {a,b}}^{\mathrm{sys,p2p}}\right)^2}\;,\\\label{eq:v2}
\end{eqnarray}
where $\sigma_{\mathrm {a,b}}^{\mathrm {stat}}$ and $\sigma_{\mathrm {a,b}}^{\mathrm {sys,p2p}}$ denote the statistical and point to point systematic uncertainties for $v_{1,1}(\pT^{\mathrm a},\pT^{\mathrm b})$, respectively (see Section~\ref{sec:m2}). The $v_1^{\mathrm{Fit}}(\pT)$ function is defined via a smooth interpolation of its values at $m$ discrete $\pT$ points, $v_1^{\mathrm{Fit}}(p_{\mathrm T,i})|_{i=1}^m$, and these together with the parameter $c$ constitute a total of $m+1$ fit parameters. An interpolation procedure is used because the functional form of $v_1^{\mathrm{Fit}}(\pT)$ is, a priori, unknown, yet it is expected to vary smoothly with $\pT$.

In the default setup, the $v_{1,1}$ data used in the fit are restricted to $\pT^{\mathrm a}<6$ GeV and $\pT^{\mathrm b}<10$ GeV, giving a total of 129 data points for each centrality interval. A cubic-spline interpolation procedure is used, and the number of interpolation points is chosen to be $m=15$ (listed in Table~\ref{tab:fit}). To account for the $\pT$-correlated uncertainty ($\sigma_{\mathrm {a,b}}^{\mathrm {sys,corr}}$, see Section~\ref{sec:m2}), the least-squares minimization is repeated after varying all the $v_{1,1}$ data points up or down by $\sigma_{\mathrm {a,b}}^{\mathrm {sys,corr}}$. Since $v_{1,1}(\pT^{\mathrm a},\pT^{\mathrm b})$ is symmetric with respect to $\pT^{\mathrm a}$ and $\pT^{\mathrm b}$, data points for $\pT^{\mathrm a}<\pT^{\mathrm b}$ are correlated with those for $\pT^{\mathrm a}>\pT^{\mathrm b}$. The effect of this correlation is evaluated by repeating the fit using only $v_{1,1}$ data for $\pT^{\mathrm a}\leq\pT^{\mathrm b}$ or $v_{1,1}$ data for $\pT^{\mathrm a}\geq\pT^{\mathrm b}$. The variations from the default fit are included in the systematic uncertainties.

Figure~\ref{fig:v1fit1} shows the fit to the $v_{1,1}$ data from Fig.~\ref{fig:v1fac3} for the 0--5\% and 40--50\% centrality intervals. The $v_{1,1}$ data are plotted as a function of $\pT^{\mathrm b}$ for six intervals of $\pT^{\mathrm a}$. The seemingly-complex patterns of the $v_{1,1}$ data are well described by the two-component fit across broad ranges of $\pT^{\mathrm a}$ and $\pT^{\mathrm b}$, with the dot-dashed lines indicating the estimated contributions from the global momentum conservation. The typical $\chi^2/$DOF of the fit is between one and two depending on the centrality, as shown in Table~\ref{tab:fit}. The deviations of the data from the fit, as shown in the bottom section of each panel, are less than $10^{-4}$ for $\pT<5$ GeV. Above that $\pT$ and in more peripheral events the deviations increase, possibly reflecting the limitation of the leading-order approximation for the momentum conservation term in Eq.~\ref{eq:v1}, or the two-component assumption in general. In this paper, the study is restricted to the 0--50\% most central events, for which the first term in Eq.~\ref{eq:v1} is comparable or larger than the second term and the statistical uncertainty of the fit is not too large.

The stability of the least-square minimization procedure is checked for three types of variations: 1) the interpolation function is varied from cubic-spline interpolation to a linear interpolation, 2) the number of interpolation points is varied from 9 to 21, and 3) the $\pT$ range of the fit is varied from $\pT^{\mathrm a},\pT^{\mathrm b}<5$ GeV to $\pT^{\mathrm a},\pT^{\mathrm b}<10$ GeV. The systematic uncertainties of the minimization procedure are calculated as the RMS sum of these three variations. They are small in central events but become substantial in peripheral events, and are important sources of uncertainties at intermediate $\pT$. The statistical uncertainty of the fit dominates at high $\pT$, while the fit uncertainty from $\sigma_{\mathrm {a,b}}^{\mathrm {sys,corr}}$ dominates at low $\pT$. The total absolute uncertainty of $v_1^{\mathrm{Fit}}$ is $\delta v_1^{\mathrm{Fit}}=0.001$--0.004 for $\pT<3$ GeV and increases rapidly for higher $\pT$ due to the greater statistical uncertainty.

Figure~\ref{fig:v1fit2} shows $v_1^{\mathrm {Fit}}(\pT)$ for various centrality intervals. A significant $v_1^{\mathrm{Fit}}$ signal is observed for all cases. It reaches a maximum between 4 GeV and 5 GeV and then falls at higher $\pT$. This falloff may indicate the onset of path-length dependent jet energy loss, which correlates with the dipole asymmetry in the initial geometry similar to higher-order $v_n$. The magnitude of the $v_1^{\mathrm{Fit}}$ is large: its peak value is comparable to that for the $v_3$ shown in Fig.~\ref{fig:respt1}, and the peak value increases by about 20\% over the measured centrality range. These results imply that the rapidity-even collective $v_1$ is an important component of the two-particle correlation at intermediate $\pT$. For example, the large positive $v_{1,1}$ harmonic seen in the bottom-right panel of Fig.~\ref{fig:rec0b} is mainly due to collective $v_1$: its contribution to $v_{1,1}$ is about three times larger than the negative momentum conservation term estimated by the global fit (top panel of Fig.~\ref{fig:v1fit1}). $v_1^{\mathrm{Fit}}(\pT)$ is negative for $\pT\lesssim1.0$ GeV, confirming a generic feature expected for collective $v_1$ as suggested by hydrodynamic model calculations~\cite{Teaney:2010vd,Gardim:2011qn}. This behavior, together with the fact that $v_{1,1}$ data show little $|\Delta\eta|$ dependence for $|\Delta\eta|>2$ (Fig.~\ref{fig:v1fac1}), is consistent with a rapidity-even $v_1$ that is almost independent of $\eta$. 

If the two-component ansatz of Eq.~\ref{eq:v1} is valid, the fit parameter $c$ should be inversely proportional to multiplicity $M$ and $\langle\pT^2\rangle$ of the whole event. This ansatz is checked by calculating the product of $c$ and the charged hadron multiplicity at mid-rapidity $\frac{dN}{d\eta}_{|\eta=0}$ from~\cite{Collaboration:2011yr}, with the assumption that $\frac{dN}{d\eta}_{|\eta=0}$ is proportional to $M$. The results are summarized in Table~\ref{tab:fit} for each centrality interval. Since $\langle\pT^2\rangle$ for the whole event is expected to vary weakly with centrality (the $\langle\pT\rangle$ for charged pions at mid-rapidity only varies by $\sim5\%$ within the 0--50\% centrality interval at the LHC~\cite{Preghenella:2011vy}), the product is also expected to vary weakly with centrality. Table~\ref{tab:fit} shows that this is indeed the case, supporting the assumptions underlying Eq.~\ref{eq:v1}.

\begin{table}[!h]
\begin{center}
\begin{ruledtabular}\begin{tabular}{l c c c }
Centrality        & $\chi^2$/DOF& $c$[$\times$0.001GeV$^{-2}$]  & $c\frac{dN}{d\eta}_{|\eta=0}$[$\times$GeV$^{-2}$]\tabularnewline\hline
   0-5\%          & 159/113     & $0.24\pm0.02$   & $0.39\pm 0.04$\tabularnewline\hline
   5-10\%         & 133/113     & $0.28\pm0.02$   & $0.37\pm 0.04$\tabularnewline\hline
   10-20\%        & 165/113     & $0.35\pm0.03$   & $0.36\pm 0.04$\tabularnewline\hline
   20-30\%        & 134/113     & $0.50\pm0.04$   & $0.34\pm 0.03$\tabularnewline\hline
   30-40\%        & 188/113     & $0.75\pm0.05$   & $0.33\pm 0.03$ \tabularnewline\hline
   40-50\%        & 181/113     & $1.16\pm0.09$   & $0.32\pm 0.03$\tabularnewline\hline\hline
 \multicolumn{4}{c}{15 interpolation points used in the default fit:}  \tabularnewline
 \multicolumn{4}{c}{0.5, 0.7, 0.9, 1.1, 1.3, 1.5, 2.0, 2.5}\tabularnewline
 \multicolumn{4}{c}{3.0, 3.5, 4.5, 5.5, 6.5, 7.5, 9.0 GeV} \tabularnewline
\end{tabular}\end{ruledtabular}
\end{center}
\caption{\label{tab:fit} Quality of the fit $\chi^2$/DOF, fit parameter $c$, and corresponding multiplicity scaled values $c\frac{dN}{d\eta}_{|\eta=0}$ for various centrality intervals. The uncertainty of $c\frac{dN}{d\eta}_{|\eta=0}$ is calculated as the quadrature sum of uncertainties from $c$ and $\frac{dN}{d\eta}_{|\eta=0}$ of Ref.~\cite{Collaboration:2011yr}. The bottom row lists the 15 $\pT$ interpolation points used in the default fit.}
\end{table}
\section{Conclusion}
\label{sec:con}

In summary, differential measurements of harmonic coefficients $v_1$--$v_6$ for the azimuthal distributions of charged particles are presented for lead-lead collisions at $\sqrt{s_{\mathrm {NN}}}=2.76$ TeV, based on an integrated luminosity of approximately 8~$\mu\mathrm{b}^{-1}$ recorded by ATLAS. The $v_n$ values are measured for $n=2$--6 using an event plane method, and are also derived for $n=1$--6 from the $v_{n,n}$ measured in a two-particle correlation method.

In the event plane method, $v_2$--$v_6$ are extracted as a function of transverse momentum ($0.5<\pT<20$ GeV), pseudorapidity ($|\eta|<2.5$) and centrality (0--80\%) by correlating tracks with the event plane determined at large rapidity. The $v_n$ values exhibit a weak $\eta$ dependence, slightly decreasing toward larger $|\eta|$. The $v_n$ values show a similar $\pT$ dependence, namely, increasing with $\pT$ to a maximum around 3--4 GeV and then decreasing for higher $\pT$. The higher-order coefficients exhibit stronger $\pT$ variations. They follow an approximate scaling relation, $v_n^{1/n}(\pT)\propto v_2^{1/2}(\pT)$, except in top 5\% most central collisions. Furthermore, the coefficients $v_3$--$v_6$ show little dependence on centrality, consistent with an anisotropy primarily associated with fluctuations in the initial geometry. In contrast, $v_2$ varies strongly with centrality, reflecting an anisotropy mainly associated with a change in the average elliptic geometry with centrality. 

In the two-particle azimuthal correlation method, harmonic coefficients for the distributions of the charged particle pairs in $\Delta\phi$, $v_{n,n}=\langle\cos n\Delta\phi\rangle$, are extracted over a broad range of $\pT$, relative pseudorapidity ($|\Delta\eta|<5$) and centrality. For pairs of charged particles with a large pseudorapidity gap ($|\Delta\eta|>2$) and one particle at $\pT<3$ GeV, the $v_{2,2}$--$v_{6,6}$ are found to factorize into the product of corresponding $v_n$ harmonics, i.e. $v_{n,n}(\pT^{\mathrm a},\pT^{\mathrm b}) \approx v_n(\pT^{\mathrm a})v_n(\pT^{\mathrm b})$, in central and mid-central events. This suggests that these values of $v_{2,2}$--$v_{6,6}$ are consistent with the response of the created matter to fluctuations in the geometry of the initial state. This factorization does not describe the $v_{2,2}$--$v_{6,6}$ data in more peripheral collisions or at higher $\pT$, primarily due to autocorrelations from di-jets. The factorization relation is also found to fail for $v_{1,1}$ over the entire $\pT$ range. A detailed investigation of the $v_{1,1}(|\Delta\eta|,\pT^{\mathrm a},\pT^{\mathrm b})$ data suggests that they are consistent with the combined contributions from a rapidity-even $v_1$ component, and a global momentum conservation component that increases linearly with the $\pT^{\mathrm a}\times\pT^{\mathrm b}$ and is inversely proportional to the event multiplicity. Motivated by this observation, a two-component fit is used to extract $v_1$. The derived signal is negative at low $\pT$, changes sign at $\pT\approx1.0$ GeV, reaches a maximum at around 4--5 GeV and then decreases at higher $\pT$. The magnitude of $v_1$ at the peak is comparable to that of $v_3$.

At low $\pT$ where single-particle anisotropy is dominated by collective flow, the two-particle correlation functions are found to be well reproduced by combining the contributions of $v_2$--$v_6$ from the event plane method and the $v_{1,1}$ (Eq.~\ref{eq:reco}). This implies that the main structures of the low $\pT$ two-particle correlation for $|\Delta\eta|>2$ largely reflect higher-order collective flow ($n\ge2$), together with a $\cos\Delta\phi$ term that contains contributions from both a rapidity-even $v_1$ and the global momentum conservation. Therefore, the low $\pT$ correlation functions do not allow significant contributions from jet-induced medium responses. The fluctuations in the initial geometry together with the low viscosity of the created matter are potentially responsible for these $v_{n,n}$ coefficients. A detailed comparison of the comprehensive $v_n$ and $v_{n,n}$ data presented in this paper with viscous hydrodynamic model calculations at low $\pT$~\cite{Qin:2010pf,Schenke:2011tv,Qiu:2011iv,Staig:2011wj,Schenke:2011bn,Qiu:2011hf} and jet energy loss model calculations at high $\pT$~\cite{Bass:2008rv,Majumder:2010qh,Betz:2011tu,Horowitz:2011cv} may help elucidate the nature of these fluctuations and better constrain the transport properties of the medium.
\section*{Acknowledgements}

We thank CERN for the very successful operation of the LHC, as well as the
support staff from our institutions without whom ATLAS could not be
operated efficiently.

We acknowledge the support of ANPCyT, Argentina; YerPhI, Armenia; ARC,
Australia; BMWF, Austria; ANAS, Azerbaijan; SSTC, Belarus; CNPq and FAPESP,
Brazil; NSERC, NRC and CFI, Canada; CERN; CONICYT, Chile; CAS, MOST and NSFC,
China; COLCIENCIAS, Colombia; MSMT CR, MPO CR and VSC CR, Czech Republic;
DNRF, DNSRC and Lundbeck Foundation, Denmark; ARTEMIS and ERC, European Union;
IN2P3-CNRS, CEA-DSM/IRFU, France; GNAS, Georgia; BMBF, DFG, HGF, MPG and AvH
Foundation, Germany; GSRT, Greece; ISF, MINERVA, GIF, DIP and Benoziyo Center,
Israel; INFN, Italy; MEXT and JSPS, Japan; CNRST, Morocco; FOM and NWO,
Netherlands; RCN, Norway; MNiSW, Poland; GRICES and FCT, Portugal; MERYS
(MECTS), Romania; MES of Russia and ROSATOM, Russian Federation; JINR; MSTD,
Serbia; MSSR, Slovakia; ARRS and MVZT, Slovenia; DST/NRF, South Africa;
MICINN, Spain; SRC and Wallenberg Foundation, Sweden; SER, SNSF and Cantons of
Bern and Geneva, Switzerland; NSC, Taiwan; TAEK, Turkey; STFC, the Royal
Society and Leverhulme Trust, United Kingdom; DOE and NSF, United States of
America.

The crucial computing support from all WLCG partners is acknowledged
gratefully, in particular from CERN and the ATLAS Tier-1 facilities at
TRIUMF (Canada), NDGF (Denmark, Norway, Sweden), CC-IN2P3 (France),
KIT/GridKA (Germany), INFN-CNAF (Italy), NL-T1 (Netherlands), PIC (Spain),
ASGC (Taiwan), RAL (UK) and BNL (USA) and in the Tier-2 facilities
worldwide.

\appendix
\section{Comprehensive data plots from two-particle correlation}
\label{app:1}
\begin{figure*}[h]
\centering
\includegraphics[width=0.98\linewidth]{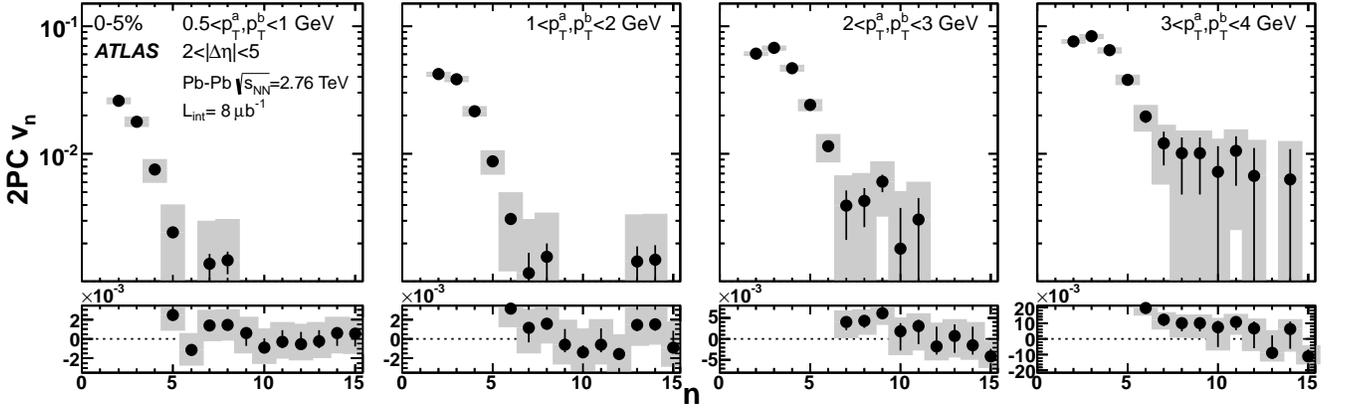}
\caption{\label{fig:ren1} $v_n$ vs. $n$ for $n\geq2$ in 0--5\% centrality interval for four fixed-$\pT$ correlations (0.5--1, 1--2, 2--3 and 3--4 GeV from left to right). The error bars and shaded bands indicate the statistical and total systematic uncertainties, respectively.}
\end{figure*}

\begin{figure*}[h]
\centering
\includegraphics[width=0.98\linewidth]{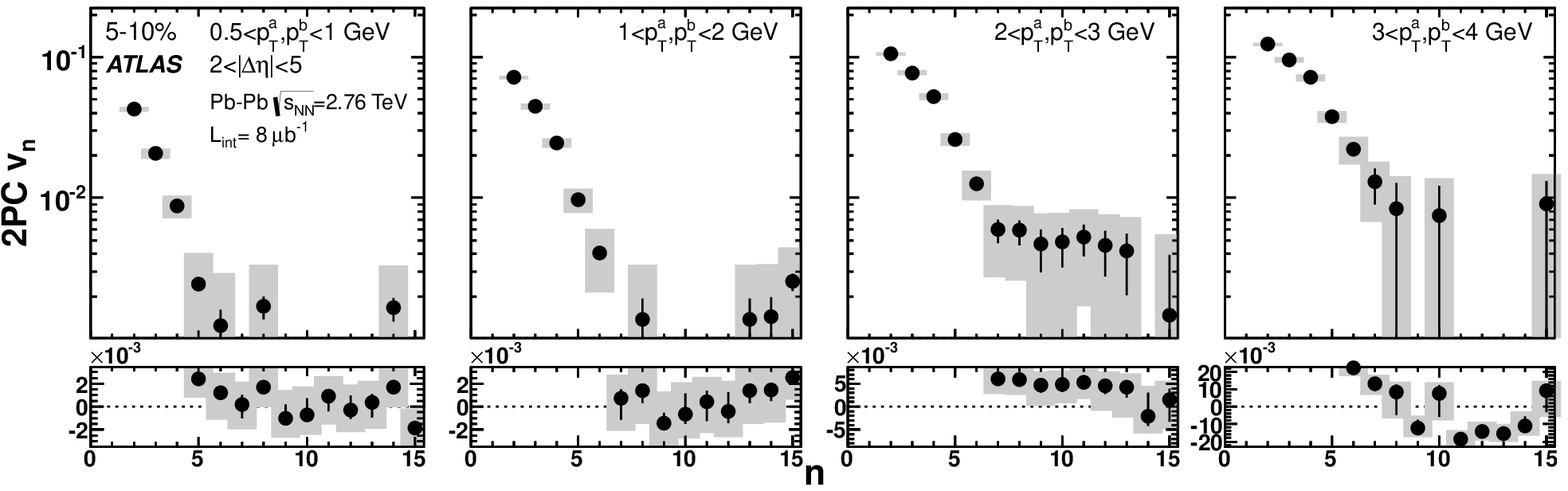}
\caption{\label{fig:ren2} $v_n$ vs. $n$ for $n\geq2$ in 5--10\% centrality interval for four fixed-$\pT$ correlations (0.5--1, 1--2, 2--3 and 3--4 GeV from left to right). The error bars and shaded bands indicate the statistical and total systematic uncertainties, respectively.}
\end{figure*}

\begin{figure*}[h]
\centering
\includegraphics[width=0.98\linewidth]{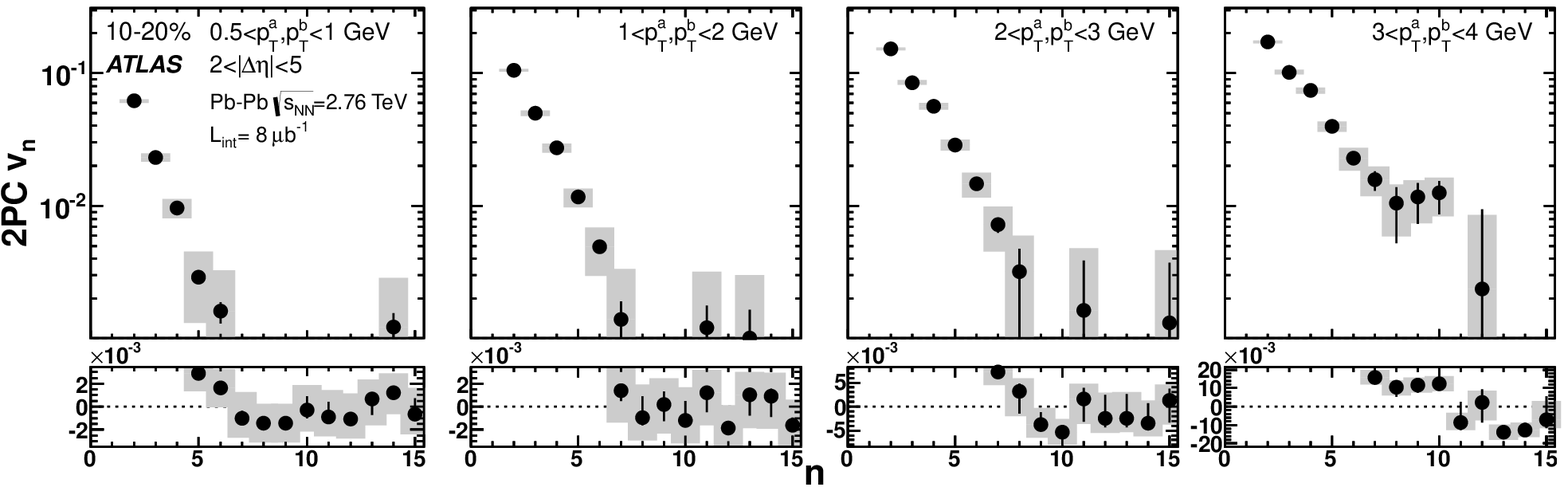}
\caption{\label{fig:ren4} $v_n$ vs. $n$ for $n\geq2$ in 10--20\% centrality interval for four fixed-$\pT$ correlations (0.5--1, 1--2, 2--3 and 3--4 GeV from left to right). The error bars and shaded bands indicate the statistical and total systematic uncertainties, respectively.}
\end{figure*}

\begin{figure*}[h]
\centering
\includegraphics[width=0.98\linewidth]{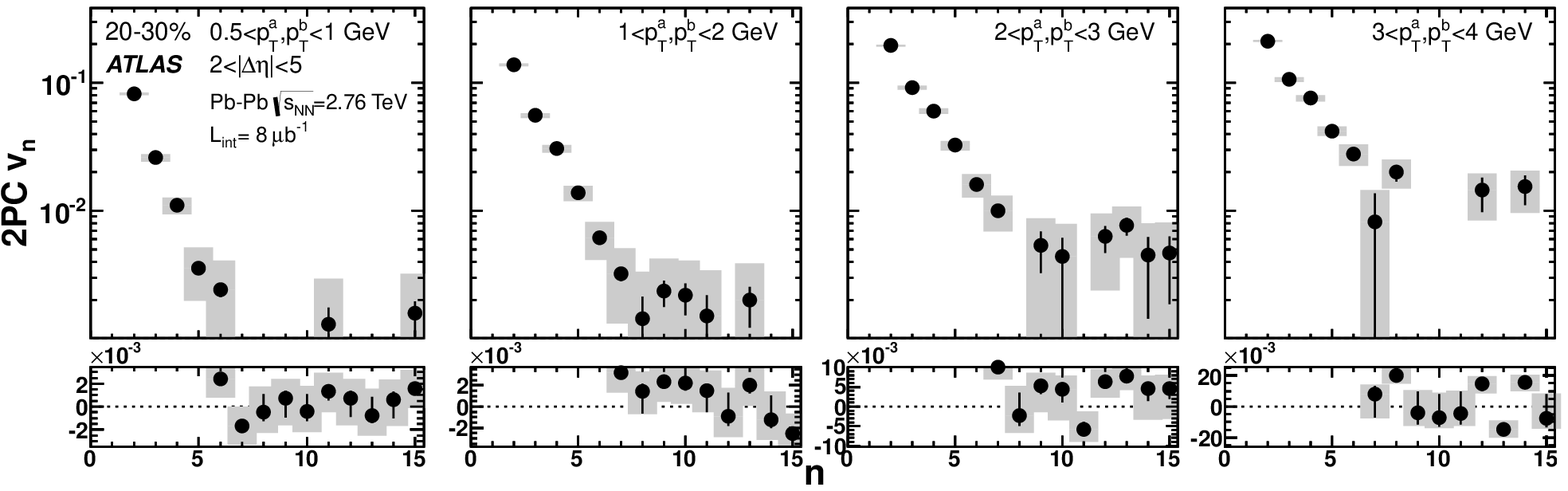}
\caption{\label{fig:ren5} $v_n$ vs. $n$ for $n\geq2$ in 20--30\% centrality interval for four fixed-$\pT$ correlations (0.5--1, 1--2, 2--3 and 3--4 GeV from left to right). The error bars and shaded bands indicate the statistical and total systematic uncertainties, respectively.}
\end{figure*}

\begin{figure*}[h]
\centering
\includegraphics[width=0.98\linewidth]{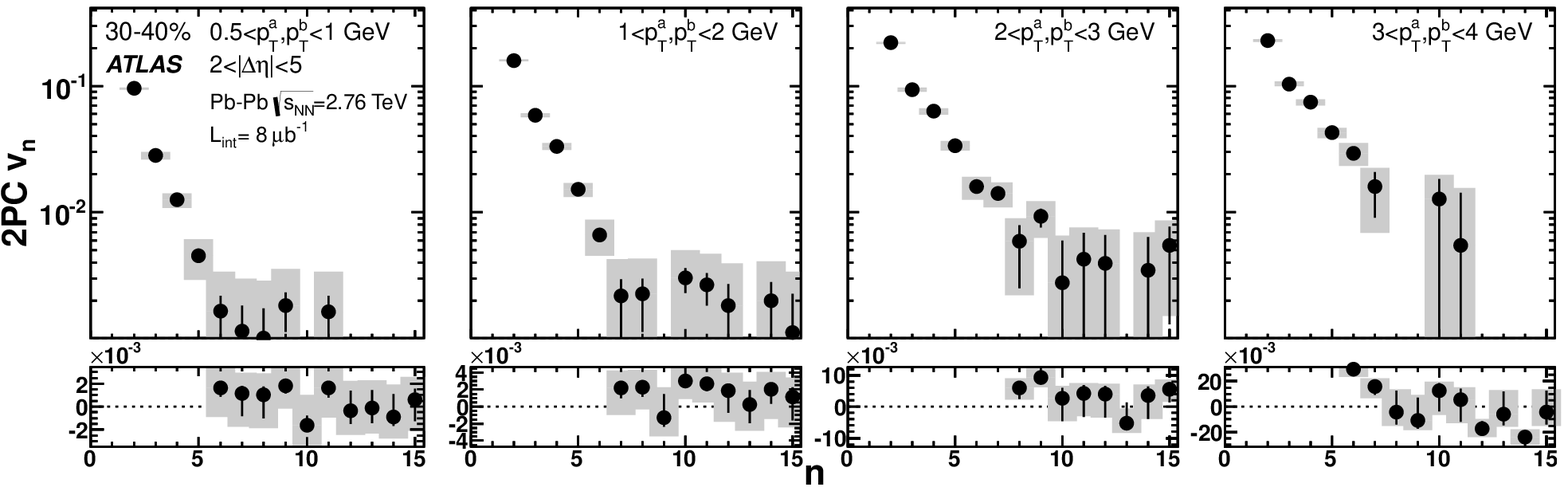}
\caption{\label{fig:ren6} $v_n$ vs. $n$ for $n\geq2$ in 30--40\% centrality interval for four fixed-$\pT$ correlations (0.5--1, 1--2, 2--3 and 3--4 GeV from left to right). The error bars and shaded bands indicate the statistical and total systematic uncertainties, respectively.}
\end{figure*}

\begin{figure*}[h]
\centering
\includegraphics[width=0.98\linewidth]{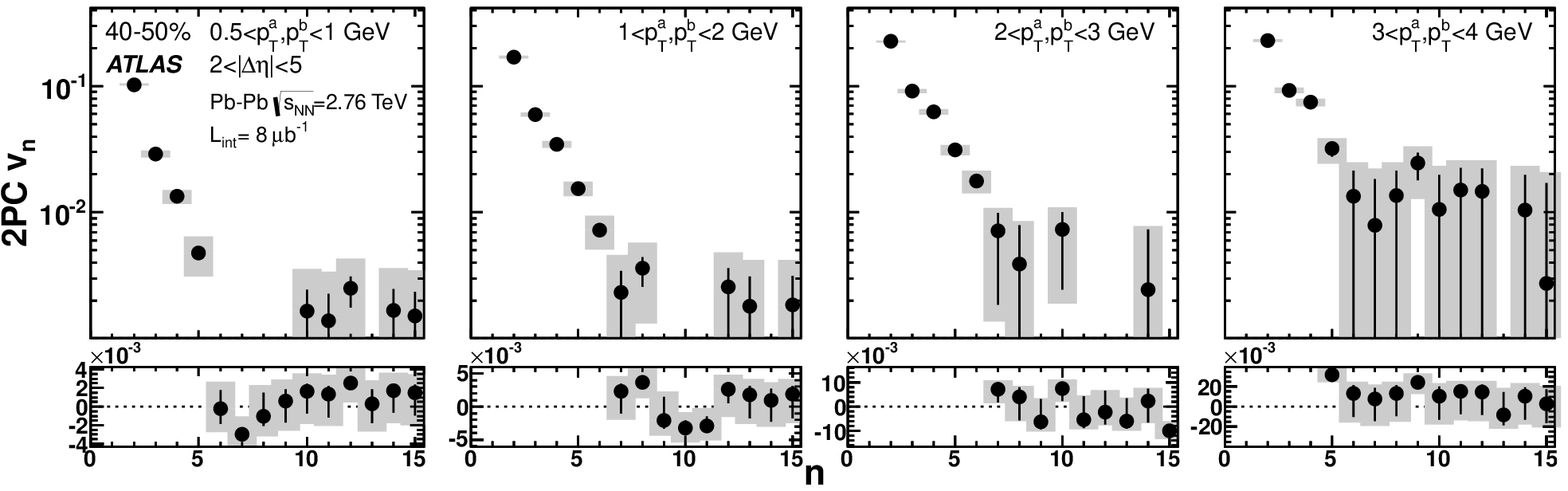}
\caption{\label{fig:ren7} $v_n$ vs. $n$ for $n\geq2$ in 40--50\% centrality interval for four fixed-$\pT$ correlations (0.5--1, 1--2, 2--3 and 3--4 GeV from left to right). The error bars and shaded bands indicate the statistical and total systematic uncertainties, respectively.}
\end{figure*}

Section~\ref{sec:re2} only presents the results from selected centrality or $\pT$ intervals. One of the primary goals of this paper is the detailed study in a wide range of centrality and $\pT$ of the $v_{n,n}$ and $v_n$, obtained from the 2PC in $\Delta\phi$ with a large pseudorapidity gap. For completeness, the $v_n$ coefficients for several fixed-$\pT$ correlations and for centrality intervals other than that shown in Fig.~\ref{fig:ren0} are presented in Figs.~\ref{fig:ren1}--\ref{fig:ren7}. 

\begin{figure*}[h]
\centering
\includegraphics[width=1\linewidth]{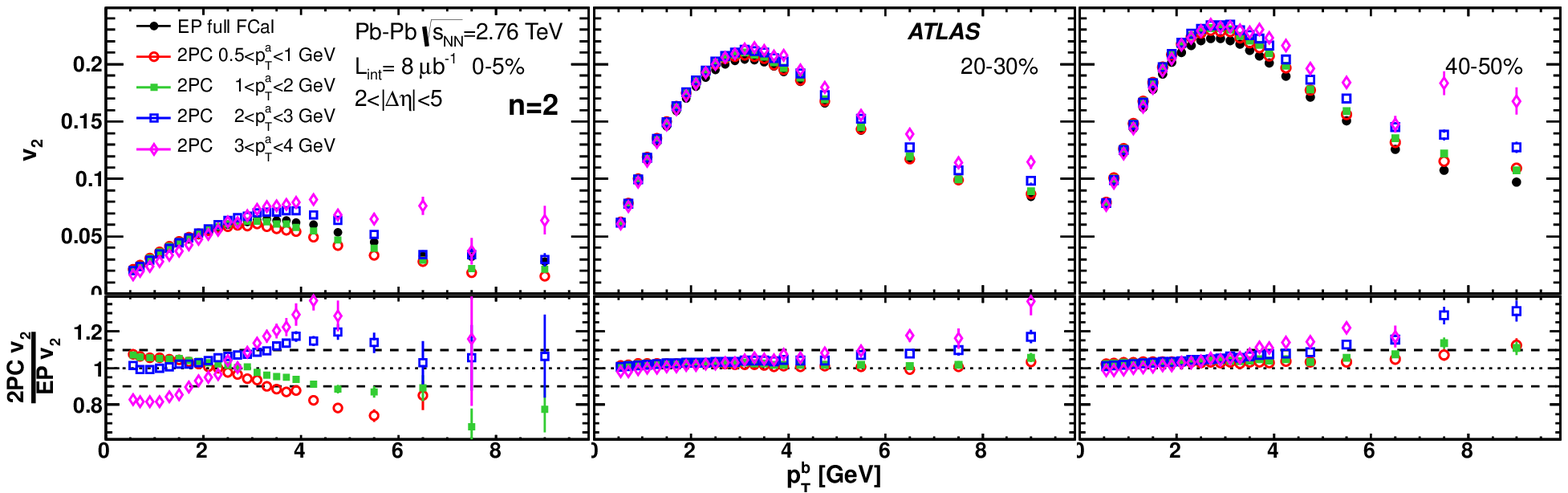}
\includegraphics[width=1\linewidth]{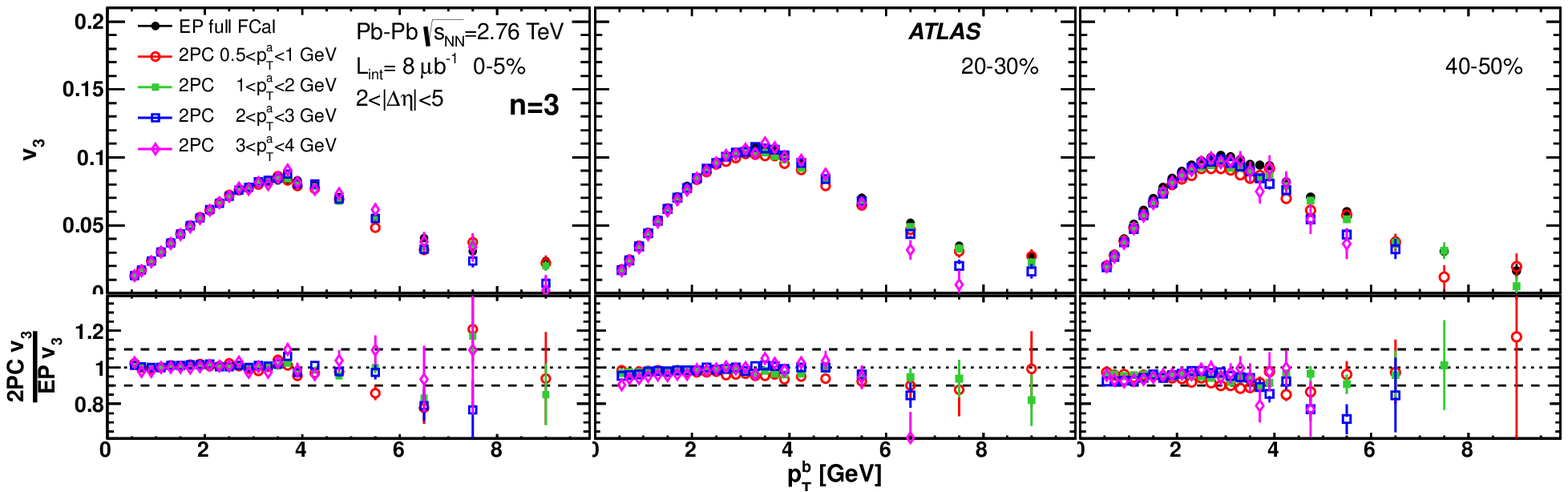}
\includegraphics[width=1\linewidth]{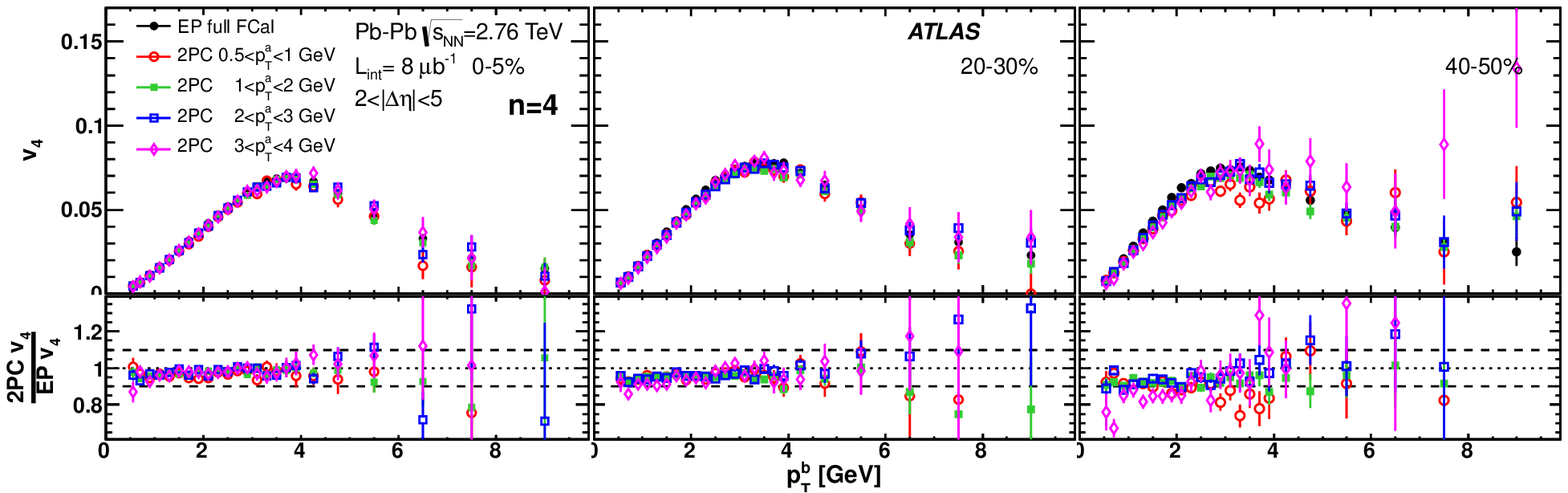}
\includegraphics[width=1\linewidth]{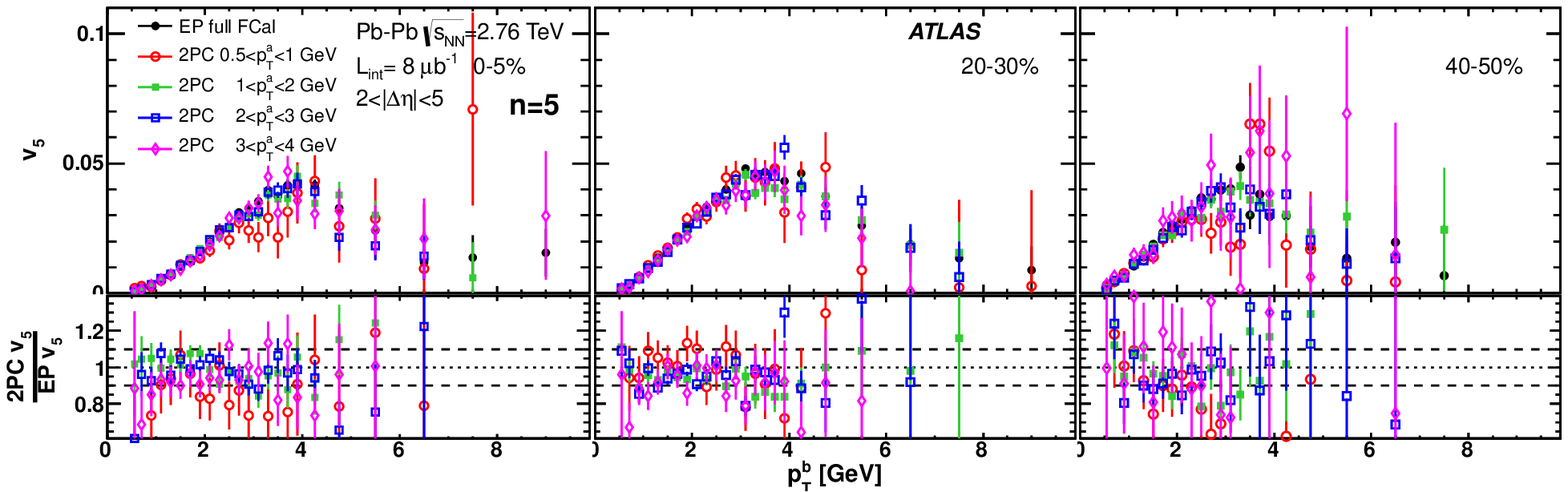}
\caption{\label{fig:disc33b} Comparison of $v_n$ as a function of $\pT$ ($v_2$--$v_5$ from top to bottom) between mixed-$\pT$ 2PC method and the EP method for 0--5\% (left panels), 20--30\% (middle panels) and 40--50\% (right panels) centrality intervals.  The error bars indicate the statistical uncertainties only. The dashed lines in the ratio plots indicate a $\pm10$\% band to guide eye.}
\end{figure*}

One important study of the factorization properties of $v_2$--$v_6$ is the comparison between the EP method and mixed-$\pT$ 2PC method. This comparison is shown in Fig.~\ref{fig:disc33} for the 0--10\% centrality interval. Figure~\ref{fig:disc33b} extends this comparison to several centrality intervals. Good consistency is observed for $v_2$--$v_6$ in non-central collisions.

\begin{sidewaysfigure*}[ht]
\vspace*{8cm}
 \includegraphics[width=1\textwidth]{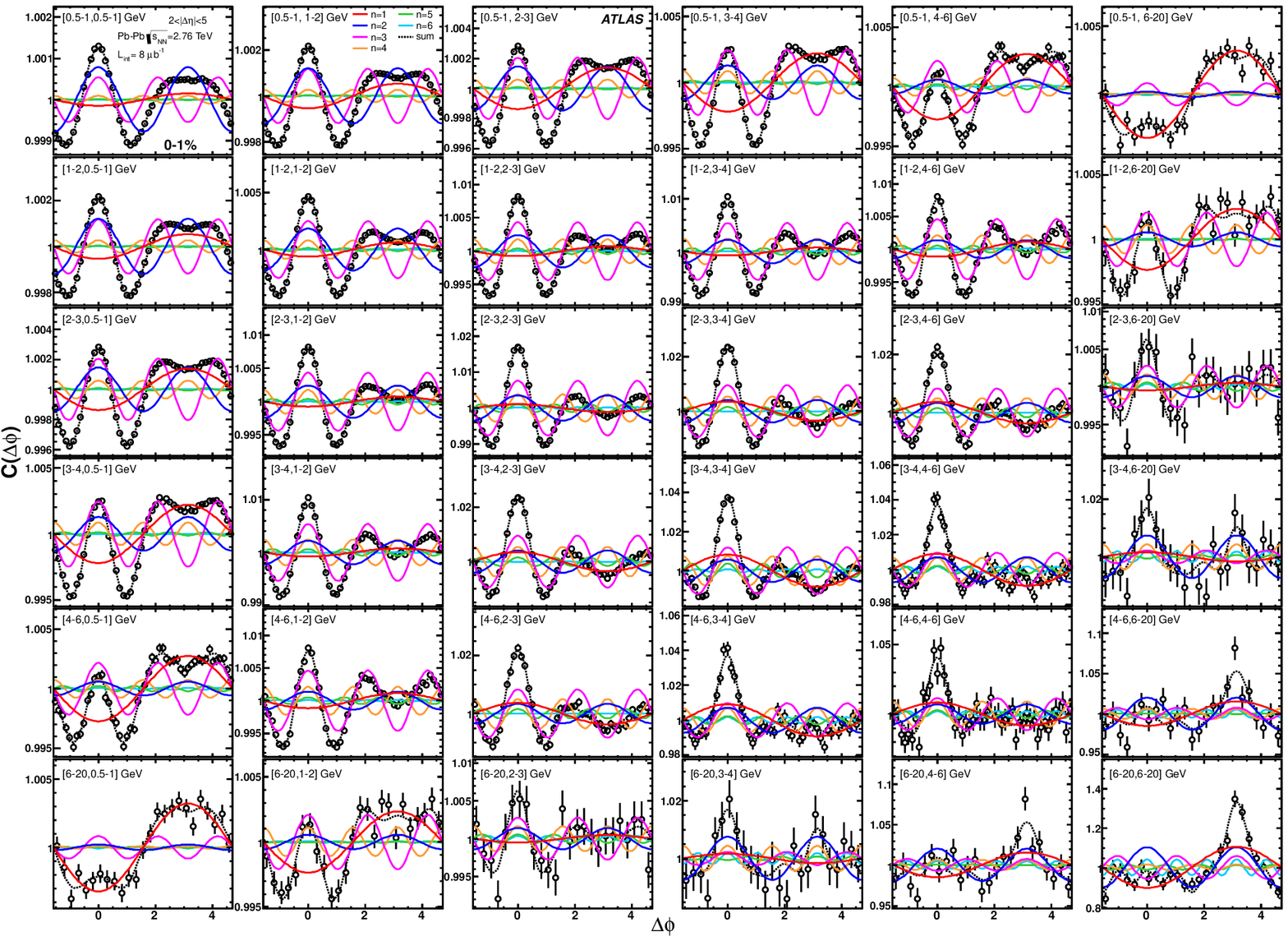}
 \caption{ \label{fig:land1} (Color online) $\Delta\phi$ correlation functions for various combinations of $\pT^{\mathrm a}$ and $\pT^{\mathrm b}$ for 0--1\% centrality interval. The superimposed solid lines (dashed lines) indicate the contributions from individual $v_{n,n}$ (sum of the first six components).}
\end{sidewaysfigure*}

\begin{sidewaysfigure*}[ht]
\vspace*{8cm}
 \includegraphics[width=1\textwidth]{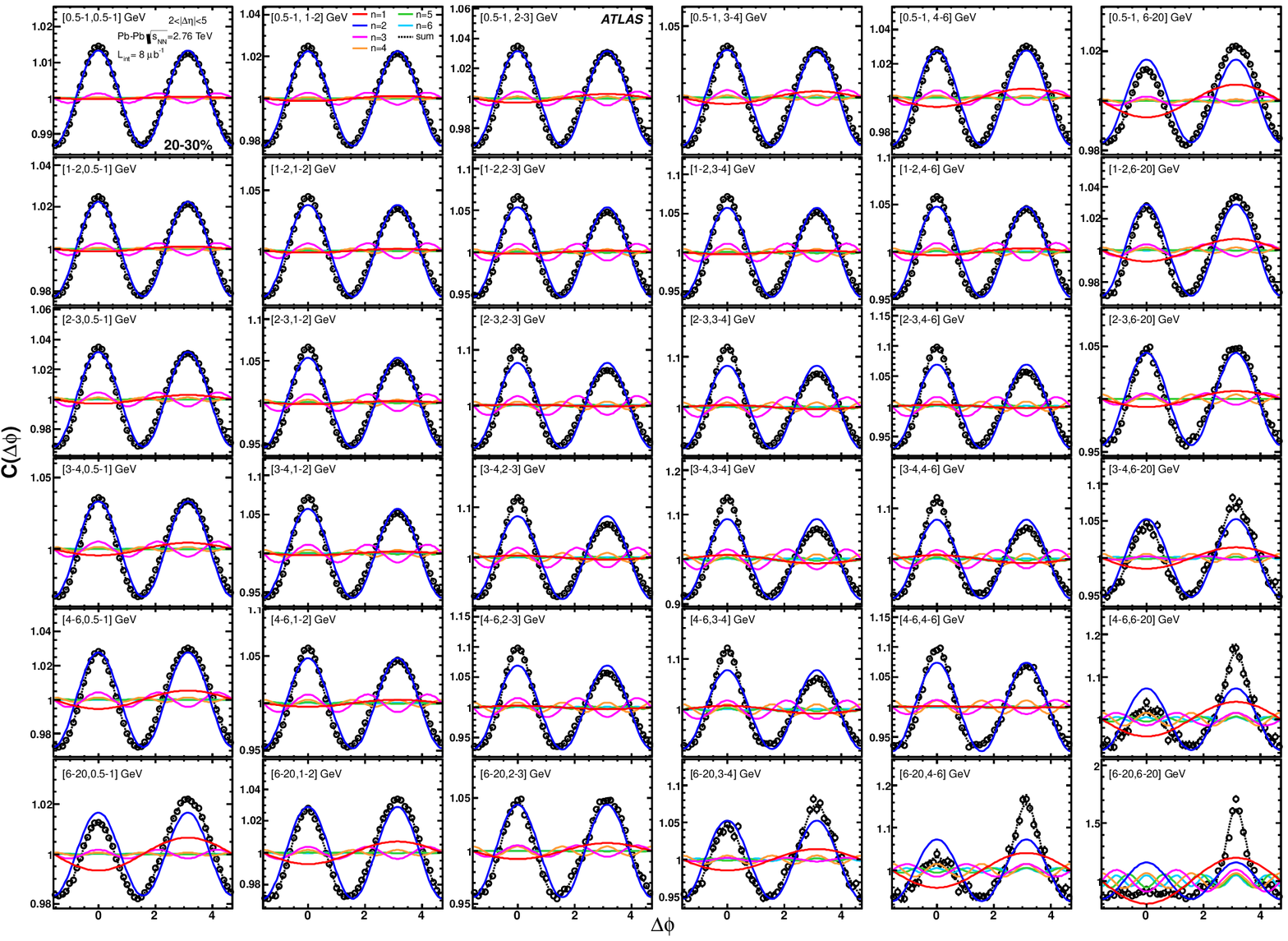}
 \caption{ \label{fig:land3} (Color online) $\Delta\phi$ correlation functions for various combinations of $\pT^{\mathrm a}$ and $\pT^{\mathrm b}$ for 20--30\% centrality interval. The superimposed solid lines (dashed lines) indicate the contributions from individual $v_{n,n}$ (sum of the first six components).}
\end{sidewaysfigure*}

\begin{sidewaysfigure*}[ht]
\vspace*{8cm}
 \includegraphics[width=1\textwidth]{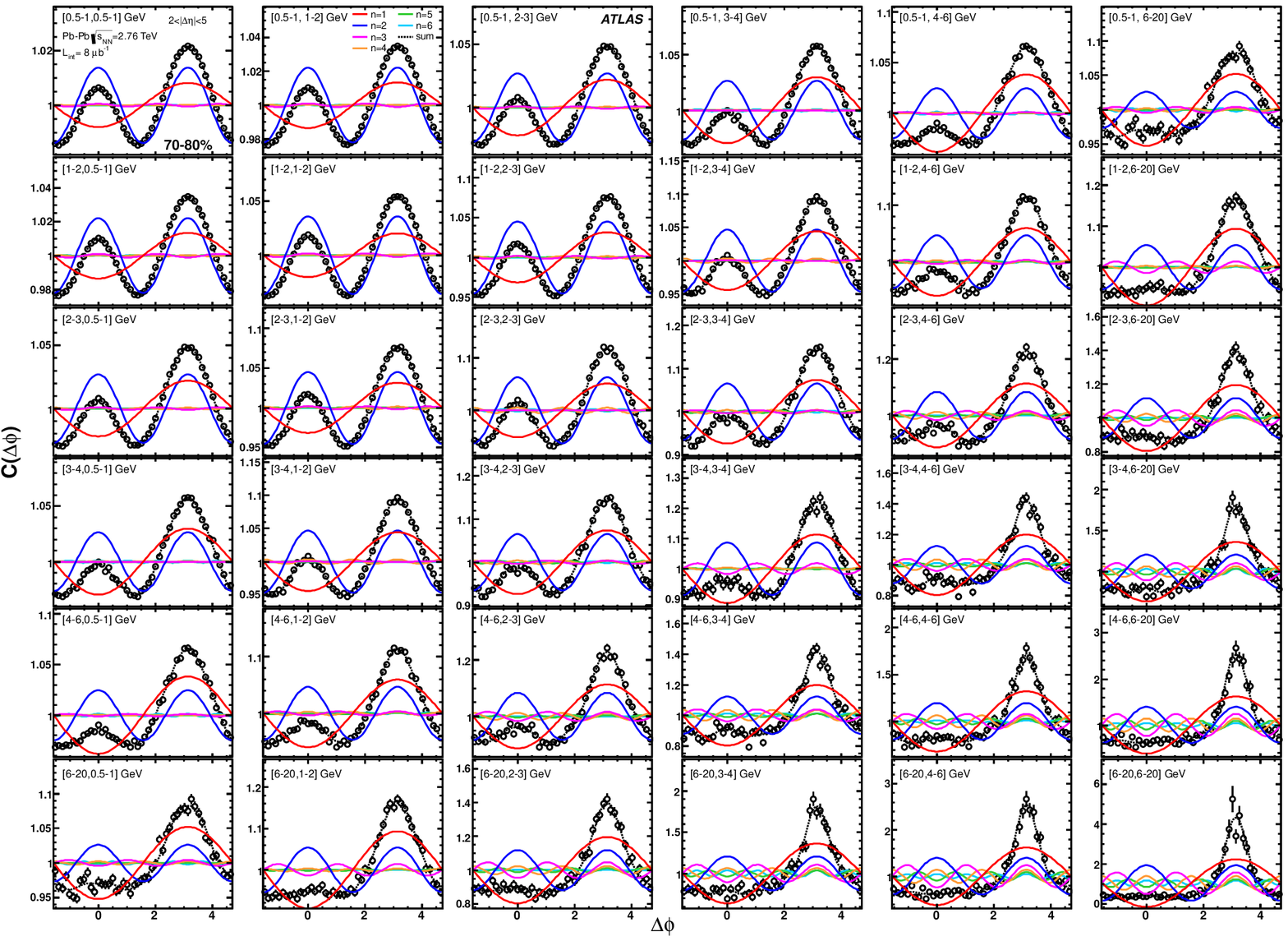}
 \caption{\label{fig:land6} (Color online) $\Delta\phi$ correlations for various combinations of $\pT^{\mathrm a}$ and $\pT^{\mathrm b}$ for 70--80\% centrality interval. The superimposed solid lines (dashed lines) indicate the contributions from individual $v_{n,n}$ (sum of the first six components).}
 \end{sidewaysfigure*}

Figures~\ref{fig:land1}--\ref{fig:land6} present the full set of the $\Delta\phi$ correlation functions with $2<|\Delta\eta|<5$ for one central, one mid-central and one peripheral centrality intervals. These correlation functions are the inputs for the $v_{n,n}$ and $v_n$ spectra. They provide the complete picture of how the structures of the correlation functions are influenced, as a function of $\pT^{\mathrm{a}}$ and $\pT^{\mathrm{b}}$, by the anisotropies related to the initial geometry and autocorrelations from the away-side jets, as well as how these influences map onto the strength and sign of the $v_{n,n}$ coefficients. Note that the correlation functions are symmetric with respect to $\pT^{\mathrm{a}}$ and $\pT^{\mathrm b}$. For example the correlation for $1<\pT^{\mathrm a}<2$ GeV and $2<\pT^{\mathrm b}<3$ GeV (denoted as [1--2,2--3] GeV) is the same as that for $2<\pT^{\mathrm a}<3$ GeV and $1<\pT^{\mathrm b}<2$ GeV (denoted as [2--3,1--2] GeV).

\begin{figure*}[h]
\includegraphics[width=1\linewidth]{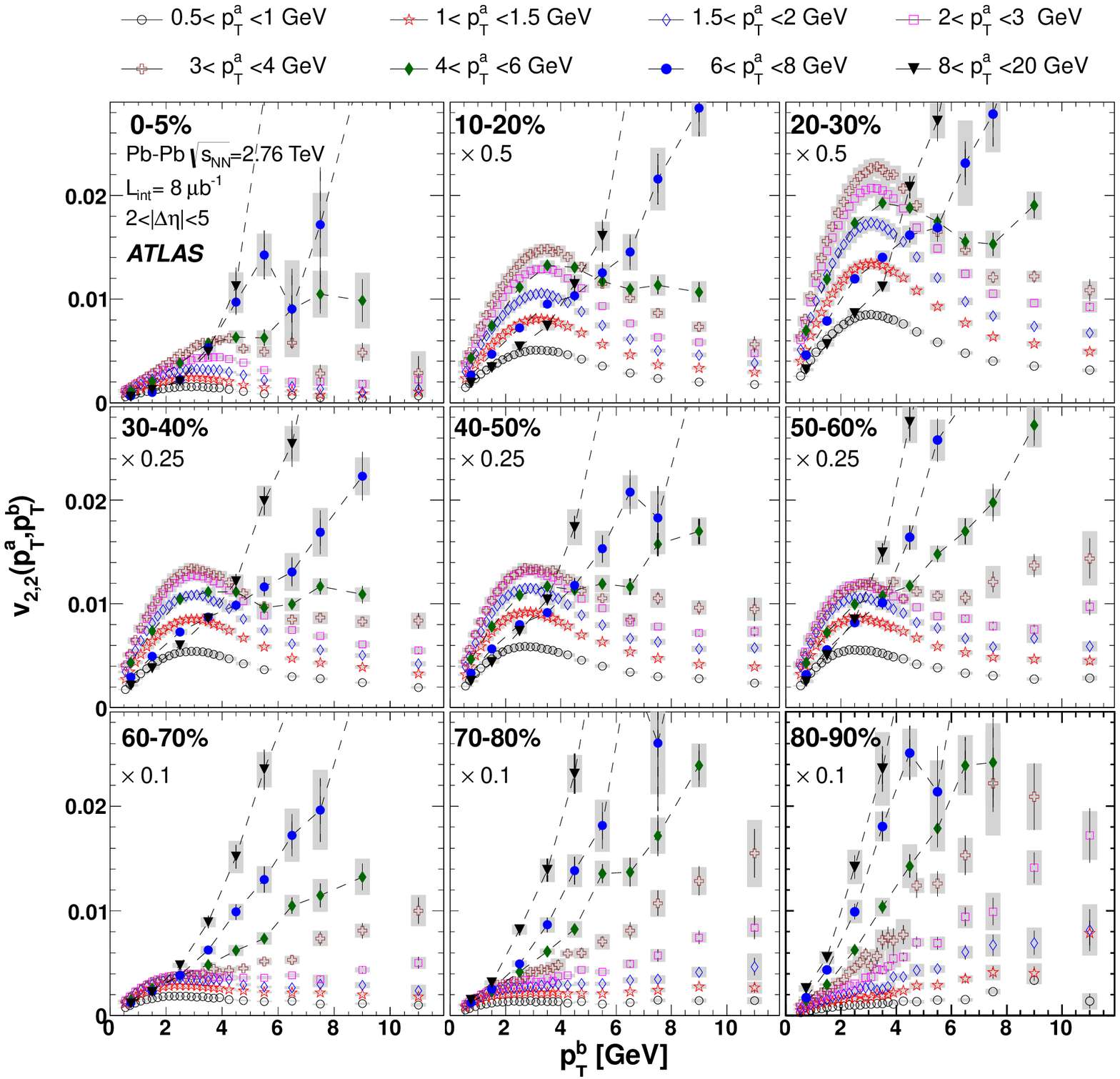}
\caption{\label{fig:v2fac} (Color online) $v_{2,2}(\pT^{\mathrm a},\pT^{\mathrm b})$ for $2<|\Delta\eta|<5$ vs. $\pT^{\mathrm b}$ for different $\pT^{\mathrm a}$ ranges. Each panel presents results in one centrality interval. The error bars and shaded bands represent statistical and systematic uncertainties, respectively. The data points for three highest $\pT^{\mathrm a}$ intervals have coarser binning in $\pT^{\mathrm b}$, hence are connected by dashed lines to guide the eye. The data in some panels are re-scaled to fit within the same vertical range.}
\end{figure*}

\begin{figure*}[t]
\includegraphics[width=1\linewidth]{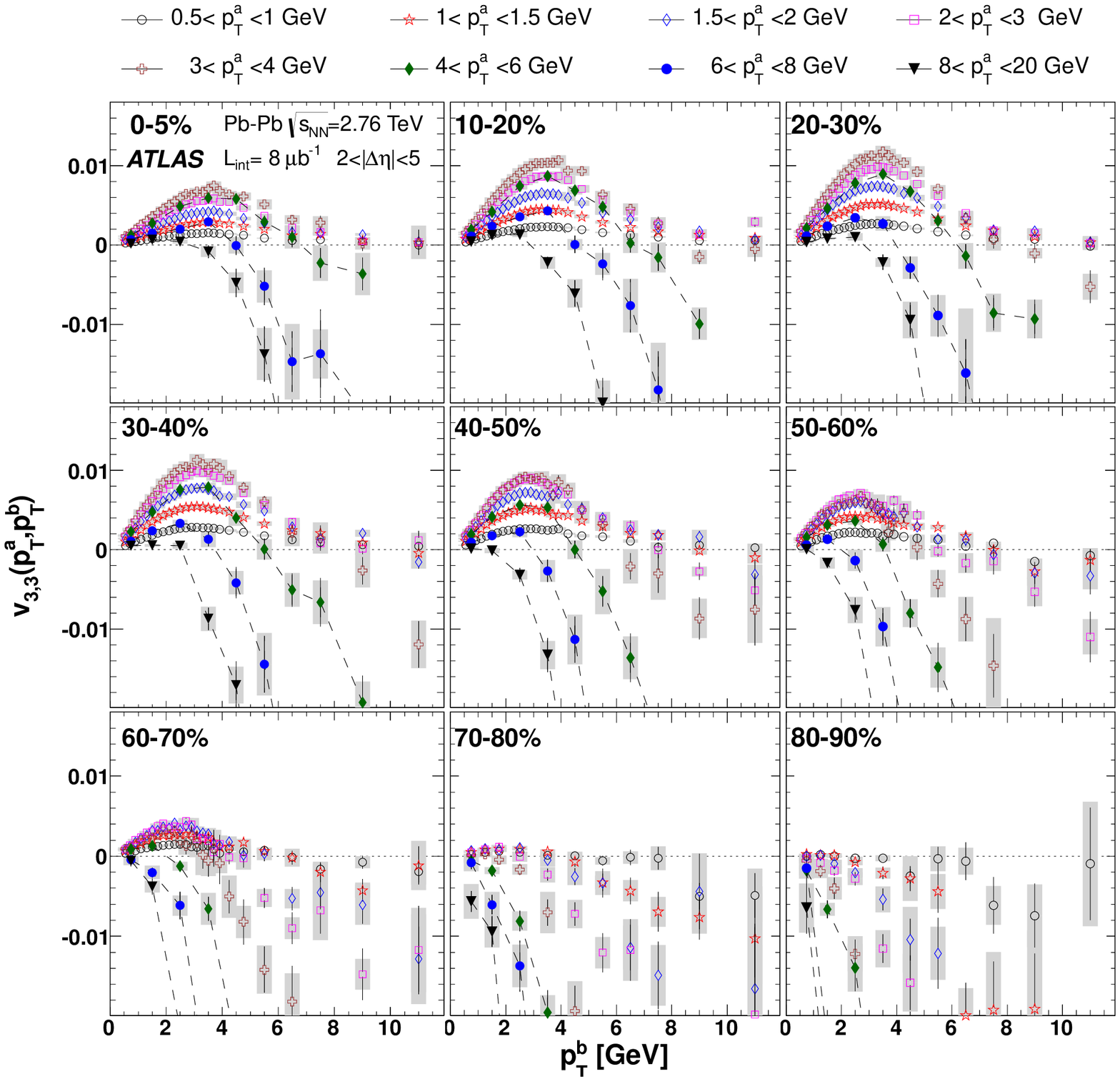}
\caption{\label{fig:v3fac} (Color online) $v_{3,3}(\pT^{\mathrm a},\pT^{\mathrm b})$ for $2<|\Delta\eta|<5$ vs. $\pT^{\mathrm b}$ for different $\pT^{\mathrm a}$ ranges. Each panel presents results in one centrality interval. The error bars and shaded bands represent statistical and systematic uncertainties, respectively. The data points for three highest $\pT^{\mathrm a}$ intervals have coarser binning in $\pT^{\mathrm b}$, hence are connected by dashed lines to guide the eye.}
\end{figure*}

\begin{figure*}[t]
\includegraphics[width=1\linewidth]{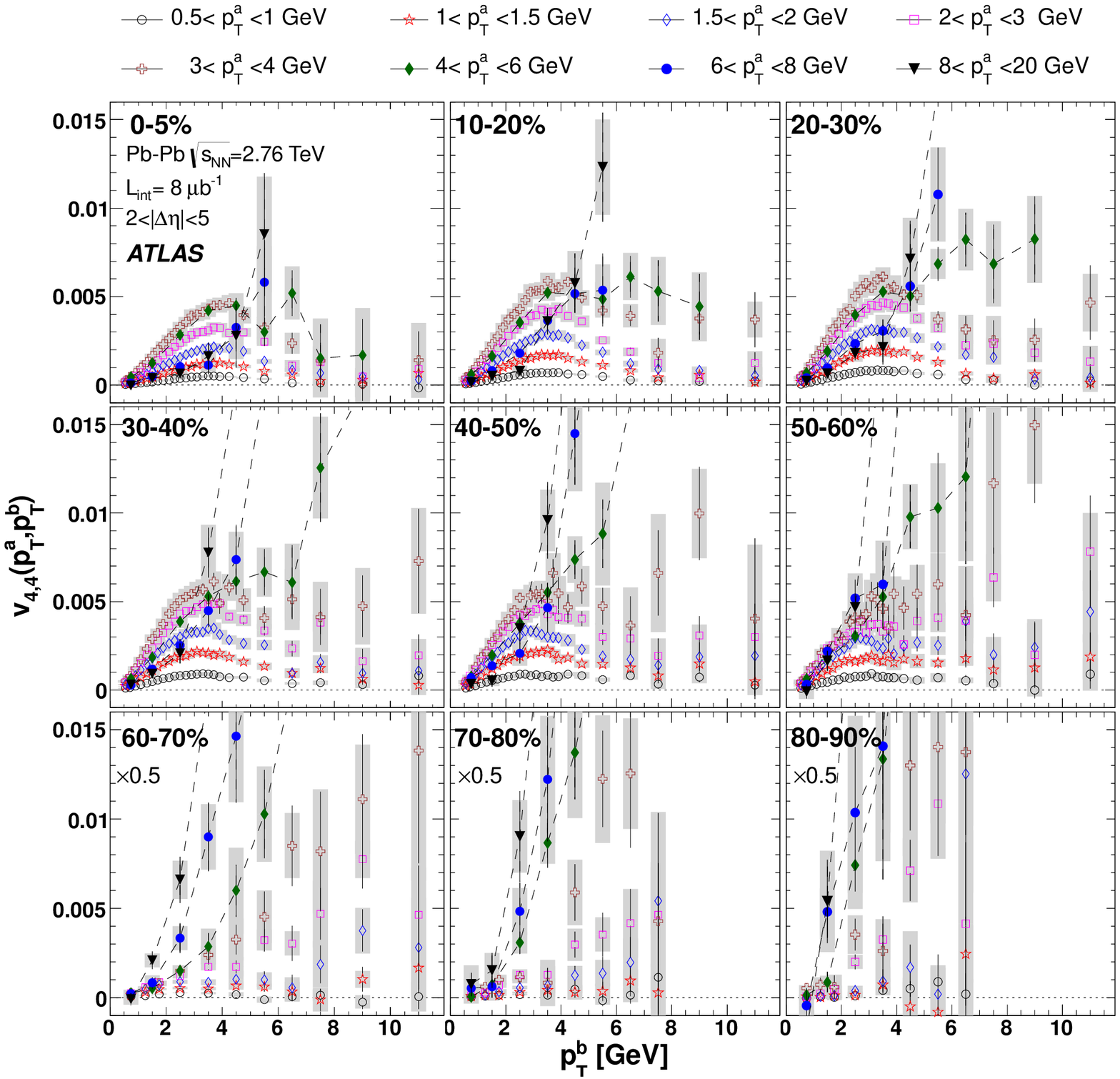}
\caption{\label{fig:v4fac} (Color online) $v_{4,4}(\pT^{\mathrm a},\pT^{\mathrm b})$ for $2<|\Delta\eta|<5$ vs. $\pT^{\mathrm b}$ for different $\pT^{\mathrm a}$ ranges. Each panel presents results in one centrality interval. The error bars and shaded bands represent statistical and systematic uncertainties, respectively. The data points for three highest $\pT^{\mathrm a}$ intervals have coarser binning in $\pT^{\mathrm b}$, hence are connected by dashed lines to guide the eye. The data in some panels are re-scaled to fit within the same vertical range.}
\end{figure*}

\begin{figure*}[h]
\centering
\includegraphics[width=0.83\linewidth]{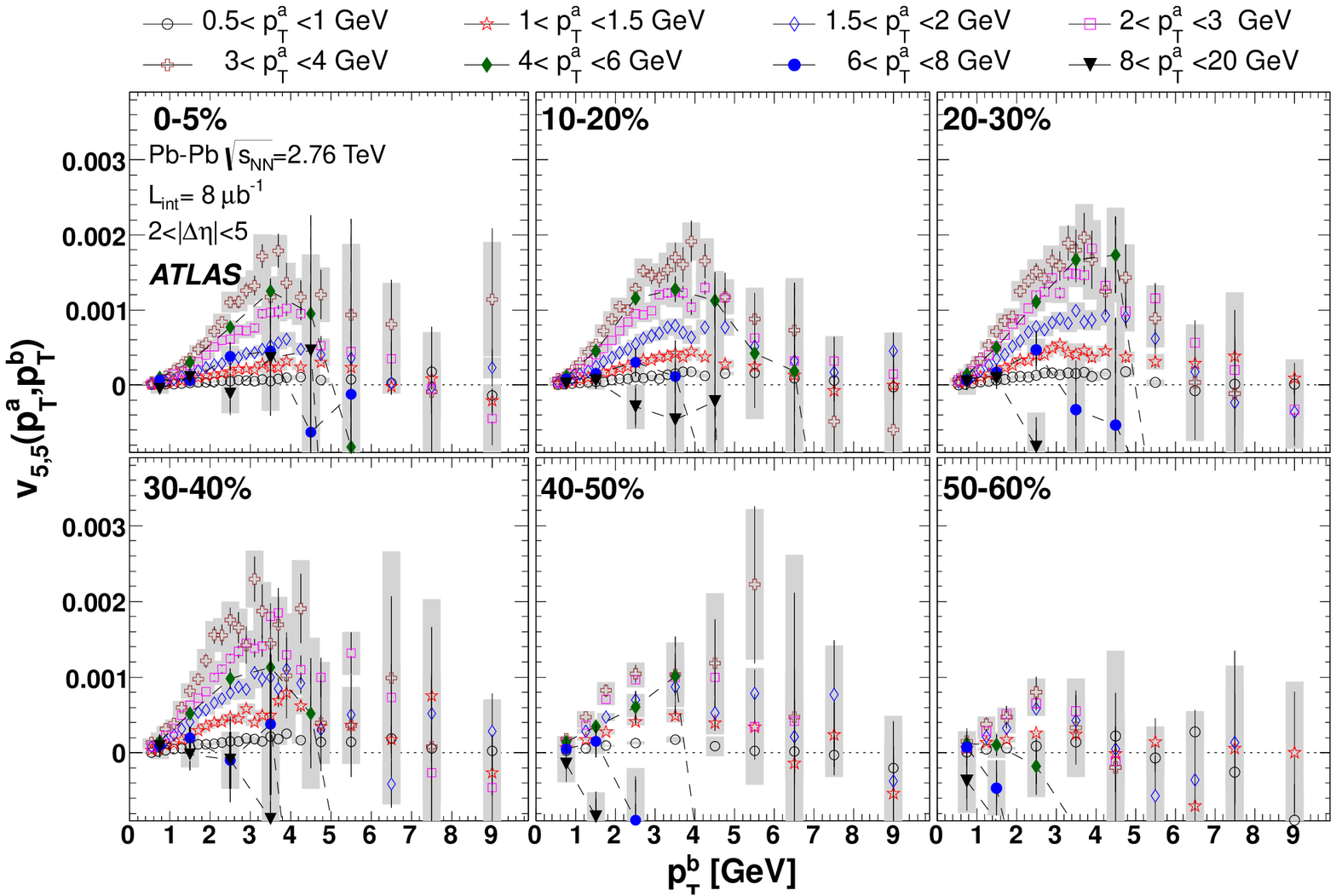}
\caption{\label{fig:v5fac} (Color online) $v_{5,5}(\pT^{\mathrm a},\pT^{\mathrm b})$ for $2<|\Delta\eta|<5$ vs. $\pT^{\mathrm b}$ for different $\pT^{\mathrm a}$ ranges. Each panel presents results in one centrality interval. The error bars and shaded bands represent statistical and systematic uncertainties, respectively. The data points for three highest $\pT^{\mathrm a}$ intervals have coarser binning in $\pT^{\mathrm b}$, hence are connected by dashed lines to guide the eye intervals.}
\end{figure*}

\begin{figure*}[h]
\centering
\includegraphics[width=0.83\linewidth]{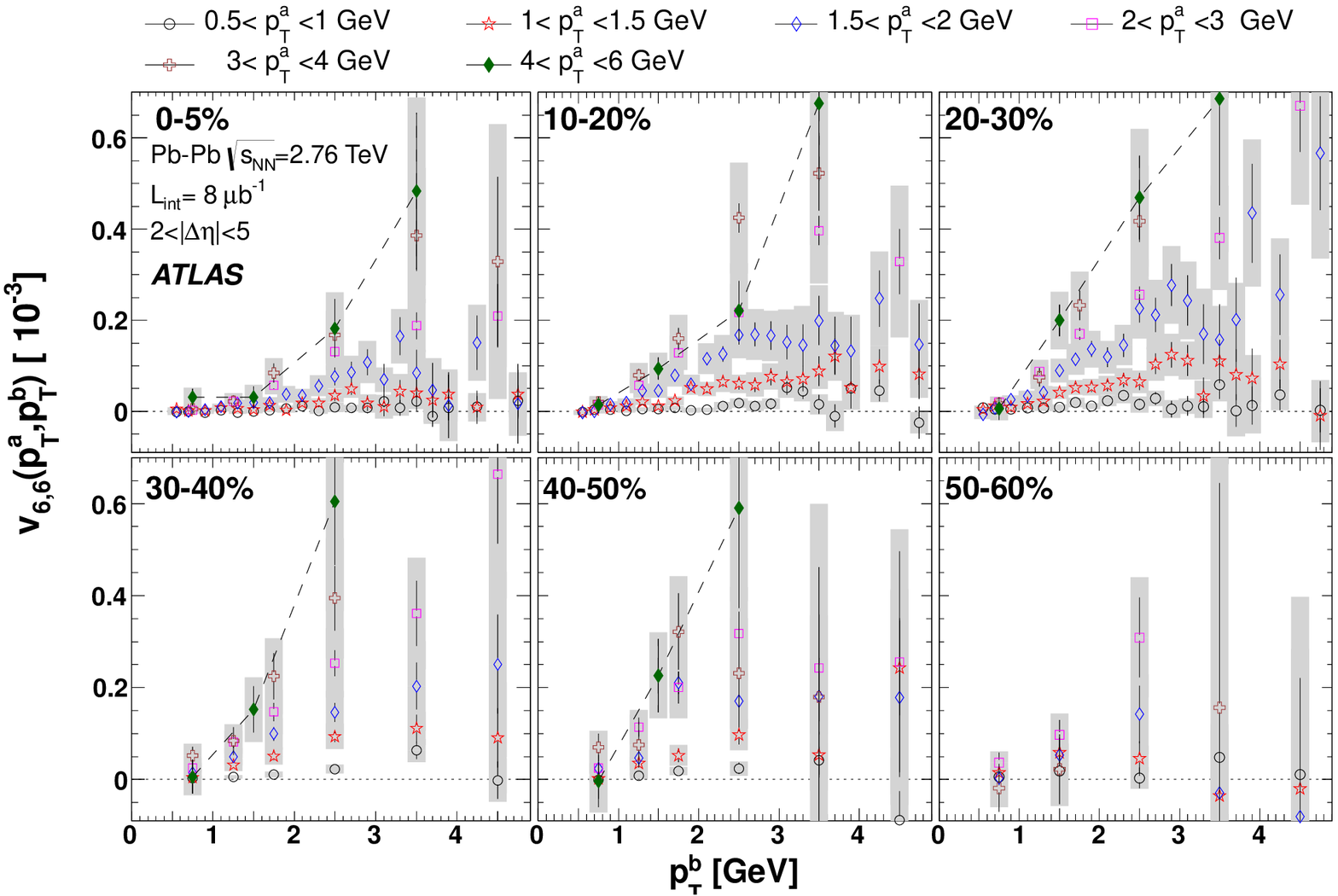}
\caption{\label{fig:v6fac} (Color online) $v_{6,6}(\pT^{\mathrm a},\pT^{\mathrm b})$ for $2<|\Delta\eta|<5$ vs. $\pT^{\mathrm b}$ for different $\pT^{\mathrm a}$ ranges. Each panel presents results in one centrality interval. The error bars and shaded bands represent statistical and systematic uncertainties, respectively. The data points for three highest $\pT^{\mathrm a}$ intervals have coarser binning in $\pT^{\mathrm b}$, hence are connected by dashed lines to guide the eye intervals.}
\end{figure*}

Figure~\ref{fig:v2fac}--\ref{fig:v6fac} shows the dependence of $v_{2,2}$--$v_{6,6}$ on $\pT^{\mathrm a}$ and $\pT^{\mathrm b}$. Together with Fig.~\ref{fig:v1fac3}, they provide a concise summary of $v_{n,n}$ spectra for all the 1-D correlation functions such as those in Fig.~\ref{fig:land1}--\ref{fig:land6} for $2<|\Delta\eta|<5$. Since the sum of the $v_{1,1}$--$v_{6,6}$ components almost exhaust the shape of the correlation functions, these data provide almost equivalent information as the correlation functions themselves over broad ranges of $\pT^{\mathrm{a}}$ and $\pT^{\mathrm{b}}$.




\onecolumngrid\clearpage
\input{atlas_authlist}
\end{document}

%% file: atlas_authlist.tex
\begin{flushleft}
{\Large The ATLAS Collaboration}

\bigskip

G.~Aad$^{\rm 48}$,
B.~Abbott$^{\rm 110}$,
J.~Abdallah$^{\rm 11}$,
S.~Abdel~Khalek$^{\rm 114}$,
A.A.~Abdelalim$^{\rm 49}$,
A.~Abdesselam$^{\rm 117}$,
O.~Abdinov$^{\rm 10}$,
B.~Abi$^{\rm 111}$,
M.~Abolins$^{\rm 87}$,
O.S.~AbouZeid$^{\rm 157}$
% ...
\bigskip

% ...
$^{10}$ Institute of Physics, Azerbaijan Academy of Sciences, Baku, Azerbaijan\\
$^{11}$ Institut de F\'isica d'Altes Energies and Departament de F\'isica de la Universitat Aut\`onoma  de Barcelona and ICREA, Barcelona, Spain\\
% ...
$^{48}$ Fakult\"{a}t f\"{u}r Mathematik und Physik, Albert-Ludwigs-Universit\"{a}t, Freiburg i.Br., Germany\\
$^{49}$ Section de Physique, Universit\'e de Gen\`eve, Geneva, Switzerland\\
% ...
$^{87}$ Department of Physics and Astronomy, Michigan State University, East Lansing MI, United States of America\\
% ...
$^{110}$ Homer L. Dodge Department of Physics and Astronomy, University of Oklahoma, Norman OK, United States of America\\
$^{111}$ Department of Physics, Oklahoma State University, Stillwater OK, United States of America\\
% ...
$^{114}$ LAL, Univ. Paris-Sud and CNRS/IN2P3, Orsay, France\\
% ...
$^{117}$ Department of Physics, Oxford University, Oxford, United Kingdom\\
% ...
$^{157}$ Department of Physics, University of Toronto, Toronto ON, Canada\\
% ...
\end{flushleft}
